\documentclass[onecolumn,showpacs,floatfix,10pt,nofootinbib]{revtex4}
\pdfoutput=1
\usepackage{epsfig}\usepackage{float}
\RequirePackage{amssymb,amsmath}
\usepackage{makeidx}
\usepackage[latin1]{inputenc}
\usepackage{graphicx}
\usepackage{indentfirst}
\usepackage{color}
\newcommand{\noi}{\noindent}
\expandafter\ifx\csname ProvidesPackage\endcsname\relax\toks0=\expandafter{%
\fi\ProvidesPackage{diagrams}[2014/12/31 v3.94 Paul Taylor's commutative
diagrams]
\toks0=\bgroup}
\ifx\diagram\isundefined\else\expandafter\endinput
\fi

\edef\cdrestoreat{
\noexpand\catcode`\noexpand\@=\the\catcode`\@
\noexpand\catcode`\noexpand\#=\the\catcode`\#
\noexpand\catcode`\noexpand\$=\the\catcode`\$
\noexpand\catcode`\noexpand\<=\the\catcode`\<
\noexpand\catcode`\noexpand\>=\the\catcode`\>
\noexpand\catcode`\noexpand\:=\the\catcode`\:
\noexpand\catcode`\noexpand\;=\the\catcode`\;
\noexpand\catcode`\noexpand\!=\the\catcode`\!
\noexpand\catcode`\noexpand\?=\the\catcode`\?
\noexpand\catcode`\noexpand\+=\the\catcode'53
}\catcode`\@=11 \catcode`\#=6 \catcode`\<=12 \catcode`\>=12 \catcode'53=12
\catcode`\:=12 \catcode`\;=12 \catcode`\!=12 \catcode`\?=12

\ifx\diagram@help@messages\CD@qK\let\diagram@help@messages y\fi

\def\cdps@Rokicki#1{\special{ps:#1}}\let\cdps@dvips\cdps@Rokicki\let
\cdps@RadicalEye\cdps@Rokicki\let\CD@HB\cdps@Rokicki\let\CD@IK\cdps@Rokicki
\let\CD@HB\cdps@Rokicki
\def\cdps@Bechtolsheim#1{\special{dvitps: Literal "#1"}}%
\let\cdps@dvitps\cdps@Bechtolsheim\let\cdps@IntegratedComputerSystems
\cdps@Bechtolsheim
\def\cdps@Clark#1{\special{dvitops: inline #1}}
\let\cdps@dvitops\cdps@Clark
\let\cdps@OzTeX\empty\let\cdps@oztex\empty\let\cdps@Trevorrow\empty
\def\cdps@Coombes#1{\special{ps-string #1}}

\count@=\year\multiply\count@12 \advance\count@\month
\ifnum\count@>24240 
\ifnum\count@>24243 
\errhelp{You may press RETURN and carry on for the time being.}\errmessage{! remain
compatible with your files. Please get it NOW.}\fi\fi

\def\CD@DE{\global\let}\def\CD@RH{\outer\def}

{\escapechar\m@ne\xdef\CD@o{\string\{}\xdef\CD@yC{\string\}}
\catcode`\&=4 \CD@DE\CD@Q=&\xdef\CD@S{\string\&}
\catcode`\$=3 \CD@DE\CD@k=$\CD@DE\CD@ND=$
\xdef\CD@nC{\string\$}\gdef\CD@LG{$$}
\catcode`\_=8 \CD@DE\CD@lJ=_
\obeylines\catcode`\^=7 \CD@DE\@super=^
\ifnum\newlinechar=10 \gdef\CD@uG{^^J}
\else\ifnum\newlinechar=13 \gdef\CD@uG{^^M}
\else\ifnum\newlinechar=-1 \gdef\CD@uG{^^J}
\else\CD@DE\CD@uG\space\expandafter\message{! input error: \noexpand
\newlinechar\space is ASCII \the\newlinechar, not LF=10 or CR=13.}
\fi\fi\fi}

\mathchardef\lessthan='30474 \mathchardef\greaterthan='30476

\ifx\tenln\CD@qK
\font\tenln=line10\relax
\fi\ifx\tenlnw\CD@qK\ifx\tenln\nullfont\let\tenlnw\nullfont\else
\font\tenlnw=linew10\relax
\fi\fi

\ifx\inputlineno\CD@qK\csname newcount\endcsname\inputlineno\inputlineno\m@ne
\fi

\def\cd@shouldnt#1{\CD@KB{* THIS (#1) SHOULD NEVER HAPPEN! *}}

\def\get@round@pair#1(#2,#3){#1{#2}{#3}}
\def\get@square@arg#1[#2]{#1{#2}}
\def\CD@AE#1{\CD@PK\let\CD@DH\CD@@E\CD@@E#1,],}
\def\CD@m{[}\def\CD@RD{]}\def\commdiag#1{{\let\enddiagram\relax\diagram[]#1%
\enddiagram}}

\def\CD@BF{{\ifx\CD@EH[\aftergroup\get@square@arg\aftergroup\CD@YH\else
\aftergroup\CD@JH\fi}}
\def\CD@CF#1#2{\def\CD@YH{#1}\def\CD@JH{#2}\futurelet\CD@EH\CD@BF}

\def\CD@KK{|}

\def\CD@PB{
\tokcase\CD@DD:\CD@y\break@args;\catcase\@super:\upper@label;\catcase\CD@lJ:%
\lower@label;\tokcase{~}:\middle@label;
\tokcase<:\CD@iF;
\tokcase>:\CD@iI;
\tokcase(:\CD@BC;
\tokcase[:\optional@;
\tokcase.:\CD@JJ;
\catcase\space:\eat@space;\catcase\bgroup:\positional@;\default:\CD@@A
\break@args;\endswitch}

\def\switch@arg{
\catcase\@super:\upper@label;\catcase\CD@lJ:\lower@label;\tokcase[:\optional@
;
\tokcase.:\CD@JJ;
\catcase\space:\eat@space;\catcase\bgroup:\positional@;\tokcase{~}:%
\middle@label;
\default:\CD@y\break@args;\endswitch}

\let\CD@tJ\relax\ifx\protect\CD@qK\let\protect\relax\fi\ifx\AtEndDocument
\CD@qK\def\CD@PG{\CD@gB}\def\CD@GF#1#2{}\else\def\CD@PG#1{\edef\CD@CH{#1}%
\expandafter\CD@oC\CD@CH\CD@OD}\def\CD@oC#1\CD@OD{\AtEndDocument{\typeout{%
\CD@tA: #1}}}\def\CD@GF#1#2{\gdef#1{#2}\AtEndDocument{#1}}\fi\def\CD@ZA#1#2{%
\def#1{\CD@PG{#2\CD@mD\CD@W}\CD@DE#1\relax}}\def\CD@uF#1\repeat{\def\CD@p{#1}%
\CD@OF}\def\CD@OF{\CD@p\relax\expandafter\CD@OF\fi}\def\CD@sF#1\repeat{\def
\CD@q{#1}\CD@PF}\def\CD@PF{\CD@q\relax\expandafter\CD@PF\fi}\def\CD@tF#1%
\repeat{\def\CD@r{#1}\CD@QF}\def\CD@QF{\CD@r\relax\expandafter\CD@QF\fi}\def
\CD@tG#1#2#3{\def#2{\let#1\iftrue}\def#3{\let#1\iffalse}#3}\if y%
\diagram@help@messages\def\CD@rG#1#2{\csname newtoks\endcsname#1#1=%
\expandafter{\csname#2\endcsname}}\else\csname newtoks\endcsname\no@cd@help
\no@cd@help{See the manual}\def\CD@rG#1#2{\let#1\no@cd@help}\fi\chardef\CD@lF
=1 \chardef\CD@lI=2 \chardef\CD@MH=5 \chardef\CD@tH=6 \chardef\CD@sH=7
\chardef\CD@PC=9 \dimendef\CD@hI=2 \dimendef\CD@hF=3 \dimendef\CD@mF=4
\dimendef\CD@mI=5 \dimendef\CD@wJ=6 \dimendef\CD@tI=8 \dimendef\CD@sI=9
\skipdef\CD@uB=1 \skipdef\CD@NF=2 \skipdef\CD@tB=3 \skipdef\CD@ZE=4 \skipdef
\countdef\CD@JC=9 \countdef\CD@gD=8 \countdef\CD@A=7 \def\sdef#1#2{\def#1{#2}%
}\def\CD@L#1{\expandafter\aftergroup\csname#1\endcsname}\def\CD@RC#1{%
\expandafter\def\csname#1\endcsname}\def\CD@sD#1{\expandafter\gdef\csname#1%
\endcsname}\def\CD@vC#1{\expandafter\edef\csname#1\endcsname}\def\CD@nF#1#2{%
\expandafter\let\csname#1\expandafter\endcsname\csname#2\endcsname}\def\CD@EE
#1#2{\expandafter\CD@DE\csname#1\expandafter\endcsname\csname#2\endcsname}%
\def\CD@AK#1{\csname#1\endcsname}\def\CD@XJ#1{\expandafter\show\csname#1%
\endcsname}\def\CD@ZJ#1{\expandafter\showthe\csname#1\endcsname}\def\CD@WJ#1{%
\expandafter\showbox\csname#1\endcsname}\def\CD@tA{Commutative Diagram}\edef
\CD@kH{\string\par}\edef\CD@dC{\string\diagram}\edef\CD@HD{\string\enddiagram
}\edef\CD@EC{\string\\}\def\CD@eF{LaTeX}\ifx\@ignoretrue\CD@qK\expandafter
\CD@tG\csname if@ignore\endcsname\ignore@true\ignore@false\def\@ignoretrue{%
\global\ignore@true}\def\@ignorefalse{\global\ignore@false}\fi

\def\CD@g{{\ifnum0=`}\fi}\def\CD@wC{\ifnum0=`{\fi}}\def\catcase#1:{\ifcat
\noexpand\CD@EH#1\CD@tJ\expandafter\CD@kC\else\expandafter\CD@dJ\fi}\def
\tokcase#1:{\ifx\CD@EH#1\CD@tJ\expandafter\CD@kC\else\expandafter\CD@dJ\fi}%
\def\CD@kC#1;#2\endswitch{#1}\def\CD@dJ#1;{}\let\endswitch\relax\def\default:%
#1;#2\endswitch{#1}\ifx\at@\CD@qK\def\at@{@}\fi\edef\CD@P{\CD@o pt\CD@yC}%
\CD@RC{\CD@P>}#1>#2>{\CD@z\rTo\sp{#1}\sb{#2}\CD@z}\CD@RC{\CD@P<}#1<#2<{\CD@z
\lTo\sp{#1}\sb{#2}\CD@z}\CD@RC{\CD@P)}#1)#2){\CD@z\rTo\sp{#1}\sb{#2}\CD@z}%
\CD@RC{\CD@P(}#1(#2({\CD@z\lTo\sp{#1}\sb{#2}\CD@z}
\def\CD@O{\def\endCD{\enddiagram}\CD@RC{\CD@P A}##1A##2A{\uTo<{##1}>{##2}%
\CD@z\CD@z}\CD@RC{\CD@P V}##1V##2V{\dTo<{##1}>{##2}\CD@z\CD@z}\CD@RC{\CD@P=}{%
\CD@z\hEq\CD@z}\CD@RC{\CD@P\CD@KK}{\vEq\CD@z\CD@z}\CD@RC{\CD@P\string\vert}{%
\vEq\CD@z\CD@z}\CD@RC{\CD@P.}{\CD@z\CD@z}\let\CD@z\CD@Q}\def\CD@IE{\let\tmp
\CD@JE\ifcat A\noexpand\CD@CH\else\ifcat=\noexpand\CD@CH\else\ifcat\relax
\noexpand\CD@CH\else\let\tmp\at@\fi\fi\fi\tmp}\def\CD@JE#1{\CD@nF{tmp}{\CD@P
\string#1}\ifx\tmp\relax\def\tmp{\at@#1}\fi\tmp}\def\CD@z{}\begingroup
\aftergroup\def\aftergroup\CD@T\aftergroup{\aftergroup\def\catcode`\@\active
\aftergroup @\endgroup{\futurelet\CD@CH\CD@IE}}\newcount\CD@uA\newcount\CD@vA
\newcount\CD@wA\newcount\CD@xA\newdimen\CD@OA\newdimen\CD@PA\CD@tG\CD@gE
\CD@@A\CD@y\CD@tG\CD@hE\CD@EA\CD@BA\newdimen\CD@RA\newdimen\CD@SA\newcount
\CD@yA\newcount\CD@zA\newdimen\CD@QA\newbox\CD@DA\CD@tG\CD@lE\CD@dA\CD@bA
\newcount\CD@LH\newcount\CD@TC\def\CD@V#1#2{\ifdim#1<#2\relax#1=#2\relax\fi}%
\def\CD@X#1#2{\ifdim#1>#2\relax#1=#2\relax\fi}\newdimen\CD@XH\CD@XH=1sp
\newdimen\CD@zC\CD@zC\z@\def\CD@cJ{\ifdim\CD@zC=1em\else\CD@nJ\fi}\def\CD@nJ{%
\CD@zC1em\def\CD@NC{\fontdimen8\textfont3 }\CD@@J\CD@NJ\setbox0=\vbox{\CD@t
\noindent\CD@k\null\penalty-9993\null\CD@ND\null\endgraf\setbox0=\lastbox
\unskip\unpenalty\setbox1=\lastbox\global\setbox\CD@IG=\hbox{\unhbox0\unskip
\unskip\unpenalty\setbox0=\lastbox}\global\setbox\CD@KG=\hbox{\unhbox1\unskip
\unpenalty\setbox1=\lastbox}}}\newdimen\CD@@I\CD@@I=1true in \divide\CD@@I300
\def\CD@zH#1{\multiply#1\tw@\advance#1\ifnum#1<\z@-\else+\fi\CD@@I\divide#1%
\tw@\divide#1\CD@@I\multiply#1\CD@@I}\def\MapBreadth{\afterassignment\CD@gI
\CD@LF}\newdimen\CD@LF\newdimen\CD@oI\def\CD@gI{\CD@oI\CD@LF\CD@V\CD@@I{4%
\CD@XH}\CD@X\CD@@I\p@\CD@zH\CD@oI\ifdim\CD@LF>\z@\CD@V\CD@oI\CD@@I\fi\CD@cJ}%
\def\CD@RJ#1{\CD@zD\count@\CD@@I#1\ifnum\count@>\z@\divide\CD@@I\count@\fi
\CD@gI\CD@NJ}\def\CD@NJ{\dimen@\CD@QC\count@\dimen@\divide\count@5\divide
\count@\CD@@I\edef\CD@OC{\the\count@}}\def\CD@AJ{\CD@QJ\z@}\def\CD@QJ#1{%
\CD@tI\axisheight\advance\CD@tI#1\relax\advance\CD@tI-.5\CD@oI\CD@zH\CD@tI
\CD@sI-\CD@tI\advance\CD@tI\CD@LF}\newdimen\CD@DC\CD@DC\z@\newdimen\CD@eJ
\CD@eJ\z@\def\CD@CJ#1{\CD@sI#1\relax\CD@tI\CD@sI\advance\CD@tI\CD@LF\relax}%
\def\horizhtdp{height\CD@tI depth\CD@sI}\def\axisheight{\fontdimen22\the
\textfont\tw@}\def\script@axisheight{\fontdimen22\the\scriptfont\tw@}\def
\ss@axisheight{\fontdimen22\the\scriptscriptfont\tw@}\def\CD@NC{0.4pt}\def
\CD@VK{\fontdimen3\textfont\z@}\def\CD@UK{\fontdimen3\textfont\z@}\newdimen
\PileSpacing\newdimen\CD@nA\CD@nA\z@\def\CD@RG{\ifincommdiag1.3em\else2em\fi}%
\newdimen\CD@YB\def\CellSize{\afterassignment\CD@kB\DiagramCellHeight}%
\newdimen\DiagramCellHeight\DiagramCellHeight-\maxdimen\newdimen
\DiagramCellWidth\DiagramCellWidth-\maxdimen\def\CD@kB{\DiagramCellWidth
\DiagramCellHeight}\def\CD@QC{3em}\newdimen\MapShortFall\def\MapsAbut{%
\MapShortFall\z@\objectheight\z@\objectwidth\z@}\newdimen\CD@iA\CD@iA\z@
\CD@tG\CD@vE\CD@aB\CD@ZB\expandafter\ifx\expandafter\iftrue\csname
ifUglyObsoleteDiagrams\endcsname\CD@ZB\else\CD@aB\fi\CD@nF{%
ifUglyObsoleteDiagrams}{relax}\newif\ifUglyObsoleteDiagrams\def\CD@nK{\CD@aB
\UglyObsoleteDiagramsfalse}\def\CD@oK{\CD@ZB\UglyObsoleteDiagramstrue}\CD@vE
\CD@nK\else\CD@oK\fi\CD@tG\CD@hK\CD@dK\CD@cK\CD@cK\def\CD@sK{\ifx\pdfoutput
\CD@qK\else\ifx\pdfoutput\relax\else\ifnum\pdfoutput>\z@\CD@pK\fi\fi\fi} \def
\CD@pK{\global\CD@dK\global\CD@aB\global\UglyObsoleteDiagramsfalse\global\let
\CD@n\empty\global\let\CD@oK\relax\global\let\CD@pK\relax\global\let\CD@sK
\relax}\def\CD@tK#1{}\ifx\pdfliteral\CD@qK\else\ifx
\pdfliteral\relax\else\let\CD@tK\pdfliteral\fi\fi\ifx\XeTeXrevision\CD@qK
\else\ifx\XeTeXrevision\relax\else\ifdim\XeTeXrevision pt<.996pt \expandafter
\message{! XeTeX version \XeTeXrevision\space does not support PDF literals,
so diagonals will not work!}\else\expandafter\message{RUNNING UNDER XETEX
\XeTeXrevision}\CD@pK\fi\fi\fi\CD@sK\def\newarrowhead{\CD@mG h\CD@BG\CD@GG>}%
\def\newarrowtail{\CD@mG t\CD@BG\CD@GG>}\def\newarrowmiddle{\CD@mG m\CD@BG
\hbox@maths\empty}\def\newarrowfiller{\CD@mG f\CD@bE\CD@MK-}\def\CD@mG#1#2#3#%
4#5#6#7#8#9{\CD@RC{r#1:#5}{#2{#6}}\CD@RC{l#1:#5}{#2{#7}}\CD@RC{d#1:#5}{#3{#8}%
}\CD@RC{u#1:#5}{#3{#9}}\CD@vC{-#1:#5}{\expandafter\noexpand\csname-#1:#4%
\endcsname\noexpand\CD@MC}\CD@vC{+#1:#5}{\expandafter\noexpand\csname+#1:#4%
\endcsname\noexpand\CD@MC}}\CD@ZA\CD@MC{\CD@eF\space diagonals are used unless
PostScript is set}\def\defaultarrowhead#1{\edef\CD@sJ{#1}\CD@@J}\def\CD@@J{%
\CD@IJ\CD@sJ<>ht\CD@IJ\CD@sJ<>th}\def\CD@IJ#1#2#3#4#5{\CD@HJ{r#4}{#3}{l#5}{#2%
}{r#4:#1}\CD@HJ{r#5}{#2}{l#4}{#3}{l#4:#1}\CD@HJ{d#4}{#3}{u#5}{#2}{d#4:#1}%
\CD@HJ{d#5}{#2}{u#4}{#3}{u#4:#1}}\def\CD@HJ#1#2#3#4#5{\begingroup\aftergroup
\CD@GJ\CD@L{#1+:#2}\CD@L{#1:#2}\CD@L{#3:#4}\CD@L{#5}\endgroup}\def\CD@GJ#1#2#%
3#4{\csname newbox\endcsname#1\def#2{\copy#1}\def#3{\copy#1}\setbox#1=\box
\voidb@x}\def\CD@sJ{}\CD@@J\def\CD@GJ#1#2#3#4{\setbox#1=#4}\ifx\tenln
\nullfont\def\CD@sJ{vee}\else\let\CD@sJ\CD@eF\fi\def\CD@xF#1#2#3{\begingroup
\aftergroup\CD@wF\CD@L{#1#2:#3#3}\CD@L{#1#2:#3}\aftergroup\CD@yF\CD@L{#1#2:#3%
-#3}\CD@L{#1#2:#3}\endgroup}\def\CD@wF#1#2{\def#1{\hbox{\rlap{#2}\kern.4%
\CD@zC#2}}}\def\CD@yF#1#2{\def#1{\hbox{\rlap{#2}\kern.4\CD@zC#2\kern-.4\CD@zC
}}}\CD@xF lh>\CD@xF rt>\CD@xF rh<\CD@xF rt<\def\CD@yF#1#2{\def#1{\hbox{\kern-%
.4\CD@zC\rlap{#2}\kern.4\CD@zC#2}}}\CD@xF rh>\CD@xF lh<\CD@xF lt>\CD@xF lt<%
\def\CD@wF#1#2{\def#1{\vbox{\vbox to\z@{#2\vss}\nointerlineskip\kern.4\CD@zC#%
2}}}\def\CD@yF#1#2{\def#1{\vbox{\vbox to\z@{#2\vss}\nointerlineskip\kern.4%
\CD@zC#2\kern-.4\CD@zC}}}\CD@xF uh>\CD@xF dt>\CD@xF dh<\CD@xF dt<\def\CD@yF#1%
#2{\def#1{\vbox{\kern-.4\CD@zC\vbox to\z@{#2\vss}\nointerlineskip\kern.4%
\CD@zC#2}}}\CD@xF dh>\CD@xF ut>\CD@xF uh<\CD@xF ut<\def\CD@BG#1{\hbox{%
\mathsurround\z@\offinterlineskip\CD@k\mkern-1.5mu{#1}\mkern-1.5mu\CD@ND}}%
\def\hbox@maths#1{\hbox{\CD@k#1\CD@ND}}\def\CD@GG#1{\hbox to\CD@LF{\setbox0=%
\hbox{\offinterlineskip\mathsurround\z@\CD@k{#1}\CD@ND}\dimen0.5\wd0\advance
\dimen0-.5\CD@oI\CD@zH{\dimen0}\kern-\dimen0\unhbox0\hss}}\def\CD@sB#1{\hbox
to2\CD@LF{\hss\offinterlineskip\mathsurround\z@\CD@k{#1}\CD@ND\hss}}\def
\CD@vF#1{\hbox{\mathsurround\z@\CD@k{#1}\CD@ND}}\def\CD@bE#1{\hbox{\kern-.15%
\CD@zC\CD@k{#1}\CD@ND\kern-.15\CD@zC}}\def\CD@MK#1{\vbox{\offinterlineskip
\kern-.2ex\CD@GG{#1}\kern-.2ex}}\def\@fillh{\xleaders\vrule\horizhtdp}\def
\@fillv{\xleaders\hrule width\CD@LF}\CD@nF{rf:-}{@fillh}\CD@nF{lf:-}{@fillh}%
\CD@nF{df:-}{@fillv}\CD@nF{uf:-}{@fillv}\CD@nF{rh:}{null}\CD@nF{rm:}{null}%
\CD@nF{rt:}{null}\CD@nF{lh:}{null}\CD@nF{lm:}{null}\CD@nF{lt:}{null}\CD@nF{dh%
:}{null}\CD@nF{dm:}{null}\CD@nF{dt:}{null}\CD@nF{uh:}{null}\CD@nF{um:}{null}%
\CD@nF{ut:}{null}\CD@nF{+h:}{null}\CD@nF{+m:}{null}\CD@nF{+t:}{null}\CD@nF{-h%
:}{null}\CD@nF{-m:}{null}\CD@nF{-t:}{null}\def\CD@@D{\hbox{\vrule height 1pt
depth-1pt width 1pt}}\CD@RC{rf:}{\CD@@D}\CD@nF{lf:}{rf:}\CD@nF{+f:}{rf:}%
\CD@RC{df:}{\CD@@D}\CD@nF{uf:}{df:}\CD@nF{-f:}{df:}\def\CD@BD{\CD@U\null
\CD@@D\null\CD@@D\null}\edef\CD@lG{\string\newarrow}\def\newarrow#1#2#3#4#5#6%
{\begingroup\edef\@name{#1}\edef\CD@oJ{#2}\edef\CD@iD{#3}\edef\CD@QG{#4}\edef
\CD@jD{#5}\edef\CD@LE{#6}\let\CD@HE\CD@sG\let\CD@FK\CD@BH\let\@x\CD@AH\ifx
\CD@oJ\CD@iD\let\CD@oJ\empty\fi\ifx\CD@LE\CD@jD\let\CD@LE\empty\fi\def\CD@LI{%
r}\def\CD@SF{l}\def\CD@IC{d}\def\CD@yJ{u}\def\CD@gH{+}\def\@m{-}\ifx\CD@iD
\CD@jD\ifx\CD@QG\CD@iD\let\CD@QG\empty\fi\ifx\CD@LE\empty\ifx\CD@iD\CD@aE\let
\@x\CD@yG\else\let\@x\CD@zG\fi\fi\else\edef\CD@a{\CD@iD\CD@oJ}\ifx\CD@a\empty
\ifx\CD@QG\CD@jD\let\CD@QG\empty\fi\fi\fi\ifmmode\aftergroup\CD@kG\else\CD@@A
\CD@oB rh{head\space\space}\CD@LE\CD@oB rf{filler}\CD@iD\CD@oB rm{middle}%
\CD@QG\ifx\CD@jD\CD@iD\else\CD@oB rf{filler}\CD@jD\fi\CD@oB rt{tail\space
\space}\CD@oJ\CD@gE\CD@HE\CD@FK\@x\CD@nG l-2+2{lu}{nw}\NorthWest\CD@nG r+2+2{%
ru}{ne}\NorthEast\CD@nG l-2-2{ld}{sw}\SouthWest\CD@nG r+2-2{rd}{se}\SouthEast
\else\aftergroup\CD@b\CD@L{r\@name}\fi\fi\endgroup}\def\CD@sG{\CD@vG\CD@LI
\CD@SF rl\Horizontal@Map}\def\CD@BH{\CD@vG\CD@IC\CD@yJ du\Vertical@Map}\def
\CD@AH{\CD@vG\CD@gH\@m+-\Vector@Map}\def\CD@yG{\CD@vG\CD@gH\@m+-\Slant@Map}%
\def\CD@zG{\CD@vG\CD@gH\@m+-\Slope@Map}\catcode`\/=\active\def\CD@vG#1#2#3#4#%
5{\CD@jG#1#3#5t:\CD@oJ/f:\CD@iD/m:\CD@QG/f:\CD@jD/h:\CD@LE//\CD@jG#2#4#5h:%
\CD@LE/f:\CD@jD/m:\CD@QG/f:\CD@iD/t:\CD@oJ//}\def\CD@jG#1#2#3#4//{\edef\CD@fG
{#2}\aftergroup\sdef\CD@L{#1\@name}\aftergroup{\aftergroup#3\CD@M#4//%
\aftergroup}}\def\CD@M#1/{\edef\CD@EH{#1}\ifx\CD@EH\empty\else\CD@L{\CD@fG#1}%
\expandafter\CD@M\fi}\catcode`\/=12 \def\CD@nG#1#2#3#4#5#6#7#8{\aftergroup
\sdef\CD@L{#6\@name}\aftergroup{\CD@L{#2\@name}\if#2#4\aftergroup\CD@CI\else
\aftergroup\CD@BI\fi\CD@L{#1\@name}%
\aftergroup(\aftergroup#3\aftergroup,\aftergroup#5\aftergroup)\aftergroup}}%
\def\CD@oB#1#2#3#4{\expandafter\ifx\csname#1#2:#4\endcsname\relax\CD@y\CD@gB{%
arrow#3 "#4" undefined}\fi}\CD@rG\CD@VE{All five components must be defined
before an arrow.}\CD@rG\CD@SE{\CD@lG, unlike \string\HorizontalMap, is a
declaration.}\def\CD@b#1{\CD@YA{Arrows \string#1 etc could not be defined}%
\CD@VE}\def\CD@kG{\CD@YA{misplaced \CD@lG}\CD@SE}\def\newdiagramgrid#1#2#3{%
\CD@RC{cdgh@#1}{#2,],}
\CD@RC{cdgv@#1}{#3,],}}
\CD@tG\ifincommdiag\incommdiagtrue\incommdiagfalse\CD@tG\CD@@F\CD@IF\CD@HF
\newcount\CD@VA\CD@VA=0 \def\CD@yH{\CD@VA6 }\def\CD@OB{\CD@VA1 \global\CD@yA1
\CD@DE\CD@YF\empty}\def\CD@YF{}\def\CD@nB#1{\relax\CD@MD\edef\CD@vJ{#1}%
\begingroup\CD@rE\else\ifcase\CD@VA\ifmmode\else\CD@YG\CD@E0\fi\or\CD@cE5\or
\CD@YG\CD@F5\or\CD@YG\CD@B5\or\CD@YG\CD@B5\or\CD@YG\CD@C5\or\CD@cE7\or\CD@YG
\CD@D7\fi\fi\endgroup\xdef\CD@YF{#1}}\def\CD@pB#1#2#3#4#5{\relax\CD@MD\xdef
\CD@vJ{#4}\begingroup\ifnum\CD@VA<#1 \expandafter\CD@cE\ifcase\CD@VA0\or#2\or
#3\else#2\fi\else\ifnum\CD@VA<6 \CD@tJ\CD@YG\CD@B#2\else\CD@YG\CD@G#2\fi\fi
\endgroup\CD@DE\CD@YF\CD@vJ\ifincommdiag\let\CD@ZD#5\else\let\CD@ZD\CD@LK\fi}%
\def\CD@yI{\global\CD@yA=\ifnum\CD@VA<5 1\else2\fi\relax}\def\CD@OI{\CD@VA
\CD@yA}\def\CD@cE#1{\aftergroup\CD@VA\aftergroup#1\aftergroup\relax}\def
\CD@HH{\def\CD@nB##1{\relax}\let\CD@pB\CD@FH\let\CD@yH\relax\let\CD@OB\relax
\let\CD@yI\relax\let\CD@OI\relax}\def\CD@FH#1#2#3#4#5{\ifincommdiag\let\CD@ZD
#5\else\xdef\CD@vJ{#4}\let\CD@ZD\CD@LK\fi}\def\CD@YG#1{\aftergroup#1%
\aftergroup\relax\CD@cE}\def\CD@B{\CD@YE\CD@S\CD@ME\CD@Q}\def\CD@G{\CD@YE{%
\CD@yC\CD@S}\CD@XE\CD@QD\CD@Q}\def\CD@F{\CD@YE{*\CD@S}\CD@RE\clubsuit\CD@Q}%
\def\CD@C{\CD@YE{\CD@S*\CD@S}\CD@RE\CD@Q\clubsuit\CD@Q}\def\CD@D{\CD@YE\CD@EC
\CD@TE\\}\def\CD@E{\CD@YE\CD@nC\CD@QE\CD@k}\def\CD@LK{\CD@YA{\CD@vJ\space
ignored \CD@dH}\CD@WE}\def\CD@FE{}\def\CD@d{\CD@YA{maps must never be enclosed
in braces}\CD@OE}\def\CD@dH{outside diagram}\def\CD@FC{\string\HonV, \string
\VonH\space and \string\HmeetV}\CD@rG\CD@ME{The way that horizontal and
vertical arrows are terminated implicitly means\CD@uG that they cannot be
mixed with each other or with \CD@FC.}\CD@rG\CD@XE{\string\pile\space is for
parallel horizontal arrows; verticals can just be put together in\CD@uG a cell%
. \CD@FC\space are not meaningful in a \string\pile.}\CD@rG\CD@RE{The
horizontal maps must point to an object, not each other (I've put in\CD@uG one
which you're unlikely to want). Use \string\pile\space if you want them
parallel.}\CD@rG\CD@TE{Parallel horizontal arrows must be in separate layers
of a \string\pile.}\CD@rG\CD@QE{Horizontal arrows may be used \CD@dH s, but
must still be in maths.}\CD@rG\CD@WE{Vertical arrows, \CD@FC\space\CD@dH s don%
't know where\CD@uG where to terminate.}\CD@rG\CD@OE{This prevents them from
stretching correctly.}\def\CD@YE#1{\CD@YA{"#1" inserted \ifx\CD@YF\empty
before \CD@vJ\else between \CD@YF\ifx\CD@YF\CD@vJ s\else\space and \CD@vJ\fi
\fi}}\count@=\year\multiply\count@12 \advance\count@\month\ifnum\count@>24247
\expandafter\endinput\fi\def
\Horizontal@Map{\CD@nB{horizontal map}\CD@LC\CD@TJ\CD@qD}\def\CD@TJ{\CD@GB-%
9999 \let\CD@ZD\CD@XD\ifincommdiag\else\CD@cJ\ifinpile\else\skip2\z@ plus 1.5%
\CD@VK minus .5\CD@UK\skip4\skip2 \fi\fi\let\CD@kD\@fillh\CD@nF{fill@dot}{rf:%
.}}\def\Vector@Map{\CD@HK4}\def\Slant@Map{\CD@HK{\CD@EF255\else6\fi}}\def
\Slope@Map{\CD@HK\CD@OC}\def\CD@HK#1#2#3#4#5#6{\CD@LC\def\CD@WK{2}\def\CD@aK{%
2}\def\CD@ZK{1}\def\CD@bK{1}\let\Horizontal@Map\CD@nI\def\CD@OG{#1}\def\CD@NI
{\CD@U#2#3#4#5#6}}\def\CD@nI{\CD@TJ\CD@JB\let\CD@ZD\CD@TD\CD@qD}\CD@tG\CD@pE
\CD@rA\CD@qA\CD@rA\def\cds@missives{\CD@rA}\def\CD@TD{\CD@vE\let\CD@OG\CD@OC
\CD@x\CD@zE\CD@WF\fi\setbox0\hbox{\incommdiagfalse\CD@HI}\CD@pE\CD@aD\else
\global\CD@YC\CD@bD\fi\ifvoid6 \ifvoid7 \CD@eE\fi\fi\CD@zE\else\CD@BD\global
\CD@YC\let\CD@CG\CD@IH\CD@YD\fi\else\CD@NI\CD@MI\global\CD@YC\CD@YD\fi}\def
\CD@LC{\begingroup\dimen1=\MapShortFall\dimen2=\CD@RG\dimen5=\MapShortFall
\setbox3=\box\voidb@x\setbox6=\box\voidb@x\setbox7=\box\voidb@x\CD@pD
\mathsurround\z@\skip2\z@ plus1fill minus 1000pt\skip4\skip2 \CD@TB}\CD@tG
\CD@tE\CD@UB\CD@TB\def\CD@U#1#2#3#4#5{\let\CD@oJ#1\let\CD@iD#2\let\CD@QG#3%
\let\CD@jD#4\let\CD@LE#5\CD@TB\ifx\CD@iD\CD@jD\CD@UB\fi}\def\CD@qD#1#2#3#4#5{%
\CD@U#1#2#3#4#5\CD@tD}\def\Vertical@Map{\CD@pB433{vertical map}\CD@cD\CD@LC
\CD@GB-9995 \let\CD@kD\@fillv\CD@nF{fill@dot}{df:.}\CD@qD}\def\break@args{%
\def\CD@tD{\CD@ZD}\CD@ZD\endgroup\aftergroup\CD@FE}\def\CD@MJ{\setbox1=\CD@oJ
\setbox5=\CD@LE\ifvoid3 \ifx\CD@QG\null\else\setbox3=\CD@QG\fi\fi\CD@@G2%
\CD@iD\CD@@G4\CD@jD}\def\CD@pF#1{\ifvoid1\else\CD@oF1#1\fi\ifvoid2\else\CD@oF
2#1\fi\ifvoid3\else\CD@oF3#1\fi\ifvoid4\else\CD@oF4#1\fi\ifvoid5\else\CD@oF5#%
1\fi} \def\CD@oF#1#2{\setbox#1\vbox{\offinterlineskip\box#1\dimen@\prevdepth
\advance\dimen@-#2\relax\setbox0\null\dp0\dimen@\ht0-\dimen@\box0}}\def\CD@@G
#1#2{\ifx#2\CD@kD\setbox#1=\box\voidb@x\else\setbox#1=#2\def#2{\xleaders\box#%
1}\fi}\CD@ZA\CD@BK{\string\HorizontalMap, \string\VerticalMap\space and
\string\DiagonalMap\CD@uG are obsolete - use \CD@lG\space to pre-define maps}%
\def\HorizontalMap#1#2#3#4#5{\CD@BK\CD@nB{old horizontal map}\CD@LC\CD@TJ\def
\CD@oJ{\CD@UH{#1}}\CD@SH\CD@iD{#2}\def\CD@QG{\CD@UH{#3}}\CD@SH\CD@jD{#4}\def
\CD@LE{\CD@UH{#5}}\CD@tD}\def\VerticalMap#1#2#3#4#5{\CD@BK\CD@pB433{vertical
map}\CD@cD\CD@LC\CD@GB-9995 \let\CD@kD\@fillv\def\CD@oJ{\CD@GG{#1}}\CD@VH
\CD@iD{#2}\def\CD@QG{\CD@GG{#3}}\CD@VH\CD@jD{#4}\def\CD@LE{\CD@GG{#5}}\CD@tD}%
\def\DiagonalMap#1#2#3#4#5{\CD@BK\CD@LC\def\CD@OG{4}\let\CD@kD\CD@qK\let
\CD@ZD\CD@YD\def\CD@WK{2}\def\CD@aK{2}\def\CD@ZK{1}\def\CD@bK{1}\def\CD@QG{%
\CD@vF{#3}}\ifPositiveGradient\let\mv\raise\def\CD@oJ{\CD@vF{#5}}\def\CD@iD{%
\CD@vF{#4}}\def\CD@jD{\CD@vF{#2}}\def\CD@LE{\CD@vF{#1}}\else\let\mv\lower\def
\CD@oJ{\CD@vF{#1}}\def\CD@iD{\CD@vF{#2}}\def\CD@jD{\CD@vF{#4}}\def\CD@LE{%
\CD@vF{#5}}\fi\CD@tD}\def\CD@aE{-}\def\CD@AD{\empty}\def\CD@SH{\CD@EG\CD@bE
\CD@aE\@fillh}\def\CD@VH{\CD@EG\CD@MK\CD@KK\@fillv}\def\CD@EG#1#2#3#4#5{\def
\CD@CH{#5}\ifx\CD@CH#2\let#4#3\else\let#4\null\ifx\CD@CH\empty\else\ifx\CD@CH
\CD@AD\else\let#4\CD@CH\fi\fi\fi}\def\CD@UH#1{\hbox{\mathsurround\z@
\offinterlineskip\def\CD@CH{#1}\ifx\CD@CH\empty\else\ifx\CD@CH\CD@AD\else
\CD@k\mkern-1.5mu{\CD@CH}\mkern-1.5mu\CD@ND\fi\fi}}\def\CD@yD#1#2{\setbox#1=%
\hbox\bgroup\setbox0=\hbox{\CD@k\labelstyle()\CD@ND}
\setbox1=\null\ht1\ht0\dp1\dp0\box1 \kern.1\CD@zC\CD@k\bgroup\labelstyle
\aftergroup\CD@LD\CD@xD}\def\CD@LD{\CD@ND\kern.1\CD@zC\egroup\CD@tD}\def
\CD@xD{\futurelet\CD@EH\CD@mJ}\def\CD@mJ{
\catcase\bgroup:\CD@v;\catcase\egroup:\missing@label;\catcase\space:\CD@TF;%
\tokcase[:\CD@XF;
\default:\CD@zJ;\endswitch}\def\CD@v{\let\CD@MD\CD@c\let\CD@CH}\def\CD@zJ#1{%
\let\CD@UF\egroup{\let\actually@braces@missing@around@macro@in@label\CD@ZH
\let\CD@MD\CD@xC\let\CD@UF\CD@VF#1%
\actually@braces@missing@around@macro@in@label}\CD@UF}\def
\actually@braces@missing@around@macro@in@label{\let\CD@CH=}\def\missing@label
{\egroup\CD@YA{missing label}\CD@PE}\def\CD@xC{\egroup\missing@label}\outer
\def\CD@ZH{}\def\CD@UF{}\def\CD@VF{\CD@wC\CD@UF}\def\CD@MD{}\def\CD@XF{\let
\CD@N\CD@xD\get@square@arg\CD@AE}\CD@rG\CD@PE{The text which has just been
read is not allowed within map labels.}\def\CD@c{\egroup\CD@YA{missing \CD@yC
\space inserted after label}\CD@PE}\def\upper@label{\CD@oD\CD@yD6}\def
\lower@label{\def\positional@{\CD@@A\break@args}\CD@yD7}\def\middle@label{%
\CD@yD3}\CD@tG\CD@yE\CD@pD\CD@oD\def\CD@iF{\ifPositiveGradient\CD@tJ
\expandafter\upper@label\else\expandafter\lower@label\fi}\def\CD@iI{%
\ifPositiveGradient\CD@tJ\expandafter\lower@label\else\expandafter
\upper@label\fi}\def\positional@{\CD@gB{labels as positional arguments are
obsolete}\CD@yE\CD@tJ\expandafter\upper@label\else\expandafter\lower@label\fi
-}\def\CD@tD{\futurelet\CD@EH\switch@arg}\def\eat@space{\afterassignment
\CD@tD\let\CD@EH= }\def\CD@TF{\afterassignment\CD@xD\let\CD@EH= }\def\CD@BC{%
\get@round@pair\CD@uD}\def\CD@uD#1#2{\def\CD@WK{#1}\def\CD@aK{#2}\CD@tD}\def
\optional@{\let\CD@N\CD@tD\get@square@arg\CD@AE}\def\CD@JJ.{\CD@sC\CD@tD}\def
\CD@sC{\let\CD@iD\fill@dot\let\CD@jD\fill@dot\def\CD@MI{\let\CD@iD\dfdot\let
\CD@jD\dfdot}}\def\CD@MI{}\def\CD@@E#1,{\CD@nH#1,\begingroup\ifx\@name\CD@RD
\CD@FF\aftergroup\CD@e\fi\aftergroup\CD@jC\else\expandafter\def\expandafter
\CD@RF\expandafter{\csname\@name\endcsname}\expandafter\CD@vD\CD@RF\CD@KD\ifx
\CD@RF\empty\aftergroup\CD@pC\expandafter\aftergroup\csname\CD@FB\@name
\endcsname\expandafter\aftergroup\csname\CD@FB @\@name\endcsname\else\gdef
\CD@GE{#1}\CD@gB{\string\relax\space inserted before `[\CD@GE'}\message{(I was
trying to read this as a \CD@tA\ option.)}\aftergroup\CD@H\fi\fi\endgroup}%
\def\CD@vD#1#2\CD@KD{\def\CD@RF{#2}}\def\CD@jC{\let\CD@CH\CD@N\let\CD@N\relax
\CD@CH}\def\CD@H#1],{
\CD@jC\relax\def\CD@RF{#1}\ifx\CD@RF\empty\def\CD@RF{[\CD@GE]}%
\else\def\CD@RF{[\CD@GE,#1]}
\fi\CD@RF}\def\CD@pC#1#2{\ifx#2\CD@qK\ifx#1\CD@qK\CD@gB{option `\@name'
undefined}\else#1\fi\else\CD@FF\expandafter#2\CD@GK\CD@PK\else\CD@QK\fi\fi
\CD@DH}\CD@tG\CD@FF\CD@QK\CD@PK\def\CD@nH#1,{\CD@FF\ifx\CD@GK\CD@qK\CD@e\else
\expandafter\CD@oH\CD@GK,#1,(,),(,)[]%
\fi\fi\CD@FF\else\CD@mH#1==,\fi}\def\CD@e{\CD@gB{option `\@name' needs (x,y)
value}\CD@PK\let\@name\empty}\def\CD@mH#1=#2=#3,{\def\@name{#1}\def\CD@GK{#2}%
\def\CD@RF{#3}\ifx\CD@RF\empty\let\CD@GK\CD@qK\fi}%
\def\CD@oH#1(#2,#3)#4,(#5,#6)#7[]{\def\CD@GK{{#2}{#3}}\def\CD@RF{#1#4#5#6}%
\ifx\CD@RF\empty\def\CD@RF{#7}\ifx\CD@RF\empty\CD@e\fi\else\CD@e\fi}\def
\CD@FB{cds@}\let\CD@N\relax\def\CD@zD#1{\ifx\CD@GK\CD@qK\CD@gB{option `\@name
' needs a value}\else#1\CD@GK\relax\fi}\def\CD@BE#1#2{\ifx\CD@GK\CD@qK#1#2%
\relax\else#1\CD@GK\relax\fi}\def\cds@@showpair#1#2{\message{x=#1,y=#2}}\def
\cds@@diagonalbase#1#2{\edef\CD@ZK{#1}\edef\CD@bK{#2}}\def\CD@DI#1{\def\CD@CH
{#1}\CD@nF{@x}{cdps@#1}\ifx\CD@CH\empty\CD@f\CD@CH{cannot be used}\else\ifx
\CD@CH\relax\CD@f\CD@CH{unknown}\else\let\CD@IK\@x\fi\fi}\def\CD@f#1#2{\CD@gB
{PostScript translator `#1' #2}}\def\CD@PH{}\def\CD@PJ{\CD@fA\edef\CD@PH{%
\noexpand\CD@KB{\@name\space ignored within maths}}}\def\diagramstyle{\CD@cJ
\let\CD@N\relax\CD@CF\CD@AE\CD@AE}\CD@tG\CD@sE
\CD@SB\CD@RB\CD@tG\CD@qE\CD@EB\CD@DB\CD@tG\CD@oE\CD@pA\CD@oA\CD@tG\CD@iE
\CD@HA\CD@GA\CD@HA\CD@tG\CD@jE\CD@JA\CD@IA\CD@tG\CD@kE\CD@LA\CD@KA\CD@tG
\CD@EF\CD@DK\CD@CK\CD@tG\CD@rE\CD@JB\CD@IB\CD@tG\CD@mE\CD@gA\CD@fA\CD@tG
\CD@nE\CD@kA\CD@jA\CD@tG\CD@AF\CD@iG\CD@hG\CD@RC{cds@ }{}\CD@RC{cds@}{}\CD@RC
{cds@1em}{\CellSize1\CD@zC}\CD@RC{cds@1.5em}{\CellSize1.5\CD@zC}\CD@RC{cds@2%
em}{\CellSize2\CD@zC}\CD@RC{cds@2.5em}{\CellSize2.5\CD@zC}\CD@RC{cds@3em}{%
\CellSize3\CD@zC}\CD@RC{cds@3.5em}{\CellSize3.5\CD@zC}\CD@RC{cds@4em}{%
\CellSize4\CD@zC}\CD@RC{cds@4.5em}{\CellSize4.5\CD@zC}\CD@RC{cds@5em}{%
\CellSize5\CD@zC}\CD@RC{cds@6em}{\CellSize6\CD@zC}\CD@RC{cds@7em}{\CellSize7%
\CD@zC}\CD@RC{cds@8em}{\CellSize8\CD@zC}\def\cds@abut{\MapsAbut\dimen1\z@
\dimen5\z@}\def\cds@alignlabels{\CD@IA\CD@KA}\def\cds@amstex{\ifincommdiag
\CD@O\else\def\CD{\diagram[amstex]}
\fi\CD@T\catcode`\@\active}\def\cds@b{\let\CD@dB\CD@bB}\def\cds@balance{\let
\CD@hA\CD@AA}\let\cds@bottom\cds@b\def\cds@center{\cds@vcentre\cds@nobalance}%
\let\cds@centre\cds@center\def\cds@centerdisplay{\CD@HA\CD@PJ\cds@balance}%
\let\cds@centredisplay\cds@centerdisplay\def\cds@crab{\CD@BE\CD@DC{.5%
\PileSpacing}}\CD@RC{cds@crab-}{\CD@DC-.5\PileSpacing}\CD@RC{cds@crab+}{%
\CD@DC.5\PileSpacing}\CD@RC{cds@crab++}{\CD@DC1.5\PileSpacing}\CD@RC{cds@crab%
--}{\CD@DC-1.5\PileSpacing}\def\cds@defaultsize{\CD@BE{\let\CD@QC}{3em}\CD@NJ
}\def\cds@displayoneliner{\CD@DB}\let\cds@dotted\CD@sC\def\cds@dpi{\CD@RJ{1%
truein}}\def\cds@dpm{\CD@RJ{100truecm}}\let\CD@XA\CD@qK\def\cds@eqno{\let
\CD@XA\CD@GK\let\CD@EJ\empty}\def\cds@fixed{\CD@qA}\CD@tG\CD@fE\CD@J\CD@I\def
\cds@flushleft{\CD@I\CD@GA\CD@PJ\cds@nobalance\CD@BE\CD@nA\CD@nA}\def\cds@gap
{\CD@AJ\setbox3=\null\ht3=\CD@tI\dp3=\CD@sI\CD@BE{\wd3=}\MapShortFall} \def
\cds@grid{\ifx\CD@GK\CD@qK\let\h@grid\relax\let\v@grid\relax\else\CD@nF{%
h@grid}{cdgh@\CD@GK}\CD@nF{v@grid}{cdgv@\CD@GK}\ifx\h@grid\relax\CD@gB{%
unknown grid `\CD@GK'}\else\CD@WB\fi\fi}\let\h@grid\relax\let\v@grid\relax
\def\cds@gridx{\ifx\CD@GK\CD@qK\else\cds@grid\fi\let\CD@CH\h@grid\let\h@grid
\v@grid\let\v@grid\CD@CH}\def\cds@h{\CD@zD\DiagramCellHeight}\def\cds@hcenter
{\let\CD@hA\CD@aA}\let\cds@hcentre\cds@hcenter\def\cds@heads{\CD@BE{\let
\CD@sJ}\CD@sJ\CD@@J\CD@vE\else\ifx\CD@sJ\CD@eF\else\CD@MC\fi\fi}\let
\cds@height\cds@h\let\cds@hmiddle\cds@balance\def\cds@htriangleheight{\CD@BE
\DiagramCellHeight\DiagramCellHeight\DiagramCellWidth1.73205%
\DiagramCellHeight}\def\cds@htrianglewidth{\CD@BE\DiagramCellWidth
\DiagramCellWidth\DiagramCellHeight.57735\DiagramCellWidth}\CD@tG\CD@zE\CD@eE
\CD@dE\CD@eE\def\cds@hug{\CD@eE} \def\cds@inline{\CD@gA\let\CD@PH\empty}\def
\cds@inlineoneliner{\CD@EB}\CD@RC{cds@l>}{\CD@zD{\let\CD@RG}\dimen2=\CD@RG}%
\def\cds@labelstyle{\CD@zD{\let\labelstyle}}\def\cds@landscape{\CD@kA}\def
\cds@large{\CellSize5\CD@zC}\let\CD@EJ\empty\def\CD@FJ{\refstepcounter{%
equation}\def\CD@XA{\hbox{\@eqnnum}}}\def\cds@LaTeXeqno{\let\CD@EJ\CD@FJ}\def
\cds@lefteqno{\CD@pA}\def\cds@leftflush{\cds@flushleft\CD@J}\def
\cds@leftshortfall{\CD@zD{\dimen1 }}\def\cds@lowershortfall{%
\ifPositiveGradient\cds@leftshortfall\else\cds@rightshortfall\fi}\def
\cds@loose{\CD@VB}\def\cds@midhshaft{\CD@JA}\def\cds@midshaft{\CD@JA}\def
\cds@midvshaft{\CD@LA}\def\cds@moreoptions{\CD@@A}\let\cds@nobalance
\cds@hcenter\def\cds@nohcheck{\CD@HH}\def\cds@nohug{\CD@dE} \def
\cds@nooptions{\def\CD@aC{\CD@WD}}\let\cds@noorigin\cds@nobalance\def
\cds@nopixel{\CD@@I4\CD@XH\CD@cJ}\def\cds@UO{\CD@oK\global\let\CD@n\empty}%
\def\cds@UglyObsolete{\cds@UO\let\cds@PS\empty}\def\CD@rK#1{\CD@gB{option `#1%
' renamed as `UglyObsolete'}}\def\cds@noPostScript{\CD@rK{noPostScript}}\def
\cds@noPS{\CD@rK{noPostScript}}\def\cds@notextflow{\CD@RB}\def\cds@noTPIC{%
\CD@CK}\def\cds@objectstyle{\CD@zD{\let\objectstyle}}\def\cds@origin{\let
\CD@hA\CD@iB}\def\cds@p{\CD@zD\PileSpacing}\let\cds@pilespacing\cds@p\def
\cds@pixelsize{\CD@zD\CD@@I\CD@gI}\def\cds@portrait{\CD@jA}\def
\cds@PostScript{\CD@nK\global\let\CD@n\empty\CD@BE\CD@DI\empty}\def\cds@PS{%
\CD@nK\global\let\CD@n\empty}\CD@GF\CD@n{\typeout{\CD@tA: try the PostScript
option for better results}}\def\cds@repositionpullbacks{\let\make@pbk\CD@fH
\let\CD@qH\CD@pH}\def\cds@righteqno{\CD@oA}\def\cds@rightshortfall{\CD@zD{%
\dimen5 }}\def\cds@ruleaxis{\CD@zD{\let\axisheight}}\def\cds@cmex{\let\CD@GG
\CD@sB\let\CD@QJ\CD@CJ}\def\cds@s{\cds@height\DiagramCellWidth
\DiagramCellHeight}\def\cds@scriptlabels{\let\labelstyle\scriptstyle}\def
\cds@shortfall{\CD@zD\MapShortFall\dimen1\MapShortFall\dimen5\MapShortFall}%
\def\cds@showfirstpass{\CD@BE{\let\CD@nD}\z@}\def\cds@silent{\def\CD@KB##1{}%
\def\CD@gB##1{}}\let\cds@size\cds@s\def\cds@small{\CellSize2\CD@zC}\def
\cds@snake{\CD@BE\CD@eJ\z@}\def\cds@t{\let\CD@dB\CD@fB}\def\cds@textflow{%
\CD@SB\CD@PJ}\def\cds@thick{\let\CD@rF\tenlnw\CD@LF\CD@NC\CD@BE\MapBreadth{2%
\CD@LF}\CD@@J}\def\cds@thin{\let\CD@rF\tenln\CD@BE\MapBreadth{\CD@NC}\CD@@J}%
\def\cds@tight{\CD@WB}\let\cds@top\cds@t\def\cds@TPIC{\CD@DK}\def
\cds@uppershortfall{\ifPositiveGradient\cds@rightshortfall\else
\cds@leftshortfall\fi}\def\cds@vcenter{\let\CD@dB\CD@cB}\let\cds@vcentre
\cds@vcenter\def\cds@vtriangleheight{\CD@BE\DiagramCellHeight
\DiagramCellHeight\DiagramCellWidth.577035\DiagramCellHeight}\def
\cds@vtrianglewidth{\CD@BE\DiagramCellWidth\DiagramCellWidth
\DiagramCellHeight1.73205\DiagramCellWidth}\def\cds@vmiddle{\let\CD@dB\CD@eB}%
\def\cds@w{\CD@zD\DiagramCellWidth}\let\cds@width\cds@w\def\diagram{\relax
\protect\CD@bC}\def\enddiagram{\protect\CD@SG}\def\CD@bC{\CD@g\CD@uI
\incommdiagtrue\edef\CD@wI{\the\CD@NB}\global\CD@NB\z@\boxmaxdepth\maxdimen
\everycr{}\CD@sK\everymath{}\everyhbox{}\ifx\pdfsyncstop\CD@qK\else
\pdfsyncstop\fi\CD@aC}\def\CD@aC{\CD@y\let\CD@N\CD@ZC\CD@CF\CD@AE\CD@WD}\def
\CD@ZC{\CD@gE\expandafter\CD@aC\else\expandafter\CD@WD\fi}\def\CD@WD{\let
\CD@EH\relax\CD@nE\CD@vE\else\CD@hK\else\CD@KB{landscape ignored without
PostScript}\CD@jA\fi\fi\fi\CD@EJ\setbox2=\vbox\bgroup\CD@JF\CD@VD}\def\CD@cH{%
\CD@nE\CD@fB\else\CD@dB\fi\CD@hA\nointerlineskip\setbox0=\null\ht0-\CD@pI\dp0%
\CD@pI\wd0\CD@kI\box0 \global\CD@QA\CD@kF\global\CD@yA\CD@XB\ifx\CD@NK\CD@qK
\global\CD@RA\CD@kF\else\global\CD@RA\CD@NK\fi\egroup\CD@zF\CD@nE\setbox2=%
\hbox to\dp2{\vrule height\wd2 depth\CD@QA width\z@\global\CD@QA\ht2\ht2\z@
\dp2\z@\wd2\z@\CD@hK\CD@tK{q 0 1 -1 0 0 0 cm}\else\global\CD@iG\CD@IK{0 1
bturn}\fi\box2\CD@gK\hss}\CD@DB\fi\ifnum\CD@yA=1 \else\CD@DB\fi\global
\@ignorefalse\CD@mE\leavevmode\fi\ifvmode\CD@TA\else\ifmmode\CD@PH\CD@GI\else
\CD@qE\CD@gA\fi\ifinner\CD@gA\fi\CD@mE\CD@GI\else\CD@sE\CD@QB\else\CD@TA\fi
\fi\fi\fi\CD@dD}\def\CD@dD{\global\CD@NB\CD@wI\relax\CD@xE\global\CD@ID\else
\aftergroup\CD@mC\fi\if@ignore\aftergroup\ignorespaces\fi\CD@wC\ignorespaces}%
\def\CD@fB{\advance\CD@pI\dimen1\relax}\def\CD@eB{\advance\CD@pI.5\dimen1%
\relax}\def\CD@bB{}\def\CD@cB{\CD@fB\advance\CD@pI\CD@YB\divide\CD@pI2
\advance\CD@pI-\axisheight\relax}\def\CD@aA{}\def\CD@iB{\CD@kF\z@}\def\CD@AA{%
\ifdim\dimen2>\CD@kF\CD@kF\dimen2 \else\dimen2\CD@kF\CD@kI\dimen0 \advance
\CD@kI\dimen2 \fi}\def\CD@QB{\skip0\z@\relax\loop\skip1\lastskip\ifdim\skip1>%
\z@\unskip\advance\skip0\skip1 \repeat\vadjust{\prevdepth\dp\strutbox\penalty
\predisplaypenalty\vskip\abovedisplayskip\CD@UA\penalty\postdisplaypenalty
\vskip\belowdisplayskip}\ifdim\skip0=\z@\else\hskip\skip0 \global\@ignoretrue
\fi}\def\CD@TA{\CD@LG\kern-\displayindent\CD@UA\CD@LG\global\@ignoretrue}\def
\CD@UA{\hbox to\hsize{\CD@fE\ifdim\CD@RA=\z@\else\advance\CD@QA-\CD@RA\setbox
2=\hbox{\kern\CD@RA\box2}\fi\fi\setbox1=\hbox{\ifx\CD@XA\CD@qK\else\CD@k
\CD@XA\CD@ND\fi}\CD@oE\CD@iE\else\advance\CD@QA\wd1 \fi\wd1\z@\box1 \fi\dimen
0\wd2 \advance\dimen0\wd1 \advance\dimen0-\hsize\ifdim\dimen0>-\CD@nA\CD@HA
\fi\advance\dimen0\CD@QA\ifdim\dimen0>\z@\CD@KB{wider than the page by \the
\dimen0 }\CD@HA\fi\CD@iE\hss\else\CD@V\CD@QA\CD@nA\fi\CD@GI\hss\kern-\wd1\box
1 }}\def\CD@GI{\CD@AF\CD@@F\else\CD@SC\global\CD@hG\fi\fi\kern\CD@QA\box2 }%
\CD@tG\CD@wE\CD@YC\CD@XC\def\CD@JF{\CD@cJ\ifdim\DiagramCellHeight=-\maxdimen
\DiagramCellHeight\CD@QC\fi\ifdim\DiagramCellWidth=-\maxdimen
\DiagramCellWidth\CD@QC\fi\global\CD@XC\CD@IF\let\CD@FE\empty\let\CD@z\CD@Q
\let\overprint\CD@eH\let\CD@s\CD@rJ\let\enddiagram\CD@ED\let\\\CD@cC\let\par
\CD@jH\let\CD@MD\empty\let\switch@arg\CD@PB\let\shift\CD@iA\baselineskip
\DiagramCellHeight\lineskip\z@\lineskiplimit\z@\mathsurround\z@\tabskip\z@
\CD@OB}\def\CD@VD{\penalty-123 \begingroup\CD@jA\aftergroup\CD@K\halign
\bgroup\global\advance\CD@NB1 \vadjust{\penalty1}\global\CD@FA\z@\CD@OB\CD@j#%
#\CD@DD\CD@Q\CD@Q\CD@OI\CD@j##\CD@DD\cr}\def\CD@ED{\CD@MD\CD@GD\crcr\egroup
\global\CD@JD\endgroup}\def\CD@j{\global\advance\CD@FA1 \futurelet\CD@EH\CD@i
}\def\CD@i{\ifx\CD@EH\CD@DD\CD@tJ\hskip1sp plus 1fil \relax\let\CD@DD\relax
\CD@vI\else\hfil\CD@k\objectstyle\let\CD@FE\CD@d\fi}\def\CD@DD{\CD@MD\relax
\CD@yI\CD@vI\global\CD@QA\CD@iA\penalty-9993 \CD@ND\hfil\null\kern-2\CD@QA
\null}\def\CD@cC{\cr}\def\across#1{\span\omit\mscount=#1 \global\advance
\CD@FA\mscount\global\advance\CD@FA\m@ne\CD@sF\ifnum\mscount>2 \CD@fJ\repeat
\ignorespaces}\def\CD@fJ{\relax\span\omit\advance\mscount\m@ne}\def\CD@qJ{%
\ifincommdiag\ifx\CD@iD\@fillh\ifx\CD@jD\@fillh\ifdim\dimen3>\z@\else\ifdim
\dimen2>93\CD@@I\ifdim\dimen2>18\p@\ifdim\CD@LF>\z@\count@\CD@bJ\advance
\count@\m@ne\ifnum\count@<\z@\count@20\let\CD@aJ\CD@uJ\fi\xdef\CD@bJ{\the
\count@}\fi\fi\fi\fi\fi\fi\fi}\def\CD@cG#1{\vrule\horizhtdp width#1\dimen@
\kern2\dimen@}\def\CD@uJ{\rlap{\dimen@\CD@@I\CD@V\dimen@{.182\p@}\CD@zH
\dimen@\advance\CD@tI\dimen@\CD@cG0\CD@cG0\CD@cG2\CD@cG6\CD@cG6\CD@cG2\CD@cG0%
\CD@cG0\CD@cG2\CD@cG6\CD@cG0\CD@cG0\CD@cG2\CD@cG2\CD@cG6\CD@cG0\CD@cG0\CD@cG2%
\CD@cG6\CD@cG2\CD@cG2\CD@cG0\CD@cG0}}\def\CD@bJ{10}\def\CD@aJ{}\def\CD@XD{%
\CD@gE\CD@TB\fi\CD@x\CD@WF\CD@HI}\def\CD@x{\CD@QJ\CD@DC\CD@MJ\ifdim\CD@DC=\z@
\else\CD@pF\CD@DC\fi\ifvoid3 \setbox3=\null\ht3\CD@tI\dp3\CD@sI\else\CD@V{\ht
3}\CD@tI\CD@V{\dp3}\CD@sI\fi\dimen3=.5\wd3 \ifdim\dimen3=\z@\CD@tE\else\dimen
3-\CD@XH\fi\else\CD@TB\fi\CD@V{\dimen2}{\wd7}\CD@V{\dimen2}{\wd6}\CD@qJ
\advance\dimen2-2\dimen3 \dimen4.5\dimen2 \dimen2\dimen4 \advance\dimen2%
\CD@eJ\advance\dimen4-\CD@eJ\advance\dimen2-\wd1 \advance\dimen4-\wd5 \ifvoid
2 \else\CD@V{\ht3}{\ht2}\CD@V{\dp3}{\dp2}\CD@V{\dimen2}{\wd2}\fi\ifvoid4 \else
\CD@V{\ht3}{\ht4}\CD@V{\dp3}{\dp4}\CD@V{\dimen4}{\wd4}\fi\advance\skip2\dimen
2 \advance\skip4\dimen4 \CD@tE\advance\skip2\skip4 \dimen0\dimen5 \advance
\dimen0\wd5 \skip3-\skip4 \advance\skip3-\dimen0 \let\CD@jD\empty\else\skip3%
\z@\relax\dimen0\z@\fi}\def\CD@WF{\offinterlineskip\lineskip.2\CD@zC\ifvoid6
\else\setbox3=\vbox{\hbox to2\dimen3{\hss\box6\hss}\box3}\fi\ifvoid7 \else
\setbox3=\vtop{\box3 \hbox to2\dimen3{\hss\box7\hss}}\fi}\def\CD@HI{\kern
\dimen1 \box1 \CD@aJ\CD@iD\hskip\skip2 \kern\dimen0 \ifincommdiag\CD@jE
\penalty1\fi\kern\dimen3 \penalty\CD@GB\hskip\skip3 \null\kern-\dimen3 \else
\hskip\skip3 \fi\box3 \CD@jD\hskip\skip4 \box5 \kern\dimen5}\def\CD@MF{\ifnum
\CD@LH>\CD@TC\CD@V{\dimen1}\objectheight\CD@V{\dimen5}\objectheight\else\CD@V
{\dimen1}\objectwidth\CD@V{\dimen5}\objectwidth\fi}\def\CD@Y{\begingroup
\ifdim\dimen7=\z@\kern\dimen8 \else\ifdim\dimen6=\z@\kern\dimen9 \else\dimen5%
\dimen6 \dimen6\dimen9 \CD@KJ\dimen4\dimen2 \CD@dG{\dimen4}\dimen6\dimen5
\dimen7\dimen8 \CD@KJ\CD@iC{\dimen2}\ifdim\dimen2<\dimen4 \kern\dimen2 \else
\kern\dimen4 \fi\fi\fi\endgroup}\def\CD@jJ{\CD@JI\setbox\z@\hbox{\lower
\axisheight\hbox to\dimen2{\CD@DF\ifPositiveGradient\dimen8\ht\CD@MH\dimen9%
\CD@mI\else\dimen8\dp3 \dimen9\dimen1 \fi\else\dimen8 \ifPositiveGradient
\objectheight\else\z@\fi\dimen9\objectwidth\fi\advance\dimen8
\ifPositiveGradient-\fi\axisheight\CD@Y\unhbox\z@\CD@DF\ifPositiveGradient
\dimen8\dp3 \dimen9\dimen0 \else\dimen8\ht\CD@MH\dimen9\CD@mF\fi\else\dimen8
\ifPositiveGradient\z@\else\objectheight\fi\dimen9\objectwidth\fi\advance
\dimen8 \ifPositiveGradient\else-\fi\axisheight\CD@Y}}}\def\CD@bD{\dimen6
\CD@aK\DiagramCellHeight\dimen7 \CD@WK\DiagramCellWidth\CD@jJ
\ifPositiveGradient\advance\dimen7-\CD@ZK\DiagramCellWidth\else\dimen7 \CD@ZK
\DiagramCellWidth\dimen6\z@\fi\advance\dimen6-\CD@bK\DiagramCellHeight\CD@mK
\setbox0=\rlap{\kern-\dimen7 \lower\dimen6\box\z@}\ht0\z@\dp0\z@\raise
\axisheight\box0 }\def\CD@mK{\setbox0\hbox{\ht\z@\z@\dp\z@\z@\wd\z@\z@\CD@hK
\expandafter\CD@tK{q \CD@eK\space\CD@lK\space\CD@kK\space\CD@eK\space0 0 cm}%
\else\global\CD@iG\CD@eD{\the\CD@TC\space\ifPositiveGradient\else-\fi\the
\CD@LH\space bturn}\fi\box\z@\CD@gK}}\def\CD@vB{\advance\CD@hF-\CD@mI\CD@wJ
\CD@hF\advance\CD@wJ\CD@hI\ifvoid\CD@sH\ifdim\CD@wJ<.1em\ifnum\CD@gD=\@m\else
\CD@aG h\CD@wJ<.1em:objects overprint:\CD@FA\CD@gD\fi\fi\else\ifhbox\CD@sH
\CD@SK\else\CD@TK\fi\advance\CD@wJ\CD@mI\CD@bH{-\CD@mI}{\box\CD@sH}{\CD@wJ}%
\z@\fi\CD@hF-\CD@mF\CD@gD\CD@FA\CD@hI\z@}\def\CD@SK{\setbox\CD@sH=\hbox{%
\unhbox\CD@sH\unskip\unpenalty}\setbox\CD@tH=\hbox{\unhbox\CD@tH\unskip
\unpenalty}\setbox\CD@sH=\hbox to\CD@wJ{\CD@OA\wd\CD@sH\unhbox\CD@sH\CD@PA
\lastkern\unkern\ifdim\CD@PA=\z@\CD@UB\advance\CD@OA-\wd\CD@tH\else\CD@TB\fi
\ifnum\lastpenalty=\z@\else\CD@JA\unpenalty\fi\kern\CD@PA\ifdim\CD@hF<\CD@OA
\CD@JA\fi\ifdim\CD@hI<\wd\CD@tH\CD@JA\fi\CD@jE\CD@hI\CD@wJ\advance\CD@hI-%
\CD@OA\advance\CD@hI\wd\CD@tH\ifdim\CD@hI<2\wd\CD@tH\CD@aG h\CD@hI<2\wd\CD@tH
:arrow too short:\CD@FA\CD@gD\fi\divide\CD@hI\tw@\CD@hF\CD@wJ\advance\CD@hF-%
\CD@hI\fi\CD@tE\kern-\CD@hI\fi\hbox to\CD@hI{\unhbox\CD@tH}\CD@HG}}\CD@tG
\ifinpile\inpiletrue\inpilefalse\inpilefalse\def\pile{\protect\CD@UJ\protect
\CD@uH}\def\CD@uH#1{\CD@l#1\CD@QD}\def\CD@UJ{\CD@nB{pile}\setbox0=\vtop
\bgroup\aftergroup\CD@lD\inpiletrue\let\CD@FE\empty\let\pile\CD@KF\let\CD@QD
\CD@PD\let\CD@GD\CD@FD\CD@yH\baselineskip.5\PileSpacing\lineskip.1\CD@zC
\relax\lineskiplimit\lineskip\mathsurround\z@\tabskip\z@\let\\\CD@wH}\def
\CD@l{\CD@DE\CD@YF\empty\halign\bgroup\hfil\CD@k\let\CD@FE\CD@d\let\\\CD@vH##%
\CD@MD\CD@ND\hfil\CD@Q\CD@R##\cr}\CD@rG\CD@NE{pile only allows one column.}%
\CD@rG\CD@UE{you left it out!}\def\CD@R{\CD@QD\CD@Q\relax\CD@YA{missing \CD@yC
\space inserted after \string\pile}\CD@NE}\def\CD@PD{\CD@MD\crcr\egroup
\egroup}\def\CD@GD{\CD@MD}\def\CD@FD{\CD@MD\relax\CD@QD\CD@YA{missing \CD@yC
\space inserted between \string\pile\space and \CD@HD}\CD@UE}\def\CD@QD{%
\CD@MD}\def\CD@lD{\vbox{\dimen1\dp0 \unvbox0 \setbox0=\lastbox\advance\dimen1%
\dp0 \nointerlineskip\box0 \nointerlineskip\setbox0=\null\dp0.5\dimen1\ht0-%
\dp0 \box0}\ifincommdiag\CD@tJ\penalty-9998 \fi\xdef\CD@YF{pile}}\def\CD@vH{%
\cr}\def\CD@wH{\noalign{\skip@\prevdepth\advance\skip@-\baselineskip
\prevdepth\skip@}}\def\CD@KF#1{#1}\def\CD@TK{\setbox\CD@sH=\vbox{\unvbox
\CD@sH\setbox1=\lastbox\setbox0=\box\voidb@x\CD@tF\setbox\CD@sH=\lastbox
\ifhbox\CD@sH\CD@rC\repeat\unvbox0 \global\CD@QA\CD@ZE}\CD@ZE\CD@QA}\def
\CD@rC{\CD@jE\setbox\CD@sH=\hbox{\unhbox\CD@sH\unskip\setbox\CD@sH=\lastbox
\unskip\unhbox\CD@sH}\ifdim\CD@wJ<\wd\CD@sH\CD@aG h\CD@wJ<\wd\CD@sH:arrow in
pile too short:\CD@FA\CD@gD\else\setbox\CD@sH=\hbox to\CD@wJ{\unhbox\CD@sH}%
\fi\else\CD@gJ\fi\setbox0=\vbox{\box\CD@sH\nointerlineskip\ifvoid0 \CD@tJ\box
1 \else\vskip\skip0 \unvbox0 \fi}\skip0=\lastskip\unskip}\def\CD@gJ{\penalty7
\noindent\unhbox\CD@sH\unskip\setbox\CD@sH=\lastbox\unskip\unhbox\CD@sH
\endgraf\setbox\CD@tH=\lastbox\unskip\setbox\CD@tH=\hbox{\CD@JG\unhbox\CD@tH
\unskip\unskip\unpenalty}\ifcase\prevgraf\cd@shouldnt P\or\ifdim\CD@wJ<\wd
\CD@tH\CD@aG h\CD@wJ<\wd\CD@sH:object in pile too wide:\CD@FA\CD@gD\setbox
\CD@sH=\hbox to\CD@wJ{\hss\unhbox\CD@tH\hss}\else\setbox\CD@sH=\hbox to\CD@wJ
{\hss\kern\CD@hF\unhbox\CD@tH\kern\CD@hI\hss}\fi\or\setbox\CD@sH=\lastbox
\unskip\CD@SK\else\cd@shouldnt Q\fi\unskip\unpenalty}\def\CD@cD{\CD@MJ\ifvoid
3 \setbox3=\null\ht3\axisheight\dp3-\ht3 \dimen3.5\CD@LF\else\dimen4\dp3
\dimen3.5\wd3 \setbox3=\CD@GG{\box3}\dp3\dimen4 \ifdim\ht3=-\dp3 \else\CD@TB
\fi\fi\dimen0\dimen3 \advance\dimen0-.5\CD@LF\setbox0\null\ht0\ht3\dp0\dp3\wd
0\wd3 \ifvoid6\else\setbox6\hbox{\unhbox6\kern\dimen0\kern2pt}\dimen0\wd6 \fi
\ifvoid7\else\setbox7\hbox{\kern2pt\kern\dimen3\unhbox7}\dimen3\wd7 \fi
\setbox3\hbox{\ifvoid6\else\kern-\dimen0\unhbox6\fi\unhbox3 \ifvoid7\else
\unhbox7\kern-\dimen3\fi}\ht3\ht0\dp3\dp0\wd3\wd0 \CD@tE\dimen4=\ht\CD@MH
\advance\dimen4\dp5 \advance\dimen4\dimen1 \let\CD@jD\empty\else\dimen4\ht3
\fi\setbox0\null\ht0\dimen4 \offinterlineskip\setbox8=\vbox spread2ex{\kern
\dimen5 \box1 \CD@iD\vfill\CD@tE\else\kern\CD@eJ\fi\box0}\ht8=\z@\setbox9=%
\vtop spread2ex{\kern-\ht3 \kern-\CD@eJ\box3 \CD@jD\vfill\box5 \kern\dimen1}%
\dp9=\z@\hskip\dimen0plus.0001fil \box9 \kern-\CD@LF\box8 \CD@kE\penalty2 \fi
\CD@tE\penalty1 \fi\kern\PileSpacing\kern-\PileSpacing\kern-.5\CD@LF\penalty
\CD@GB\null\kern\dimen3}\def\CD@cI{\ifhbox\CD@VA\CD@KB{clashing verticals}\ht
\CD@MH.5\dp\CD@VA\dp\CD@MH-\ht5 \CD@yB\ht\CD@MH\z@\dp\CD@MH\z@\fi\dimen1\dp
\CD@VA\CD@xA\prevgraf\unvbox\CD@VA\CD@wA\lastpenalty\unpenalty\setbox\CD@VA=%
\null\setbox\CD@lI=\hbox{\CD@JG\unhbox\CD@lI\unskip\unpenalty\dimen0\lastkern
\unkern\unkern\unkern\kern\dimen0 \CD@HG}\setbox\CD@lF=\hbox{\unhbox\CD@lF
\dimen0\lastkern\unkern\unkern\global\CD@QA\lastkern\unkern\kern\dimen0 }%
\CD@tF\ifnum\CD@xA>4 \CD@zI\repeat\unskip\unskip\advance\CD@mF.5\wd\CD@VA
\advance\CD@mF\wd\CD@lF\advance\CD@mI.5\wd\CD@VA\advance\CD@mI\wd\CD@lI\ifnum
\CD@FA=\CD@lA\CD@OA.5\wd\CD@VA\edef\CD@NK{\the\CD@OA}\fi\setbox\CD@VA=\hbox{%
\kern-\CD@mF\box\CD@lF\unhbox\CD@VA\box\CD@lI\kern-\CD@mI\penalty\CD@wA
\penalty\CD@NB}\ht\CD@VA\dimen1 \dp\CD@VA\z@\wd\CD@VA\CD@tB\CD@vB}\def\CD@zI{%
\ifdim\wd\CD@lF<\CD@QA\setbox\CD@lF=\hbox to\CD@QA{\CD@JG\unhbox\CD@lF}\fi
\advance\CD@xA\m@ne\setbox\CD@VA=\hbox{\box\CD@lF\unhbox\CD@VA}\unskip\setbox
\CD@lF=\lastbox\setbox\CD@lF=\hbox{\unhbox\CD@lF\unskip\unpenalty\dimen0%
\lastkern\unkern\unkern\global\CD@QA\lastkern\unkern\kern\dimen0 }}\def\CD@yB
{\dimen1\dp\CD@VA\ifhbox\CD@VA\CD@xB\else\CD@zB\fi\setbox\CD@VA=\vbox{%
\penalty\CD@NB}\dp\CD@VA-\dp\CD@MH\wd\CD@VA\CD@tB}\def\CD@zB{\unvbox\CD@VA
\CD@wA\lastpenalty\unpenalty\ifdim\dimen1<\ht\CD@MH\CD@aG v\dimen1<\ht\CD@MH:%
rows overprint:\CD@NB\CD@wA\fi}\def\CD@xB{\dimen0=\ht\CD@VA\setbox\CD@VA=%
\hbox\bgroup\advance\dimen1-\ht\CD@MH\unhbox\CD@VA\CD@xA\lastpenalty
\unpenalty\CD@wA\lastpenalty\unpenalty\global\CD@RA-\lastkern\unkern\setbox0=%
\lastbox\CD@tF\setbox\CD@VA=\hbox{\box0\unhbox\CD@VA}\setbox0=\lastbox\ifhbox
0 \CD@kJ\repeat\global\CD@SA-\lastkern\unkern\global\CD@QA\CD@JK\unhbox\CD@VA
\egroup\CD@JK\CD@QA\CD@bH{\CD@SA}{\box\CD@VA}{\CD@RA}{\dimen1}}\def\CD@kJ{%
\setbox0=\hbox to\wd0\bgroup\unhbox0 \unskip\unpenalty\dimen7\lastkern\unkern
\ifnum\lastpenalty=1 \unpenalty\CD@UB\else\CD@TB\fi\ifnum\lastpenalty=2
\unpenalty\dimen2.5\dimen0\advance\dimen2-.5\dimen1\advance\dimen2-%
\axisheight\else\dimen2\z@\fi\setbox0=\lastbox\dimen6\lastkern\unkern\setbox1%
=\lastbox\setbox0=\vbox{\unvbox0 \CD@tE\kern-\dimen1 \else\ifdim\dimen2=\z@
\else\kern\dimen2 \fi\fi}\ifdim\dimen0<\ht0 \CD@aG v\dimen0<\ht0:upper part of
vertical too short:{\CD@tE\CD@NB\else\CD@wA\fi}\CD@xA\else\setbox0=\vbox to%
\dimen0{\unvbox0}\fi\setbox1=\vtop{\unvbox1}\ifdim\dimen1<\dp1 \CD@aG v\dimen
1<\dp1:lower part of vertical too short:\CD@NB\CD@wA\else\setbox1=\vtop to%
\dimen1{\ifdim\dimen2=\z@\else\kern-\dimen2 \fi\unvbox1 }\fi\box1 \kern\dimen
6 \box0 \kern\dimen7 \CD@HG\global\CD@QA\CD@JK\egroup\CD@JK\CD@QA\relax}%
\countdef\CD@u=14 \newcount\CD@CA\newcount\CD@XB\newcount\CD@NB\let\CD@LB
\insc@unt\newcount\CD@FA\newcount\CD@lA\let\CD@mA\CD@XB\newcount\CD@MB\CD@tG
\CD@DF\CD@bI\CD@aI\CD@aI\def\CD@nD{-1}\def\CD@K{\ifnum\CD@nD<\z@\else
\begingroup\scrollmode\showboxdepth\CD@nD\showboxbreadth\maxdimen\showlists
\endgroup\fi\CD@bI\CD@zF\CD@CA=\CD@u\advance\CD@CA1 \CD@XB=\CD@CA\ifnum\CD@NB
=1 \CD@JA\fi\advance\CD@XB\CD@NB\dimen1\z@\skip0\z@\count@=\insc@unt\advance
\count@\CD@u\divide\count@2 \ifnum\CD@XB>\count@\CD@KB{The diagram has too
many rows! It can't be reformatted.}\else\CD@NG\CD@WI\fi\CD@cH}\def\CD@NG{%
\CD@NB\CD@CA\CD@uF\ifnum\CD@NB<\CD@XB\setbox\CD@NB\box\voidb@x\advance\CD@NB1%
\relax\repeat\CD@NB\CD@CA\skip\z@\z@\CD@uF\CD@GB\lastpenalty\unpenalty\ifnum
\CD@GB>\z@\CD@KE\repeat\ifnum\CD@GB=-123 \CD@tJ\unpenalty\else\cd@shouldnt D%
\fi\ifx\v@grid\relax\else\CD@NB\CD@XB\advance\CD@NB\m@ne\expandafter\CD@VJ
\v@grid\fi\CD@MB\CD@mA\CD@tB\z@\CD@XG\ifx\h@grid\relax\else\expandafter\CD@LJ
\h@grid\fi\count@\CD@XB\advance\count@\m@ne\CD@YB\ht\count@}\def\CD@KE{%
\ifcase\CD@GB\or\CD@MG\else\CD@uA-\lastpenalty\unpenalty\CD@vA\lastpenalty
\unpenalty\setbox0=\lastbox\CD@WG\fi\CD@wD}\def\CD@wD{\skip1\lastskip\unskip
\advance\skip0\skip1 \ifdim\skip1=\z@\else\expandafter\CD@wD\fi}\def\CD@MG{%
\setbox0=\lastbox\CD@pI\dp0 \advance\CD@pI\skip\z@\skip\z@\z@\advance\CD@NF
\CD@pI\CD@uE\ifnum\CD@NB>\CD@CA\CD@NF\DiagramCellHeight\CD@pI\CD@NF\advance
\CD@pI-\CD@qI\fi\fi\CD@qI\ht0 \CD@NF\CD@qI\setbox\CD@NB\hbox{\unhbox\CD@NB
\unhbox0}\dp\CD@NB\CD@pI\ht\CD@NB\CD@qI\advance\CD@NB1 }\def\CD@WG{\ifnum
\CD@uA<\z@\advance\CD@uA\CD@XB\ifnum\CD@uA<\CD@CA\CD@UG\else\CD@OA\dp\CD@uA
\CD@PA\ht\CD@uA\setbox\CD@uA\hbox{\box\z@\penalty\CD@vA\penalty\CD@GB\unhbox
\CD@uA}\dp\CD@uA\CD@OA\ht\CD@uA\CD@PA\fi\else\CD@UG\fi}\def\CD@UG{\CD@KB{%
diagonal goes outside diagram (lost)}}\def\CD@fI{\advance\CD@uA\CD@XB\ifnum
\CD@uA<\CD@CA\CD@UG\else\ifnum\CD@uA=\CD@NB\CD@VG\else\ifnum\CD@uA>\CD@NB
\cd@shouldnt M\else\CD@OA\dp\CD@uA\CD@PA\ht\CD@uA\setbox\CD@uA\hbox{\box\z@
\penalty\CD@vA\penalty\CD@GB\unhbox\CD@uA}\dp\CD@uA\CD@OA\ht\CD@uA\CD@PA\fi
\fi\fi}\def\CD@WI{\CD@t\CD@AJ\setbox\CD@PC=\hbox{\CD@k A\@super f\CD@lJ f%
\CD@ND}\CD@ZE\z@\CD@JK\z@\CD@kI\z@\CD@kF\z@\CD@NB=\CD@XB\CD@NF\z@\CD@uB\z@
\CD@uF\ifnum\CD@NB>\CD@CA\advance\CD@NB\m@ne\CD@qI\ht\CD@NB\CD@pI\dp\CD@NB
\advance\CD@NF\CD@qI\CD@rI\advance\CD@uB\CD@NF\CD@KC\CD@ZI\CD@w\ht\CD@NB
\CD@qI\dp\CD@NB\CD@pI\nointerlineskip\box\CD@NB\CD@NF\CD@pI\setbox\CD@NB\null
\ht\CD@NB\CD@uB\repeat\CD@wB\nointerlineskip\box\CD@NB\CD@gG\CD@ZE
\DiagramCellWidth{width}\CD@gG\CD@JK\DiagramCellHeight{height}\CD@VA\CD@LB
\advance\CD@VA-\CD@lA\advance\CD@VA\m@ne\advance\CD@VA\CD@mA\dimen0\wd\CD@VA
\CD@tI\axisheight\dimen1\CD@uB\advance\dimen1-\CD@YB\dimen2\CD@kI\advance
\dimen2-\dimen0 \advance\CD@XB-\CD@CA\advance\CD@LB-\CD@lA}\count@\year
\multiply\count@12 \advance\count@\month\ifnum\count@>24254 \loop\iftrue
\repeat\fi\def\CD@wB{\CD@qI-\CD@NF\CD@pI\CD@NF
\setbox\CD@MH=\null\dp\CD@MH\CD@NF\ht\CD@MH-\CD@NF\CD@mF\z@\CD@mI\z@\CD@lA
\CD@LB\advance\CD@lA-\CD@MB\advance\CD@lA\CD@mA\CD@FA\CD@LB\CD@VA\CD@MB\CD@sF
\ifnum\CD@FA>\CD@lA\advance\CD@FA\m@ne\advance\CD@VA\m@ne\CD@tB\wd\CD@VA
\setbox\CD@FA=\box\voidb@x\CD@yB\repeat\CD@w\ht\CD@NB\CD@qI\dp\CD@NB\CD@pI}%
\def\CD@gG#1#2#3{\ifdim#1>.01\CD@zC\CD@PA#2\relax\advance\CD@PA#1\relax
\advance\CD@PA.99\CD@zC\count@\CD@PA\divide\count@\CD@zC\CD@KB{increase cell #%
3 to \the\count@ em}\fi}\def\CD@rI{\CD@FA=\CD@LB\penalty4 \noindent\unhbox
\CD@NB\CD@sF\unskip\setbox0=\lastbox\ifhbox0 \advance\CD@FA\m@ne\setbox\CD@FA
\hbox to\wd0{\null\penalty-9990\null\unhbox0}\repeat\CD@lA\CD@FA\advance
\CD@FA\CD@MB\advance\CD@FA-\CD@mA\ifnum\CD@FA<\CD@LB\count@\CD@FA\advance
\count@\m@ne\dimen0=\wd\count@\count@\CD@MB\advance\count@\m@ne\CD@tB\wd
\count@\CD@sF\ifnum\CD@FA<\CD@LB\CD@DJ\CD@XG\dimen0\wd\CD@FA\advance\CD@FA1
\repeat\fi\CD@sF\CD@GB\lastpenalty\unpenalty\ifnum\CD@GB>\z@\CD@vA
\lastpenalty\unpenalty\CD@VG\repeat\endgraf\unskip\ifnum\lastpenalty=4
\unpenalty\else\cd@shouldnt S\fi}\def\CD@VG{\advance\CD@vA\CD@lA\advance
\CD@vA\m@ne\setbox0=\lastbox\ifnum\CD@vA<\CD@LB\setbox\CD@vA\hbox{\box0%
\penalty\CD@GB\unhbox\CD@vA}\else\CD@UG\fi}\def\CD@bG{}\CD@tG\CD@uE\CD@WB
\CD@VB\def\CD@DJ{\advance\dimen0\wd\CD@FA\divide\dimen0\tw@\CD@uE\dimen0%
\DiagramCellWidth\else\CD@V{\dimen0}\DiagramCellWidth\CD@pJ\fi\advance\CD@tB
\dimen0 }\def\CD@XG{\setbox\CD@MB=\vbox{}\dp\CD@MB=\CD@uB\wd\CD@MB\CD@tB
\advance\CD@MB1 }\def\CD@LJ#1,{\def\CD@GK{#1}\ifx\CD@GK\CD@RD\else\advance
\CD@tB\CD@GK\DiagramCellWidth\CD@XG\expandafter\CD@LJ\fi}\def\CD@VJ#1,{\def
\CD@GK{#1}\ifx\CD@GK\CD@RD\else\ifnum\CD@NB>\CD@CA\CD@NF\CD@GK
\DiagramCellHeight\advance\CD@NF-\dp\CD@NB\advance\CD@NB\m@ne\ht\CD@NB\CD@NF
\fi\expandafter\CD@VJ\fi}\def\CD@pJ{\CD@wE\CD@OA\dimen0 \advance\CD@OA-%
\DiagramCellWidth\ifdim\CD@OA>2\MapShortFall\CD@KB{badly drawn diagonals (see
manual)}\let\CD@pJ\empty\fi\else\let\CD@pJ\empty\fi}\def\CD@KC{\CD@VA\CD@mA
\CD@sF\ifnum\CD@VA<\CD@MB\dimen0\dp\CD@VA\advance\dimen0\CD@NF\dp\CD@VA\dimen
0 \advance\CD@VA1 \repeat}\def\CD@bH#1#2#3#4{\ifnum\CD@FA<\CD@LB\CD@OA=#1%
\relax\setbox\CD@FA=\hbox{\setbox0=#2\dimen7=#4\relax\dimen8=#3\relax\ifhbox
\CD@FA\unhbox\CD@FA\advance\CD@OA-\lastkern\unkern\fi\ifdim\CD@OA=\z@\else
\kern-\CD@OA\fi\raise\dimen7\box0 \kern-\dimen8 }\ifnum\CD@FA=\CD@lA\CD@V
\CD@kF\CD@OA\fi\else\cd@shouldnt O\fi}\def\CD@w{\setbox\CD@NB=\hbox{\CD@FA
\CD@lA\CD@VA\CD@mA\CD@PA\z@\relax\CD@sF\ifnum\CD@FA<\CD@LB\CD@tB\wd\CD@VA
\relax\CD@eI\advance\CD@FA1 \advance\CD@VA1 \repeat}\CD@V\CD@kI{\wd\CD@NB}\wd
\CD@NB\z@}\def\CD@eI{\ifhbox\CD@FA\CD@OA\CD@tB\relax\advance\CD@OA-\CD@PA
\relax\ifdim\CD@OA=\z@\else\kern\CD@OA\fi\CD@PA\CD@tB\advance\CD@PA\wd\CD@FA
\relax\unhbox\CD@FA\advance\CD@PA-\lastkern\unkern\fi}\def\CD@ZI{\setbox
\CD@sH=\box\voidb@x\CD@VA=\CD@MB\CD@FA\CD@LB\CD@VA\CD@mA\advance\CD@VA\CD@FA
\advance\CD@VA-\CD@lA\advance\CD@VA\m@ne\CD@tB\wd\CD@VA\count@\CD@LB\advance
\count@\m@ne\CD@hF.5\wd\count@\advance\CD@hF\CD@tB\CD@A\m@ne\CD@gD\@m\CD@sF
\ifnum\CD@FA>\CD@lA\advance\CD@FA\m@ne\advance\CD@hF-\CD@tB\CD@PI\wd\CD@VA
\CD@tB\advance\CD@hF\CD@tB\advance\CD@VA\m@ne\CD@tB\wd\CD@VA\repeat\CD@mF
\CD@kF\CD@mI-\CD@mF\CD@vB}\newcount\CD@GB\def\CD@s{}\def\CD@t{\mathsurround
\z@\hsize\z@\rightskip\z@ plus1fil minus\maxdimen\parfillskip\z@\linepenalty
9000 \looseness0 \hfuzz\maxdimen\hbadness10000 \clubpenalty0 \widowpenalty0
\displaywidowpenalty0 \interlinepenalty0 \predisplaypenalty0
\postdisplaypenalty0 \interdisplaylinepenalty0 \interfootnotelinepenalty0
\floatingpenalty0 \brokenpenalty0 \everypar{}\leftskip\z@\parskip\z@
\parindent\z@\pretolerance10000 \tolerance10000 \hyphenpenalty10000
\exhyphenpenalty10000 \binoppenalty10000 \relpenalty10000 \adjdemerits0
\doublehyphendemerits0 \finalhyphendemerits0 \baselineskip\z@\CD@IA\prevdepth
\z@}\newbox\CD@KG\newbox\CD@IG\def\CD@JG{\unhcopy\CD@KG}\def\CD@HG{\unhcopy
\CD@IG}\def\CD@iJ{\hbox{}\penalty1\nointerlineskip}\def\CD@PI{\penalty5
\noindent\setbox\CD@MH=\null\CD@mF\z@\CD@mI\z@\ifnum\CD@FA<\CD@LB\ht\CD@MH\ht
\CD@FA\dp\CD@MH\dp\CD@FA\unhbox\CD@FA\skip0=\lastskip\unskip\else\CD@OK\skip0%
=\z@\fi\endgraf\ifcase\prevgraf\cd@shouldnt Y \or\cd@shouldnt Z \or\CD@RI\or
\CD@XI\else\CD@QI\fi\unskip\setbox0=\lastbox\unskip\unskip\unpenalty\noindent
\unhbox0\setbox0\lastbox\unpenalty\unskip\unskip\unpenalty\setbox0\lastbox
\CD@tF\CD@GB\lastpenalty\unpenalty\ifnum\CD@GB>\z@\setbox\z@\lastbox\CD@lB
\repeat\endgraf\unskip\unskip\unpenalty}\def\CD@YJ{\CD@uA\CD@XB\advance\CD@uA
-\CD@NB\CD@vA\CD@FA\advance\CD@vA-\CD@lA\advance\CD@vA1 \expandafter\message{%
prevgraf=\the\prevgraf at (\the\CD@uA,\the\CD@vA)}}\def\CD@XI{\CD@CE\setbox
\CD@lI=\lastbox\setbox\CD@lI=\hbox{\unhbox\CD@lI\unskip\unpenalty}\unskip
\ifdim\ht\CD@lI>\ht\CD@PC\setbox\CD@MH=\copy\CD@lI\else\ifdim\dp\CD@lI>\dp
\CD@PC\setbox\CD@MH=\copy\CD@lI\else\CD@FG\CD@lI\fi\fi\advance\CD@mF.5\wd
\CD@lI\advance\CD@mI.5\wd\CD@lI\setbox\CD@lI=\hbox{\unhbox\CD@lI\CD@HG}\CD@bH
\CD@mF{\box\CD@lI}\CD@mI\z@\CD@yB\CD@vB}\def\CD@CE{\ifnum\CD@A>0 \advance
\dimen0-\CD@tB\CD@iA-.5\dimen0 \CD@A-\CD@A\else\CD@A0 \CD@iA\z@\fi\setbox
\CD@MH=\lastbox\setbox\CD@MH=\hbox{\unhbox\CD@MH\unskip\unskip\unpenalty
\setbox0=\lastbox\global\CD@QA\lastkern\unkern}\advance\CD@iA-.5\CD@QA\unskip
\setbox\CD@MH=\null\CD@mI\CD@iA\CD@mF-\CD@iA}\def\CD@Z{\ht\CD@MH\CD@tI\dp
\CD@MH\CD@sI}\def\CD@FG#1{\setbox\CD@MH=\hbox{\CD@V{\ht\CD@MH}{\ht#1}\CD@V{%
\dp\CD@MH}{\dp#1}\CD@V{\wd\CD@MH}{\wd#1}\vrule height\ht\CD@MH depth\dp\CD@MH
width\wd\CD@MH}}\def\CD@QI{\CD@CE\CD@Z\setbox\CD@lI=\lastbox\unskip\setbox
\CD@lF=\lastbox\unskip\setbox\CD@lF=\hbox{\unhbox\CD@lF\unskip\global\CD@yA
\lastpenalty\unpenalty}\advance\CD@yA9999 \ifcase\CD@yA\CD@VI\or\CD@YI\or
\CD@TI\or\CD@dI\or\CD@cI\or\CD@SI\else\cd@shouldnt9\fi}\def\CD@VI{\CD@FG
\CD@lI\CD@UI\setbox\CD@sH=\box\CD@lF\setbox\CD@tH=\box\CD@lI}\def\CD@YI{%
\CD@FG\CD@lF\setbox\CD@lI\hbox{\penalty8 \unhbox\CD@lI\unskip\unpenalty\ifnum
\lastpenalty=8 \else\CD@xH\fi}\CD@UI\setbox\CD@lF=\hbox{\unhbox\CD@lF\unskip
\unpenalty\global\setbox\CD@DA=\lastbox}\ifdim\wd\CD@lF=\z@\else\CD@xH\fi
\setbox\CD@sH=\box\CD@DA}\def\CD@xH{\CD@KB{extra material in \string\pile
\space cell (lost)}}\def\CD@UI{\CD@yB\ifvoid\CD@sH\else\CD@KB{Clashing
horizontal arrows}\CD@mI.5\CD@hF\CD@mF-\CD@mI\CD@vB\CD@mI\z@\CD@mF\z@\fi
\CD@hI\CD@hF\advance\CD@hI-\CD@mI\CD@hF-\CD@mF\CD@JC\CD@FA}\def\CD@RI{\setbox
0\lastbox\unskip\CD@iA\z@\CD@Z\ifdim\skip0>\z@\CD@tJ\CD@A0 \else\ifnum\CD@A<1
\CD@A0 \dimen0\CD@tB\fi\advance\CD@A1 \fi}\def\VonH{\CD@MA46\VonH{.5\CD@LF}}%
\def\HonV{\CD@MA57\HonV{.5\CD@LF}}\def\HmeetV{\CD@MA44\HmeetV{-\MapShortFall}%
}\def\CD@MA#1#2#3#4{\CD@pB34#1{\string#3}\CD@SD\CD@GB-999#2 \dimen0=#4\CD@tI
\dimen0\advance\CD@tI\axisheight\CD@sI\dimen0\advance\CD@sI-\axisheight\CD@CF
\CD@HC\CD@ZD}\def\CD@HC#1{\setbox0=\hbox{\CD@k#1\CD@ND}\dimen0.5\wd0 \CD@tI
\ht0 \CD@sI\dp0 \CD@ZD}\def\CD@SD{\setbox0=\null\ht0=\CD@tI\dp0=\CD@sI\wd0=%
\dimen0 \copy0\penalty\CD@GB\box0 }\def\CD@TI{\CD@GC\CD@yB}\def\CD@dI{\CD@GC
\CD@vB}\def\CD@SI{\CD@GC\CD@yB\CD@vB}\def\CD@GC{\setbox\CD@lI=\hbox{\unhbox
\CD@lI}\setbox\CD@lF=\hbox{\unhbox\CD@lF\global\setbox\CD@DA=\lastbox}\ht
\CD@MH\ht\CD@DA\dp\CD@MH\dp\CD@DA\advance\CD@mF\wd\CD@DA\advance\CD@mI\wd
\CD@lI}\CD@tG\ifPositiveGradient\CD@CI\CD@BI\CD@CI\CD@tG\ifClimbing\CD@rB
\CD@qB\CD@rB\newcount\DiagonalChoice\DiagonalChoice\m@ne\ifx\tenln\nullfont
\CD@tJ\def\CD@qF{\CD@KH\ifPositiveGradient/\else\CD@k\backslash\CD@ND\fi}%
\else\def\CD@qF{\CD@rF\char\count@}\fi\let\CD@rF\tenln\def\Use@line@char#1{%
\hbox{#1\CD@rF\ifPositiveGradient\else\advance\count@64 \fi\char\count@}}\def
\CD@cF{\Use@line@char{\count@\CD@TC\multiply\count@8\advance\count@-9\advance
\count@\CD@LH}}\def\CD@ZF{\Use@line@char{\ifcase\DiagonalChoice\CD@gF\or
\CD@fF\or\CD@fF\else\CD@gF\fi}}\def\CD@gF{\ifnum\CD@TC=\z@\count@'33 \else
\count@\CD@TC\multiply\count@\sixt@@n\advance\count@-9\advance\count@\CD@LH
\advance\count@\CD@LH\fi}\def\CD@fF{\count@'\ifcase\CD@LH55\or\ifcase\CD@TC66%
\or22\or52\or61\or72\fi\or\ifcase\CD@TC66\or25\or22\or63\or52\fi\or\ifcase
\CD@TC66\or16\or36\or22\or76\fi\or\ifcase\CD@TC66\or27\or25\or67\or22\fi\fi
\relax}\def\CD@uC#1{\hbox{#1\setbox0=\Use@line@char{#1}\ifPositiveGradient
\else\raise.3\ht0\fi\copy0 \kern-.7\wd0 \ifPositiveGradient\raise.3\ht0\fi
\box0}}\def\CD@jF#1{\hbox{\setbox0=#1\kern-.75\wd0 \vbox to.25\ht0{%
\ifPositiveGradient\else\vss\fi\box0 \ifPositiveGradient\vss\fi}}}\def\CD@jI#%
1{\hbox{\setbox0=#1\dimen0=\wd0 \vbox to.25\ht0{\ifPositiveGradient\vss\fi
\box0 \ifPositiveGradient\else\vss\fi}\kern-.75\dimen0 }}\CD@RC{+h:>}{%
\Use@line@char\CD@fF}\CD@RC{-h:>}{\Use@line@char\CD@gF}\CD@nF{+t:<}{-h:>}%
\CD@nF{-t:<}{+h:>}\CD@RC{+t:>}{\CD@jF{\Use@line@char\CD@fF}}\CD@RC{-t:>}{%
\CD@jI{\Use@line@char\CD@gF}}\CD@nF{+h:<}{-t:>}\CD@nF{-h:<}{+t:>}\CD@RC{+h:>>%
}{\CD@uC\CD@fF}\CD@RC{-h:>>}{\CD@uC\CD@gF}\CD@nF{+t:<<}{-h:>>}\CD@nF{-t:<<}{+%
h:>>}\CD@nF{+h:>->}{+h:>>}\CD@nF{-h:>->}{-h:>>}\CD@nF{+t:<-<}{-h:>>}\CD@nF{-t%
:<-<}{+h:>>}\CD@RC{+t:>>}{\CD@jF{\CD@uC\CD@fF}}\CD@RC{-t:>>}{\CD@jI{\CD@uC
\CD@gF}}\CD@nF{+h:<<}{-t:>>}\CD@nF{-h:<<}{+t:>>}\CD@nF{+t:>->}{+t:>>}\CD@nF{-%
t:>->}{-t:>>}\CD@nF{+h:<-<}{-t:>>}\CD@nF{-h:<-<}{+t:>>}\CD@RC{+f:-}{\CD@EF
\null\else\CD@cF\fi}\CD@nF{-f:-}{+f:-}\def\CD@tC#1#2{\vbox to#1{\vss\hbox to#%
2{\hss.\hss}\vss}}\def\hfdot{\CD@tC{2\axisheight}{.5em}}%
\def\vfdot{\CD@tC{1ex}\z@}
\def\CD@bF{\hbox{\dimen0=.3\CD@zC\dimen1\dimen0 \ifnum\CD@LH>\CD@TC\CD@iC{%
\dimen1}\else\CD@dG{\dimen0}\fi\CD@tC{\dimen0}{\dimen1}}}\newarrowfiller{.}%
\hfdot\hfdot\vfdot\vfdot\def\dfdot{\CD@bF\CD@CK}\CD@RC{+f:.}{\dfdot}\CD@RC{-f%
:.}{\dfdot}\def\CD@@K#1{\hbox\bgroup\def\CD@CH{#1\egroup}\afterassignment
\CD@CH
\count@='}\def\lnchar{\CD@@K\CD@qF}\def\CD@dF#1{\setbox#1=\hbox{\dimen5\dimen
#1 \setbox8=\box#1 \dimen1\wd8 \count@\dimen5 \divide\count@\dimen1 \ifnum
\count@=0 \box8 \ifdim\dimen5<.95\dimen1 \CD@gB{diagonal line too short}\fi
\else\dimen3=\dimen5 \advance\dimen3-\dimen1 \divide\dimen3\count@\dimen4%
\dimen3 \CD@dG{\dimen4}\ifPositiveGradient\multiply\dimen4\m@ne\fi\dimen6%
\dimen1 \advance\dimen6-\dimen3 \loop\raise\count@\dimen4\copy8 \ifnum\count@
>0 \kern-\dimen6 \advance\count@\m@ne\repeat\fi}}\def\CD@CG#1{\CD@EF\CD@xJ{#1%
}\else\CD@dF{#1}\fi}\def\CD@IH#1{}\newdimen\objectheight\objectheight1.8ex
\newdimen\objectwidth\objectwidth1em \def\CD@YD{\dimen6=\CD@aK
\DiagramCellHeight\dimen7=\CD@WK\DiagramCellWidth\CD@KJ\ifnum\CD@LH>0 \ifnum
\CD@TC>0 \CD@aF\else\aftergroup\CD@VC\fi\else\aftergroup\CD@UC\fi}\def\CD@VC{%
\CD@YA{diagonal map is nearly vertical}\CD@NA}\def\CD@UC{\CD@YA{diagonal map
is nearly horizontal}\CD@NA}\CD@rG\CD@NA{Use an orthogonal map instead}\def
\CD@aF{\CD@MJ\dimen3\dimen7\dimen7\dimen6\CD@iC{\dimen7}\advance\dimen3-%
\dimen7 \CD@MF\ifnum\CD@LH>\CD@TC\advance\dimen6-\dimen1\advance\dimen6-%
\dimen5 \CD@iC{\dimen1}\CD@iC{\dimen5}\else\dimen0\dimen1\advance\dimen0%
\dimen5\CD@dG{\dimen0}\advance\dimen6-\dimen0 \fi\dimen2.5\dimen7\advance
\dimen2-\dimen1 \dimen4.5\dimen7\advance\dimen4-\dimen5 \ifPositiveGradient
\dimen0\dimen5 \advance\dimen1-\CD@WK\DiagramCellWidth\advance\dimen1 \CD@ZK
\DiagramCellWidth\setbox6=\llap{\unhbox6\kern.1\ht2}\setbox7=\rlap{\kern.1\ht
2\unhbox7}\else\dimen0\dimen1 \advance\dimen1-\CD@ZK\DiagramCellWidth\setbox7%
=\llap{\unhbox7\kern.1\ht2}\setbox6=\rlap{\kern.1\ht2\unhbox6}\fi\setbox6=%
\vbox{\box6\kern.1\wd2}\setbox7=\vtop{\kern.1\wd2\box7}\CD@dG{\dimen0}%
\advance\dimen0-\axisheight\advance\dimen0-\CD@bK\DiagramCellHeight\dimen5-%
\dimen0 \advance\dimen0\dimen6 \advance\dimen1.5\dimen3 \ifdim\wd3>\z@\ifdim
\ht3>-\dp3\CD@TB\fi\fi\dimen3\dimen2 \dimen7\dimen2\advance\dimen7\dimen4
\ifvoid3 \else\CD@tE\else\dimen8\ht3\advance\dimen8-\axisheight\CD@iC{\dimen8%
}\CD@X{\dimen8}{.5\wd3}\dimen9\dp3\advance\dimen9\axisheight\CD@iC{\dimen9}%
\CD@X{\dimen9}{.5\wd3}\ifPositiveGradient\advance\dimen2-\dimen9\advance
\dimen4-\dimen8 \else\advance\dimen4-\dimen9\advance\dimen2-\dimen8 \fi\fi
\advance\dimen3-.5\wd3 \fi\dimen9=\CD@aK\DiagramCellHeight\advance\dimen9-2%
\DiagramCellHeight\CD@tE\advance\dimen2\dimen4 \CD@CG{2}\dimen2-\dimen0%
\advance\dimen2\dp2 \else\CD@CG{2}\CD@CG{4}\ifPositiveGradient\dimen2-\dimen0%
\advance\dimen2\dp2 \dimen4\dimen5\advance\dimen4-\ht4 \else\dimen4-\dimen0%
\advance\dimen4\dp4 \dimen2\dimen5\advance\dimen2-\ht2 \fi\fi\setbox0=\hbox to%
\z@{\kern\dimen1 \ifvoid1 \else\ifPositiveGradient\advance\dimen0-\dp1 \lower
\dimen0 \else\advance\dimen5-\ht1 \raise\dimen5 \fi\rlap{\unhbox1}\fi\raise
\dimen2\rlap{\unhbox2}\ifvoid3 \else\lower.5\dimen9\rlap{\kern\dimen3\unhbox3%
}\fi\kern.5\dimen7 \lower.5\dimen9\box6 \lower.5\dimen9\box7 \kern.5\dimen7
\CD@tE\else\raise\dimen4\llap{\unhbox4}\fi\ifvoid5 \else\ifPositiveGradient
\advance\dimen5-\ht5 \raise\dimen5 \else\advance\dimen0-\dp5 \lower\dimen0 \fi
\llap{\unhbox5}\fi\hss}\ht0=\axisheight\dp0=-\ht0\box0 }\def\NorthWest{\CD@BI
\CD@rB\DiagonalChoice0 }\def\NorthEast{\CD@CI\CD@rB\DiagonalChoice1 }\def
\SouthWest{\CD@CI\CD@qB\DiagonalChoice3 }\def\SouthEast{\CD@BI\CD@qB
\DiagonalChoice2 }\def\CD@aD{\vadjust{\CD@uA\CD@FA\advance\CD@uA
\ifPositiveGradient\else-\fi\CD@ZK\relax\CD@vA\CD@NB\advance\CD@vA-\CD@bK
\relax\hbox{\advance\CD@uA\ifPositiveGradient-\fi\CD@WK\advance\CD@vA\CD@aK
\hbox{\box6 \kern\CD@DC\kern\CD@eJ\penalty1 \box7 \box\z@}\penalty\CD@uA
\penalty\CD@vA}\penalty\CD@uA\penalty\CD@vA\penalty104}}\def\CD@eH#1{\relax
\vadjust{\hbox@maths{#1}\penalty\CD@FA\penalty\CD@NB\penalty\tw@}}\def\CD@lB{%
\ifcase\CD@GB\or\or\CD@bH{.5\wd0}{\box0}{.5\wd0}\z@\or\unhbox\z@\setbox\z@
\lastbox\CD@bH{.5\wd0}{\box0}{.5\wd0}\z@\unpenalty\unpenalty\setbox\z@
\lastbox\or\CD@TG\else\advance\CD@GB-100 \ifnum\CD@GB<\z@\cd@shouldnt B\fi
\setbox\z@\hbox{\kern\CD@mF\copy\CD@MH\kern\CD@mI\CD@uA\CD@XB\advance\CD@uA-%
\CD@NB\penalty\CD@uA\CD@uA\CD@FA\advance\CD@uA-\CD@lA\penalty\CD@uA\unhbox\z@
\global\CD@yA\lastpenalty\unpenalty\global\CD@zA\lastpenalty\unpenalty}\CD@uA
-\CD@yA\CD@vA\CD@zA\CD@fI\fi}\def\CD@TG{\unhbox\z@\setbox\z@\lastbox\CD@uA
\lastpenalty\unpenalty\advance\CD@uA\CD@mA\CD@vA\CD@XB\advance\CD@vA-%
\lastpenalty\unpenalty\dimen1\lastkern\unkern\setbox3\lastbox\dimen0\lastkern
\unkern\setbox0=\hbox to\z@{\unhbox0\setbox0\lastbox\setbox7\lastbox
\unpenalty\CD@eJ\lastkern\unkern\CD@DC\lastkern\unkern\setbox6\lastbox\dimen7%
\CD@tB\advance\dimen7-\wd\CD@uA\ifdim\dimen7<\z@\CD@CI\multiply\dimen7\m@ne
\let\mv\empty\else\CD@BI\def\mv{\raise\ht1}\kern-\dimen7 \fi\ifnum\CD@vA>%
\CD@NB\dimen6\CD@uB\advance\dimen6-\ht\CD@vA\else\dimen6\z@\fi\CD@jJ\CD@mK
\setbox1\null\ht1\dimen6\wd1\dimen7 \dimen7\dimen2 \dimen6\wd1 \CD@KJ\CD@uA
\CD@LH\CD@vA\CD@TC\dimen6\ht1 \CD@KJ\setbox2\null\divide\dimen2\tw@\advance
\dimen2\CD@eJ\CD@eG{\dimen2}\wd2\dimen2 \dimen0.5\dimen7 \advance\dimen0%
\ifPositiveGradient\else-\fi\CD@eJ\CD@dG{\dimen0}\advance\dimen0-\axisheight
\ht2\dimen0 \dimen0\CD@DC\CD@eG{\dimen0}\advance\dimen0\ht2\ht2\dimen0 \dimen
0\ifPositiveGradient-\fi\CD@DC\CD@dG{\dimen0}\advance\dimen0\wd2\wd2\dimen0
\setbox4\null\dimen0 .6\CD@zC\CD@eG{\dimen0}\ht4\dimen0 \dimen0 .2\CD@zC
\CD@dG{\dimen0}\wd4\dimen0 \dimen0\wd2 \ifvoid6\else\dimen1\ht4 \advance
\dimen1\ht2 \CD@CC6+-\raise\dimen1\rlap{\ifPositiveGradient\advance\dimen0-%
\wd6\advance\dimen0-\wd4 \else\advance\dimen0\wd4 \fi\kern\dimen0\box6}\fi
\dimen0\wd2 \ifvoid7\else\dimen1\ht4 \advance\dimen1-\ht2 \CD@CC7-+\lower
\dimen1\rlap{\ifPositiveGradient\advance\dimen0\wd4 \else\advance\dimen0-\wd7%
\advance\dimen0-\wd4 \fi\kern\dimen0\box7}\fi\mv\box0\hss}\ht0\z@\dp0\z@
\CD@bH{\z@}{\box\z@}{\z@}{\axisheight}}\def\CD@CC#1#2#3{\dimen4.5\wd#1 \ifdim
\dimen4>.25\dimen7\dimen4=.25\dimen7\fi\ifdim\dimen4>\CD@zC\dimen4.4\dimen4
\advance\dimen4.6\CD@zC\fi\CD@eG{\dimen4}\dimen5\axisheight\CD@dG{\dimen5}%
\advance\dimen4-\dimen5 \dimen5\dimen4\CD@eG{\dimen5}\advance\dimen0%
\ifPositiveGradient#2\else#3\fi\dimen5 \CD@dG{\dimen4}\advance\dimen1\dimen4 }
\def\CD@eD#1{\expandafter\CD@IK{#1}}\CD@ZA\CD@EK{output is PostScript
dependent}\def\CD@SC{\CD@IK{/bturn {gsave currentpoint currentpoint translate
4 2 roll neg exch atan rotate neg exch neg exch translate } def /eturn {%
currentpoint grestore moveto} def}}\def\CD@gK{\relax\CD@hK\CD@tK{Q}\else
\CD@IK{eturn}\fi} \def\CD@OJ#1{\count@#1\relax\multiply\count@7\advance
\count@16577\divide\count@33154 }\def\CD@fD#1{\expandafter\special{#1}} \def
\CD@xJ#1{\setbox#1=\hbox{\dimen0\dimen#1\CD@dG{\dimen0}\CD@OJ{\dimen0}\setbox
0=\null\ifPositiveGradient\count@-\count@\ht0\dimen0 \else\dp0\dimen0 \fi\box
0 \CD@uA\count@\CD@OJ\CD@LF\CD@fD{pn \the\count@}\CD@fD{pa 0 0}\CD@OJ{\dimen#%
1}\CD@fD{pa \the\count@\space\the\CD@uA}\CD@fD{fp}\kern\dimen#1}}\def\CD@JI{%
\CD@KJ\begingroup\ifdim\dimen7<\dimen6 \dimen2=\dimen6 \dimen6=\dimen7 \dimen
7=\dimen2 \count@\CD@LH\CD@LH\CD@TC\CD@TC\count@\else\dimen2=\dimen7 \fi
\ifdim\dimen6>.01\p@\CD@KI\global\CD@QA\dimen0 \else\global\CD@QA\dimen7 \fi
\endgroup\dimen2\CD@QA\CD@iK\CD@lK{\ifPositiveGradient\else-\fi\dimen6}\CD@iK
\CD@kK{\ifPositiveGradient-\fi\dimen6}\CD@iK\CD@eK{\dimen7}}\def\CD@KI{\CD@hJ
\ifdim\dimen7>1.73\dimen6 \divide\dimen2 4 \multiply\CD@TC2 \else\dimen2=0.%
353553\dimen2 \advance\CD@LH-\CD@TC\multiply\CD@TC4 \fi\dimen0=4\dimen2 \CD@ZG
4\CD@ZG{-2}\CD@ZG2\CD@ZG{-2.5}}\def\CD@AI{\begingroup\count@\dimen0 \dimen2 45%
pt \divide\count@\dimen2 \ifdim\dimen0<\z@\advance\count@\m@ne\fi\ifodd
\count@\advance\count@1\CD@@A\else\CD@y\fi\advance\dimen0-\count@\dimen2
\CD@gE\multiply\dimen0\m@ne\fi\ifnum\count@<0 \multiply\count@-7 \fi\dimen3%
\dimen1 \dimen6\dimen0 \dimen7 3754936sp \ifdim\dimen0<6\p@\def\CD@OG{4000}%
\fi\CD@KJ\dimen2\dimen3\CD@dG{\dimen2}\CD@hJ\multiply\CD@TC-6 \dimen0\dimen2
\CD@ZG1\CD@ZG{0.3}\dimen1\dimen0 \dimen2\dimen3 \dimen0\dimen3 \CD@ZG3\CD@ZG{%
1.5}\CD@ZG{0.3}\divide\count@2 \CD@gE\multiply\dimen1\m@ne\fi\ifodd\count@
\dimen2\dimen1\dimen1\dimen0\dimen0-\dimen2 \fi\divide\count@2 \ifodd\count@
\multiply\dimen0\m@ne\multiply\dimen1\m@ne\fi\global\CD@QA\dimen0\global
\CD@RA\dimen1\endgroup\dimen6\CD@QA\dimen7\CD@RA}\def\CD@OC{255}\let\CD@OG
\CD@OC\def\CD@KJ{\begingroup\ifdim\dimen7<\dimen6 \dimen9\dimen7\dimen7\dimen
6\dimen6\dimen9\CD@@A\else\CD@y\fi\dimen2\z@\dimen3\CD@XH\dimen4\CD@XH\dimen0%
\z@\dimen8=\CD@OG\CD@XH\CD@lC\global\CD@yA\dimen\CD@gE0\else3\fi\global\CD@zA
\dimen\CD@gE3\else0\fi\endgroup\CD@LH\CD@yA\CD@TC\CD@zA}\def\CD@lC{\count@
\dimen6 \divide\count@\dimen7 \advance\dimen6-\count@\dimen7 \dimen9\dimen4
\advance\dimen9\count@\dimen0 \ifdim\dimen9>\dimen8 \CD@@C\else\CD@AC\ifdim
\dimen6>\z@\dimen9\dimen6 \dimen6\dimen7 \dimen7\dimen9 \expandafter
\expandafter\expandafter\CD@lC\fi\fi}\def\CD@@C{\ifdim\dimen0=\z@\ifdim\dimen
9<2\dimen8 \dimen0\dimen8 \fi\else\advance\dimen8-\dimen4 \divide\dimen8%
\dimen0 \ifdim\count@\CD@XH<2\dimen8 \count@\dimen8 \dimen9\dimen4 \advance
\dimen9\count@\dimen0 \CD@AC\fi\fi}\def\CD@AC{\dimen4\dimen0 \dimen0\dimen9
\advance\dimen2\count@\dimen3 \dimen9\dimen2 \dimen2\dimen3 \dimen3\dimen9 }%
\def\CD@ZG#1{\CD@dG{\dimen2}\advance\dimen0 #1\dimen2 }\def\CD@dG#1{\divide#1%
\CD@TC\multiply#1\CD@LH}\def\CD@eG#1{\divide#1\CD@vA\multiply#1\CD@uA}\def
\CD@iC#1{\divide#1\CD@LH\multiply#1\CD@TC}\def\CD@hJ{\dimen6\CD@LH\CD@XH
\multiply\dimen6\CD@LH\dimen7\CD@TC\CD@XH\multiply\dimen7\CD@TC\CD@KJ}\def
\CD@iK#1#2{\begingroup\dimen@#2\relax\loop\ifdim\dimen2<.4\maxdimen\multiply
\dimen2\tw@\multiply\dimen@\tw@\repeat\divide\dimen2\@cclvi\divide\dimen@
\dimen2\relax\multiply\dimen@\@cclvi\expandafter\CD@jK\the\dimen@\endgroup
\let#1\CD@fK}{\catcode`p=12 \catcode`0=12 \catcode`.=12 \catcode`t=12 \gdef
\CD@jK#1pt{\gdef\CD@fK{#1}}}\ifx\errorcontextlines\CD@qK\CD@tJ\let\CD@GH
\relax\else\def\CD@GH{\errorcontextlines\m@ne}\fi\ifnum\inputlineno<0 \let
\CD@CD\empty\let\CD@W\empty\let\CD@mD\relax\let\CD@uI\relax\let\CD@vI\relax
\let\CD@zF\relax\else\def\CD@W{ at line \number\inputlineno}\def\CD@mD{ - first occurred%
}\def\CD@uI{\edef\CD@h{\the\inputlineno}\global\let\CD@jB\CD@h}\def\CD@h{9999%
}\def\CD@vI{\xdef\CD@jB{\the\inputlineno}}\def\CD@jB{\CD@h}\def\CD@zF{\ifnum
\CD@h<\inputlineno\edef\CD@CD{\space at lines \CD@h--\the\inputlineno}\else
\edef\CD@CD{\CD@W}\fi}\fi\let\CD@CD\empty\def\CD@YA#1#2{\CD@GH\errhelp=#2%
\expandafter\errmessage{\CD@tA: #1}}\def\CD@KB#1{\begingroup\expandafter
\message{! \CD@tA: #1\CD@CD}\ifnum\CD@XB>\CD@NB\ifnum\CD@CA>\CD@NB\else\ifnum
\CD@lA>\CD@FA\else\ifnum\CD@LB>\CD@FA\advance\CD@XB-\CD@NB\advance\CD@FA-%
\CD@lA\advance\CD@FA1\relax\expandafter\message{! (error detected at row \the
\CD@XB, column \the\CD@FA, but probably caused elsewhere)}\fi\fi\fi\fi
\endgroup}\def\CD@gB#1{{\expandafter\message{\CD@tA\space Warning: #1\CD@W}}}%
\def\CD@CB#1#2{\CD@gB{#1 \string#2 is obsolete\CD@mD}}\def\CD@AB#1{\CD@CB{%
Dimension}{#1}\CD@DE#1\CD@BB\CD@BB}\def\CD@BB{\CD@OA=}\def\CD@@B#1{\CD@CB{%
Count}{#1}\CD@DE#1\CD@OH\CD@OH}\def\CD@OH{\count@=}\def\HorizontalMapLength{%
\CD@AB\HorizontalMapLength}\def\VerticalMapHeight{\CD@AB\VerticalMapHeight}%
\def\VerticalMapDepth{\CD@AB\VerticalMapDepth}\def\VerticalMapExtraHeight{%
\CD@AB\VerticalMapExtraHeight}\def\VerticalMapExtraDepth{\CD@AB
\VerticalMapExtraDepth}\def\DiagonalLineSegments{\CD@@B\DiagonalLineSegments}%
\ifx\tenln\nullfont\CD@ZA\CD@KH{\CD@eF\space diagonal line and arrow font not
available}\else\let\CD@KH\relax\fi\def\CD@aG#1#2<#3:#4:#5#6{\begingroup\CD@PA
#3\relax\advance\CD@PA-#2\relax\ifdim.1em<\CD@PA\CD@uA#5\relax\CD@vA#6\relax
\ifnum\CD@uA<\CD@vA\count@\CD@vA\advance\count@-\CD@uA\CD@KB{#4 by \the\CD@PA
}\if#1v\let\CD@CH\CD@JK\edef\tmp{\the\CD@uA--\the\CD@vA,\the\CD@FA}\else
\advance\count@\count@\if#1l\advance\count@-\CD@A\else\if#1r\advance\count@
\CD@A\fi\fi\advance\CD@PA\CD@PA\let\CD@CH\CD@ZE\edef\tmp{\the\CD@NB,\the
\CD@uA--\the\CD@vA}\fi\divide\CD@PA\count@\ifdim\CD@CH<\CD@PA\global\CD@CH
\CD@PA\fi\fi\fi\endgroup}\CD@tG\CD@xE\CD@JD\CD@ID\CD@rG\CD@xI{See the message
above.}\CD@rG\CD@lH{Perhaps you've forgotten to end the diagram before
resuming the text, in\CD@uG which case some garbage may be added to the
diagram, but we should be ok now.\CD@uG Alternatively you've left a blank line
in the middle - TeX will now complain\CD@uG that the remaining \CD@S s are
misplaced - so please use comments for layout.}\CD@rG\CD@hD{You have already
closed too many brace pairs or environments; an \CD@HD\CD@uG command was (%
over)due.}\CD@rG\CD@hH{\CD@dC\space and \CD@HD\space commands must match.}%
\def\CD@jH{\ifnum\inputlineno=0 \else\expandafter\CD@iH\fi}\def\CD@iH{\CD@MD
\CD@GD\crcr\CD@YA{missing \CD@HD\space inserted before \CD@kH- type "h"}%
\CD@lH\enddiagram\CD@AG\CD@kH\par}\def\CD@AG#1{\edef\enddiagram{\noexpand
\CD@rD{#1\CD@W}}}\def\CD@rD#1{\CD@YA{\CD@HD\space(anticipated by #1) ignored}%
\CD@xI\let\enddiagram\CD@SG}\def\CD@SG{\CD@YA{misplaced \CD@HD\space ignored}%
\CD@hH}\def\CD@mC{\CD@YA{missing \CD@HD\space inserted.}\CD@hD\CD@AG{closing
group}}\ifx\ifx
\fi\fi

\catcode`\$=3 
\def\vboxtoz{\vbox to\z@}

\def\scriptaxis#1{\@scriptaxis{$\scriptstyle#1$}}
\def\ssaxis#1{\@ssaxis{$\scriptscriptstyle#1$}}
\def\@scriptaxis#1{\dimen0\axisheight\advance\dimen0-\ss@axisheight\raise
\dimen0\hbox{#1}}\def\@ssaxis#1{\dimen0\axisheight\advance\dimen0-%
\ss@axisheight\raise\dimen0\hbox{#1}}

\ifx\boldmath\CD@qK
\let\boldscriptaxis\scriptaxis
\def\boldscript#1{\hbox{$\scriptstyle#1$}}
\else\def\boldscriptaxis#1{\@scriptaxis{\boldmath$\scriptstyle#1$}}
\def\boldscript#1{\hbox{\boldmath$\scriptstyle#1$}}
\fi

\def\raisehook#1#2#3{\hbox{\setbox3=\hbox{#1$\scriptscriptstyle#3$}%
\dimen0\ss@axisheight
\dimen1\axisheight\advance\dimen1-\dimen0
\dimen2\ht3\advance\dimen2-\dimen0%
\advance\dimen2-0.021em\advance\dimen1 #2\dimen2%
\raise\dimen1\box3}}
\def\shifthook#1#2#3{\setbox1=\hbox{#1$\scriptscriptstyle#3$}\dimen0\wd1%
\divide\dimen0 12\CD@zH{\dimen0}
\dimen1\wd1\advance\dimen1-2\dimen0 \advance\dimen1-2\CD@oI\CD@zH{\dimen1}%
\kern#2\dimen1\box1}

\def\@cmex{\mathchar"03}

\def\make@pbk#1{\setbox\tw@\hbox to\z@{#1}\ht\tw@\z@\dp\tw@\z@\box\tw@}\def
\CD@fH#1{\overprint{\hbox to\z@{#1}}}\def\CD@qH{\kern0.11em}\def\CD@pH{\kern0%
.35em}

\def\dblvert{\def\CD@rH{\kern.5\PileSpacing}}\def\CD@rH{}

\def\SEpbk{\make@pbk{\CD@qH\CD@rH\vrule depth 2.87ex height -2.75ex width 0.%
95em \vrule height -0.66ex depth 2.87ex width 0.05em \hss}}

\def\SWpbk{\make@pbk{\hss\vrule height -0.66ex depth 2.87ex width 0.05em
\vrule depth 2.87ex height -2.75ex width 0.95em \CD@qH\CD@rH}}

\def\NEpbk{\make@pbk{\CD@qH\CD@rH\vrule depth -3.81ex height 4.00ex width 0.%
95em \vrule height 4.00ex depth -1.72ex width 0.05em \hss}}

\def\NWpbk{\make@pbk{\hss\vrule height 4.00ex depth -1.72ex width 0.05em
\vrule depth -3.81ex height 4.00ex width 0.95em \CD@qH\CD@rH}}

\def\puncture{{\setbox0\hbox{A}\vrule height.53\ht0 depth-.47\ht0 width.35\ht
0 \kern.12\ht0 \vrule height\ht0 depth-.65\ht0 width.06\ht0 \kern-.06\ht0
\vrule height.35\ht0 depth0pt width.06\ht0 \kern.12\ht0 \vrule height.53\ht0
depth-.47\ht0 width.35\ht0 }}

\def\NEclck{\overprint{\raise2.5ex\rlap{ \CD@rH$\scriptstyle\searrow$}}}
\def\NEanti{\overprint{\raise2.5ex\rlap{ \CD@rH$\scriptstyle\nwarrow$}}}
\def\NWclck{\overprint{\raise2.5ex\llap{$\scriptstyle\nearrow$ \CD@rH}}}
\def\NWanti{\overprint{\raise2.5ex\llap{$\scriptstyle\swarrow$ \CD@rH}}}
\def\SEclck{\overprint{\lower1ex\rlap{ \CD@rH$\scriptstyle\swarrow$}}}
\def\SEanti{\overprint{\lower1ex\rlap{ \CD@rH$\scriptstyle\nearrow$}}}
\def\SWclck{\overprint{\lower1ex\llap{$\scriptstyle\nwarrow$ \CD@rH}}}
\def\SWanti{\overprint{\lower1ex\llap{$\scriptstyle\searrow$ \CD@rH}}}

\def\rhvee{\mkern-10mu\greaterthan}
\def\lhvee{\lessthan\mkern-10mu}
\def\dhvee{\vboxtoz{\vss\hbox{$\vee$}\kern0pt}}
\def\uhvee{\vboxtoz{\hbox{$\wedge$}\vss}}
\newarrowhead{vee}\rhvee\lhvee\dhvee\uhvee

\def\dhlvee{\vboxtoz{\vss\hbox{$\scriptstyle\vee$}\kern0pt}}
\def\uhlvee{\vboxtoz{\hbox{$\scriptstyle\wedge$}\vss}}
\newarrowhead{littlevee}{\mkern1mu\scriptaxis\rhvee}{\scriptaxis\lhvee}%
\dhlvee\uhlvee\ifx\boldmath\CD@qK
\newarrowhead{boldlittlevee}{\mkern1mu\scriptaxis\rhvee}{\scriptaxis\lhvee}%
\dhlvee\uhlvee\else
\def\dhblvee{\vboxtoz{\vss\boldscript\vee\kern0pt}}
\def\uhblvee{\vboxtoz{\boldscript\wedge\vss}}
\newarrowhead{boldlittlevee}{\mkern1mu\boldscriptaxis\rhvee}{\boldscriptaxis
\lhvee}\dhblvee\uhblvee
\fi

\def\rhcvee{\mkern-10mu\succ}
\def\lhcvee{\prec\mkern-10mu}
\def\dhcvee{\vboxtoz{\vss\hbox{$\curlyvee$}\kern0pt}}
\def\uhcvee{\vboxtoz{\hbox{$\curlywedge$}\vss}}
\newarrowhead{curlyvee}\rhcvee\lhcvee\dhcvee\uhcvee

\def\rhvvee{\mkern-13mu\gg}
\def\lhvvee{\ll\mkern-13mu}
\def\dhvvee{\vboxtoz{\vss\hbox{$\vee$}\kern-.6ex\hbox{$\vee$}\kern0pt}}
\def\uhvvee{\vboxtoz{\hbox{$\wedge$}\kern-.6ex \hbox{$\wedge$}\vss}}
\newarrowhead{doublevee}\rhvvee\lhvvee\dhvvee\uhvvee

\def\rhtriangle{\triangleright\mkern1.2mu}
\def\lhtriangle{\triangleleft\mkern.8mu}
\def\uhtriangle{\vbox{\kern-.2ex \hbox{$\scriptscriptstyle\bigtriangleup$}%
\kern-.25ex}}
\def\dhtriangle{\vbox{\kern-.28ex \hbox{$\scriptscriptstyle\bigtriangledown$}%
\kern-.1ex}}
\def\dhblack{\vbox{\kern-.25ex\nointerlineskip\hbox{$\blacktriangledown$}}}%
\def\uhblack{\vbox{\kern-.25ex\nointerlineskip\hbox{$\blacktriangle$}}}%
\def\dhlblack{\vbox{\kern-.25ex\nointerlineskip\hbox{$\scriptstyle
\blacktriangledown$}}}
\def\uhlblack{\vbox{\kern-.25ex\nointerlineskip\hbox{$\scriptstyle
\blacktriangle$}}}
\newarrowhead{triangle}\rhtriangle\lhtriangle\dhtriangle\uhtriangle
\newarrowhead{blacktriangle}{\mkern-1mu\blacktriangleright\mkern.4mu}{%
\blacktriangleleft}\dhblack\uhblack\newarrowhead{littleblack}{\mkern-1mu%
\scriptaxis\blacktriangleright}{\scriptaxis\blacktriangleleft\mkern-2mu}%
\dhlblack\uhlblack

\def\rhla{\hbox{\setbox0=\lnchar55\dimen0=\wd0\kern-.6\dimen0\ht0\z@\raise
\axisheight\box0\kern.1\dimen0}}
\def\lhla{\hbox{\setbox0=\lnchar33\dimen0=\wd0\kern.05\dimen0\ht0\z@\raise
\axisheight\box0\kern-.5\dimen0}}
\def\dhla{\vboxtoz{\vss\rlap{\lnchar77}}}
\def\uhla{\vboxtoz{\setbox0=\lnchar66 \wd0\z@\kern-.15\ht0\box0\vss}}
\newarrowhead{LaTeX}\rhla\lhla\dhla\uhla

\def\lhlala{\lhla\kern.3em\lhla}
\def\rhlala{\rhla\kern.3em\rhla}
\def\uhlala{\hbox{\uhla\raise-.6ex\uhla}}
\def\dhlala{\hbox{\dhla\lower-.6ex\dhla}}
\newarrowhead{doubleLaTeX}\rhlala\lhlala\dhlala\uhlala

\def\hhO{\scriptaxis\bigcirc\mkern.4mu} \def\hho{{\circ}\mkern1.2mu}%
\newarrowhead{o}\hho\hho\circ\circ
\newarrowhead{O}\hhO\hhO{\scriptstyle\bigcirc}{\scriptstyle\bigcirc}

\def\rhtimes{\mkern-5mu{\times}\mkern-.8mu}\def\lhtimes{\mkern-.8mu{\times}%
\mkern-5mu}\def\uhtimes{\setbox0=\hbox{$\times$}\ht0\axisheight\dp0-\ht0%
\lower\ht0\box0 }\def\dhtimes{\setbox0=\hbox{$\times$}\ht0\axisheight\box0 }%
\newarrowhead{X}\rhtimes\lhtimes\dhtimes\uhtimes\newarrowhead+++++

\newarrowhead{Y}{\mkern-3mu\Yright}{\Yleft\mkern-3mu}\Ydown\Yup

\newarrowhead{->}\rightarrow\leftarrow\downarrow\uparrow

\newarrowhead{=>}\Rightarrow\Leftarrow{\@cmex7F}{\@cmex7E}

\newarrowhead{harpoon}\rightharpoonup\leftharpoonup\downharpoonleft
\upharpoonleft

\def\twoheaddownarrow{\rlap{$\downarrow$}\raise-.5ex\hbox{$\downarrow$}}
\def\twoheaduparrow{\rlap{$\uparrow$}\raise.5ex\hbox{$\uparrow$}}
\newarrowhead{->>}\twoheadrightarrow\twoheadleftarrow\twoheaddownarrow
\twoheaduparrow

\def\ltvee{\mkern-1mu{\lessthan}\mkern.4mu}
\newarrowtail{vee}\greaterthan\ltvee\vee\wedge

\newarrowtail{littlevee}{\scriptaxis\greaterthan}{\mkern-1mu\scriptaxis
\lessthan}{\scriptstyle\vee}{\scriptstyle\wedge}\ifx\boldmath\CD@qK
\newarrowtail{boldlittlevee}{\scriptaxis\greaterthan}{\mkern-1mu\scriptaxis
\lessthan}{\scriptstyle\vee}{\scriptstyle\wedge}\else\newarrowtail{%
boldlittlevee}{\boldscriptaxis\greaterthan}{\mkern-1mu\boldscriptaxis
\lessthan}{\boldscript\vee}{\boldscript\wedge}\fi

\newarrowtail{curlyvee}\succ{\mkern-1mu{\prec}\mkern.4mu}\curlyvee\curlywedge

\def\rttriangle{\mkern1.2mu\triangleright}
\newarrowtail{triangle}\rttriangle\lhtriangle\dhtriangle\uhtriangle
\newarrowtail{blacktriangle}\blacktriangleright{\mkern-1mu\blacktriangleleft
\mkern.4mu}\dhblack\uhblack\newarrowtail{littleblack}{\scriptaxis
\blacktriangleright\mkern-2mu}{\mkern-1mu\scriptaxis\blacktriangleleft}%
\dhlblack\uhlblack

\def\rtla{\hbox{\setbox0=\lnchar55\dimen0=\wd0\kern-.5\dimen0\ht0\z@\raise
\axisheight\box0\kern-.2\dimen0}}
\def\ltla{\hbox{\setbox0=\lnchar33\dimen0=\wd0\kern-.15\dimen0\ht0\z@\raise
\axisheight\box0\kern-.5\dimen0}}
\def\dtla{\vbox{\setbox0=\rlap{\lnchar77}\dimen0=\ht0\kern-.7\dimen0\box0%
\kern-.1\dimen0}}
\def\utla{\vbox{\setbox0=\rlap{\lnchar66}\dimen0=\ht0\kern-.1\dimen0\box0%
\kern-.6\dimen0}}
\newarrowtail{LaTeX}\rtla\ltla\dtla\utla

\def\rtvvee{\gg\mkern-3mu}
\def\ltvvee{\mkern-3mu\ll}
\def\dtvvee{\vbox{\hbox{$\vee$}\kern-.6ex \hbox{$\vee$}\vss}}
\def\utvvee{\vbox{\vss\hbox{$\wedge$}\kern-.6ex \hbox{$\wedge$}\kern\z@}}
\newarrowtail{doublevee}\rtvvee\ltvvee\dtvvee\utvvee

\def\ltlala{\ltla\kern.3em\ltla}
\def\rtlala{\rtla\kern.3em\rtla}
\def\utlala{\hbox{\utla\raise-.6ex\utla}}
\def\dtlala{\hbox{\dtla\lower-.6ex\dtla}}
\newarrowtail{doubleLaTeX}\rtlala\ltlala\dtlala\utlala

\def\utbar{\vrule height 0.093ex depth0pt width 0.4em}
\let\dtbar\utbar
\def\rtbar{\mkern1.5mu\vrule height 1.1ex depth.06ex width .04em\mkern1.5mu}%
\let\ltbar\rtbar
\newarrowtail{mapsto}\rtbar\ltbar\dtbar\utbar
\newarrowtail{|}\rtbar\ltbar\dtbar\utbar

\def\rthooka{\raisehook{}+\subset\mkern-1mu}
\def\lthooka{\mkern-1mu\raisehook{}+\supset}
\def\rthookb{\raisehook{}-\subset\mkern-2mu}
\def\lthookb{\mkern-1mu\raisehook{}-\supset}

\def\dthooka{\shifthook{}+\cap}
\def\dthookb{\shifthook{}-\cap}
\def\uthooka{\shifthook{}+\cup}
\def\uthookb{\shifthook{}-\cup}

\newarrowtail{hooka}\rthooka\lthooka\dthooka\uthooka\newarrowtail{hookb}%
\rthookb\lthookb\dthookb\uthookb

\ifx\boldmath\CD@qK\newarrowtail{boldhooka}\rthooka\lthooka\dthooka\uthooka
\newarrowtail{boldhookb}\rthookb\lthookb\dthookb\uthookb\newarrowtail{%
boldhook}\rthooka\lthooka\dthookb\uthooka\else\def\rtbhooka{\raisehook
\boldmath+\subset\mkern-1mu}
\def\ltbhooka{\mkern-1mu\raisehook\boldmath+\supset}
\def\rtbhookb{\raisehook\boldmath-\subset\mkern-2mu}
\def\ltbhookb{\mkern-1mu\raisehook\boldmath-\supset}
\def\dtbhooka{\shifthook\boldmath+\cap}
\def\dtbhookb{\shifthook\boldmath-\cap}
\def\utbhooka{\shifthook\boldmath+\cup}
\def\utbhookb{\shifthook\boldmath-\cup}
\newarrowtail{boldhooka}\rtbhooka\ltbhooka\dtbhooka\utbhooka\newarrowtail{%
boldhookb}\rtbhookb\ltbhookb\dtbhookb\utbhookb\newarrowtail{boldhook}%
\rtbhooka\ltbhooka\dtbhooka\utbhooka\fi

\def\dtsqhooka{\shifthook{}+\sqcap}
\def\ltsqhooka{\mkern-1mu\raisehook{}+\sqsupset}
\def\rtsqhooka{\raisehook{}+\sqsubset\mkern-1mu}
\def\utsqhooka{\shifthook{}+\sqcup}
\newarrowtail{sqhook}\rtsqhooka\ltsqhooka\dtsqhooka\utsqhooka

\newarrowtail{hook}\rthooka\lthookb\dthooka\uthooka\newarrowtail{C}\rthooka
\lthookb\dthooka\uthooka

\newarrowtail{o}\hho\hho\circ\circ
\newarrowtail{O}\hhO\hhO{\scriptstyle\bigcirc}{\scriptstyle\bigcirc}

\newarrowtail{X}\lhtimes\rhtimes\uhtimes\dhtimes\newarrowtail+++++

\newarrowtail{Y}\Yright\Yleft\Ydown\Yup

\newarrowtail{harpoon}\leftharpoondown\rightharpoondown\upharpoonright
\downharpoonright

\newarrowtail{<=}\Leftarrow\Rightarrow{\@cmex7E}{\@cmex7F}

\newarrowfiller{=}=={\@cmex77}{\@cmex77}
\def\vfthree{\mid\!\!\!\mid\!\!\!\mid}
\newarrowfiller{3}\equiv\equiv\vfthree\vfthree

\def\vfdashstrut{\vrule width0pt height1.3ex depth0.7ex}
\def\vfthedash{\vrule width\CD@LF height0.6ex depth 0pt}
\def\hfthedash{\CD@AJ\vrule\horizhtdp width 0.26em}
\def\hfdash{\mkern5.5mu\hfthedash\mkern5.5mu}
\def\vfdash{\vfdashstrut\vfthedash}
\newarrowfiller{dash}\hfdash\hfdash\vfdash\vfdash

\newarrowmiddle+++++

\iffalse
\newarrow{To}----{vee}
\newarrow{Arr}----{LaTeX}
\newarrow{Dotsto}....{vee}
\newarrow{Dotsarr}....{LaTeX}
\newarrow{Dashto}{}{dash}{}{dash}{vee}
\newarrow{Dasharr}{}{dash}{}{dash}{LaTeX}
\newarrow{Mapsto}{mapsto}---{vee}
\newarrow{Mapsarr}{mapsto}---{LaTeX}
\newarrow{IntoA}{hooka}---{vee}
\newarrow{IntoB}{hookb}---{vee}
\newarrow{Embed}{vee}---{vee}
\newarrow{Emarr}{LaTeX}---{LaTeX}
\newarrow{Onto}----{doublevee}
\newarrow{Dotsonarr}....{doubleLaTeX}
\newarrow{Dotsonto}....{doublevee}
\newarrow{Dotsonarr}....{doubleLaTeX}
\else
\newarrow{To}---->
\newarrow{Arr}---->
\newarrow{Dotsto}....>
\newarrow{Dotsarr}....>
\newarrow{Dashto}{}{dash}{}{dash}>
\newarrow{Dasharr}{}{dash}{}{dash}>
\newarrow{Mapsto}{mapsto}--->
\newarrow{Mapsarr}{mapsto}--->
\newarrow{IntoA}{hooka}--->
\newarrow{IntoB}{hookb}--->
\newarrow{Embed}>--->
\newarrow{Emarr}>--->
\newarrow{Onto}----{>>}
\newarrow{Dotsonarr}....{>>}
\newarrow{Dotsonto}....{>>}
\newarrow{Dotsonarr}....{>>}
\fi

\newarrow{Implies}===={=>}
\newarrow{Project}----{triangle}
\newarrow{Pto}----{harpoon}
\newarrow{Relto}{harpoon}---{harpoon}

\newarrow{Eq}=====
\newarrow{Line}-----
\newarrow{Dots}.....
\newarrow{Dashes}{}{dash}{}{dash}{}

\newarrow{SquareInto}{sqhook}--->

\newarrowhead{cmexbra}{\@cmex7B}{\@cmex7C}{\@cmex3B}{\@cmex38}
\newarrowtail{cmexbra}{\@cmex7A}{\@cmex7D}{\@cmex39}{\@cmex3A}
\newarrowmiddle{cmexbra}{\braceru\bracelu}{\bracerd\braceld}{\vcenter{%
\hbox@maths{\@cmex3D\mkern-2mu}}}
{\vcenter{\hbox@maths{\mkern2mu\@cmex3C}}}
\newarrow{@brace}{cmexbra}-{cmexbra}-{cmexbra}
\newarrow{@parenth}{cmexbra}---{cmexbra}
\def\rightBrace{\d@brace[thick,cmex]}
\def\leftBrace{\u@brace[thick,cmex]}
\def\upperBrace{\r@brace[thick,cmex]}
\def\lowerBrace{\l@brace[thick,cmex]}
\def\rightParenth{\d@parenth[thick,cmex]}
\def\leftParenth{\u@parenth[thick,cmex]}
\def\upperParenth{\r@parenth[thick,cmex]}
\def\lowerParenth{\l@parenth[thick,cmex]}

\let\hEq\rEq
\let\vEq\uEq

\def\labelstyle{
\ifincommdiag
\textstyle
\else
\scriptstyle
\fi}
\let\objectstyle\displaystyle

\newdiagramgrid{pentagon}{0.618034,0.618034,1,1,1,1,0.618034,0.618034}{1.%
17557,1.17557,1.902113,1.902113}

\newdiagramgrid{perspective}{0.75,0.75,1.1,1.1,0.9,0.9,0.95,0.95,0.75,0.75}{0%
.75,0.75,1.1,1.1,0.9,0.9}

\diagramstyle[
dpi=300,
vmiddle,nobalance,
loose,
thin,
pilespacing=10pt,%
shortfall=4pt,
]

\ifx\CD@qK\else\CD@PK\relax\CD@FF\CD@e\fi\fi

\CD@vE\CD@hK\else\fi\else\fi\cdrestoreat
\begin{document}

\title{Fission and spallation data evaluation using induced-activity method}
\author{G. S. Karapetyan}
\email{gayane@if.usp.br}
\affiliation{Instituto de Fisica, Universidade de S\~ao Paulo \\ Rua do Mat\~ao, Travessa R 187, 05508-900 S\~ao Paulo, SP, Brazil}

\begin{abstract}

The induced-activity investigations in off-line analysis performed in different experiments, concerning pre-actinide and actinide nuclei,
 are here presented and discussed. Generalized expressions for the determination of independent yields/cross sections of radioactive nuclei, formed in the targets, are derived and analysed. 
The fragment mass distribution from $^{238}$U, $^{232}$Th and $^{181}$Ta photofission at the bremsstrahlung end-point energies of 50 and 3500 MeV, and from $^{241}$Am, $^{238}$U and $^{237}$Np fission induced by 660-MeV protons, are scrutinized from the point of view of the multimodal fission approach.
The results of these studies are hence compared with theoretical model calculations using the CRISP code.
A multimodal fission option has been added to this code, which allows to account the contribution of symmetric and asymmetric (superasymmetric) fission to the total fission yield. Moreover, this work contains the general results obtained in the analysis of the isomer ratios of fission fragments from $^{238}$U and $^{232}$Th targets at the bremsstrahlung end-point energies of 50 and 3500 MeV. Moreover, the values of the average  angular momenta of primary fragments are estimated by using the statistical model calculation.  
We subsequently discuss the complex particle-induced reaction, such as heavy-ions and deuterons, by using the thick-target thick-catcher technique and the two-step vector model framework as well. This is accomplished in order to present the investigation of the main processes (fission, spallation and (multi)fragmentation) in intermediate- and high-energy ranges of the incident particle.
The set of experimental data, presented in this work, encompasses not merely the data as total production cross sections. Notwithstanding, it further covers other data, as individual yields/cross sections, charge, mass and spin distributions of the reaction fragments, as well as  kinematic features. These sources of experimental data can serve as a consistent set of benchmarking data, still necessary for the study of heavy nuclei. Besides, it is also useful for  technological applications, from astrophysics and environmental sciences to accelerator technology and accelerator-based nuclear waste transmutation and energy amplification as well.

\end{abstract}
\maketitle

\section*{1. Introduction}

Fission and spallation play a prominent role in nuclear physics, presenting paramount importance in nuclear reactions, being active fields of current investigation. Nuclear energy, for instance, which followed the discovery of fission, has been of profound importance to mankind. Besides, fission and spallation are interesting as a large scale collective motion of the nucleus, and they are   important exit channels for many nuclear reactions as well.

Fission is a slow process with respect to nuclear timescales, involving the deformation of the nucleus. The disintegration of the nucleus into two fragments of similar masses is accompanied by a complete rearrangement of the nuclear structure. The dynamical process that leads to fission determines the fragments features in the final states. A plethora of experimental investigations,  regarding 
mass and energy distributions of fragments in the fission of nuclei from Pb to No \cite{Gonnenwein},  has 
confirmed the validity of the hypothesis on the existence of independent fission modes, as first
stated by Turkevich and Niday \cite{Turkevich}. This hypothesis has been reinforced by 
theoretical works by Pashkevich \cite{Pashkevich} and Brosa $\textit{et al.}$ \cite{Brosa}. These studies  showed that multimodality of the mass and energy distributions of fission fragments is due to the valley structure of the deformation potential energy surface of a fissioning nucleus. The properties of fragments produced in the fission of actinide and pre-actinide nuclei provide important information concerning the relationship among different fission components. 

Spallation reactions, resulting from the interaction between high-energy projectiles and heavy target nuclei, can be straightforwardly described. In fact, the nucleons from a target nucleus are spalled due to the bombardment of an energetic nucleon, leaving behind fragments. Their mass and atomic numbers are smaller than the ones associated to the target nucleus itself. Serber was the first one to come up with the theoretical description of the spallation mechanism \cite{Serber}. He proposed a two stage model, consisting of a intra-nuclear cascade (INC) followed by statistical de-excitation. According to this framework, the first step is regarded in terms of collisions between the incident particle and the individual nuclear particles -- nucleon-nucleon collisions. After the INC, the residual nucleus is left into an equilibrium state, wherein the excitation energy is shared by a large number of nucleons. If the excitation energy is greater than the binding energy of particles, emission of nucleons and light fragments ($\alpha$, $d$, $t$) and other reaction channels --- as evaporation, (multi)fragmentation and fission --- are possible. Some of the nucleons ejected during the spallation reactions have sufficient energy to induce further spallation reaction with neighboring nuclei inside a thick target, leading to a multiplication of the emitted neutrons. It is therefore possible to produce intense neutron fluxes from high energy beams. These fluxes  could compete with reactor ones. Spallation reactions in heavy elements are also particularly interesting, since the fission process provides a prominent competing reaction, which can have effects on the cross sections of the other reactions. 

In this study the main aim is to  discuss  the prominent features of spallation and fission reactions in the context of the mechanism of interaction
of intermediate- and high-energy projectiles with heavy nuclei.
Some examples of fundamental research on the dynamics of projectile-nucleus interaction will be presented. The determination of the distribution in $Z$ and $A$ of the residual nuclei resulting from the interaction allows to estimate the contribution of the different reaction channels such as evaporation, (multi)fragmentation and fission, the total reaction cross section. Thereafter we will compare these contributions to previous investigations.

The nuclear fission and spallation phenomena have been studied by several methods and techniques, including direct and inverse kinematics \cite{Duijvestijn, Maslov, Titarenko, Enqvist, Stoulos, Rejmund, Michel}. In these studies the basic mechanism, responsible for the development of various processes that usually occur during the interaction between intermediate- and high-energy projectiles and heavy nuclei targets, has been determined. Particularly, the experiments performed in inverse kinematics by Schmidt $\textit{et al.}$ \cite{Schmidt} represent the comprehensive study of the fission fragment charge distributions in a large variety of actinides and pre-actinides.

On the other hand, the induced-activity method is a prominent and convenient tool to study the nuclear properties of reaction residues, based upon the $\gamma$-ray spectroscopy.
Measurements by the induced-activity method allow to construct charge and mass distributions of reaction products, as well as to determine the main kinematic characteristics of reaction fragments and to estimate their angular momenta. 

The scientific interest of the results presented in this work is evinced by the main relevant data on numerous studies of nucleon-nucleus and nucleus-nucleus interactions that have been   previously obtained in the literature.  
The investigations of the different reactions in various conditions of excitation can shed new light on the mechanism of reactions, and the features of interactions of the distinct nucleons with the nucleus can be derived. The modification of either the incident energy or  the target nucleus (or both) leads to a very rich spectrum of phenomena that we present here, which have to be described theoretically. The next argument to present the main results of investigations is to provide conditions to understand how the mechanism of nucleon-nucleus and nucleus-nucleus interactions is modified, into transitions toward more complex systems. The nuclear data for the production of residual nuclides are basic quantities for the calculation of radioactive inventories of spallation targets in spallation neutron sources and in accelerator-driven devices for energy amplification and for transmutation of nuclear waste. Finally, presenting such results has a great importance for the model and code improvement as well. A reliable and comprehensive experimental database for reactions should be extended by forthcoming further experimental investigations, in order to allow benchmarking of new developed models.

\section*{2. Induced-activity method of data analysis}

In off-line experiment, the target nucleus is initially excited by an incident beam of charged particles, photons, neutrons or heavy-ions. Hence the induced target is displayed into a special hall for measurements. After irradiation, both stable and radioactive nuclides are produced, being characterized by different half-lives and decay energies. High resolution Ge $\gamma$-ray spectrometry plays a prominent role in the experiment. $\gamma$-ray spectrometry serves to identify radionuclides and to measure their activities. It is thus necessary to identify the energies of all $\gamma$-rays of interest. The analysis of the $\gamma$-spectra, the unambiguous identification of nuclides and the selection of activities used in the final calculation of cross section turned out to be the most time-consuming step in the experiments. The identification of the reaction products is conducted through the definition of the half-lives,  energies and intensities of the nuclear $\gamma$-transitions of the radioactive fragments \cite{Firestone}. Fig. 1 presents the $\gamma$-spectrum from the irradiated $^{232}$Th sample at 50 MeV bremsstrahlung photons,  having   measurement time of 30 h. The $\gamma$-peak areas are used to calculate the cross sections of  reaction products.
A considerable amount of data has been obtained for nucleon-induced reactions on different targets by using induced-activity method \cite{Michel, Michel1}. The main goal of these studies was to provide a database for model calculations of the production of cosmogenic nuclides in extraterrestrial matter by solar and galactic cosmic ray protons \cite{Michel, Michel1}. Therefore, the detailed subjects of such studies were dedicated mainly to measure of large amount of excitation functions of the reaction residues. We derive in what follows the framework for determination of the nuclide production cross section used in our investigations.

The production of radioactive isotope is defined by the following differential equation:
\begin{eqnarray}
\frac{dN}{dt} = N_{p}N_{n}\sigma - \lambda N,
\end{eqnarray}
\noindent
where $\lambda$ is the decay constant, $\sigma$ is the cross section of a given isotope ($\lambda = \ln2/T_{1/2})$ (in 1/cm$^{2}$ units), $N$ is the  number of the produced isotope, $N_{p}$ denotes the incident particle beam intensity (min$^{-1}$), and  $N_{n}$ stands for the number of target nuclei (in 1/cm$^{2}$ units). 

This equation is easily solved if $N_{p}$ does not depend upon the irradiation time. In this case, the number of nuclei formed during irradiation is determined by following equation:
\begin{eqnarray}
N_{0}(t_{1}) = \frac{N_{p}N_{n}\sigma}{\lambda}(1-e^{-\lambda t_{1}}),
\end{eqnarray}
\noindent
where $t_{1}$ is the irradiation time.

If we denote by $t_{2}$ the time of exposure between the end of the irradiation and the beginning of the measurement, then the number of formed nuclei of an isotope under interest will decrease exponentially as:
\begin{eqnarray}
N(t_{2}) = N_{0}(t_{1})e^{-\lambda t_{2}}.
\end{eqnarray}

Hence, the number decays $\Delta N$ during the time of measurement $t_{3}$ reads
\begin{eqnarray}
\Delta{N} =  N_{0}(t_{1})e^{-\lambda t_{2}} - N_{0}(t_{1})e^{-\lambda (t_{2}+t_{3})} =
\frac{N_{p}N_{n}\sigma}{\lambda}(1-e^{-\lambda t_{1}})e^{-\lambda t_{2}}(1-e^{-\lambda t_{3}}).
\end{eqnarray}
From eq.(4) we can obtain the reaction fragment production cross section in the absence of a parent isotope -- that may  contribute to the measured cross section via $\beta^{\pm}$- decays -- which usually considered as an independent cross section (I) by using the following equation:
\begin{eqnarray}
\hspace{-0.2cm}
\sigma=\frac{\Delta{N} \lambda}{N_{d} N_{n} k \epsilon \eta (1-e^{-\lambda t_{1}}) e^{-\lambda t_{2}}(1-e^{-\lambda t_{3}})},
\end{eqnarray}
Eq.(5) also takes into account the $\gamma$-ray detection efficiency ($\epsilon$), the intensity of $\gamma$-transitions ($\eta$), the total coefficient of $\gamma$-ray absorption, both in the target and the detector materials ($k$). 

If the cross section of a given isotope includes a contribution from the $\beta^{\pm}$-decay of neighboring unstable isobars, the cross section calculation becomes more complicated (Diagram 1). 
In these cases, the measured number of isotope atoms is considered as $N = N^{(1)} + N^{(2)} + N^{(3)}$, where $N^{(1)}$ is an independent cross section, defined by eq.(5), $N^{(2)}$ is the number of nuclei formed by the decay of the parent isotopes during the time of irradiation, and $N^{(3)}$ is the number of nuclei formed by the decay of the given isotopes after irradiation. 
\begin{diagram}
A &\rTo^{f_{AB}} &B\end{diagram}
As a result, we obtain the following equation:
\begin{eqnarray}
\Delta{N(t_{3})}\!\! &=& \!\!N_{p} N_{n} k \epsilon \eta \Biggl[
\! \frac{\sigma_{A} f_{AB}\lambda_{A} \lambda_{B}}{\lambda_{B}-\lambda_{A}}\Biggl(\frac{(1\!-\!e^{-\lambda_{A} t_{1}}) e^{-\lambda_{A}
t_{2}} (1\!-\!e^{-\lambda_{A} t_{3}})} {\lambda^{2}_{A}} - \frac{(1\!- \!e^{-\lambda_{B} t_{1}}) e^{-\lambda_{B}
t_{2}} (1\!-\! e^{-\lambda_{B} t_{3}})} {\lambda^{2}_{B}}\Biggr)\nonumber\\&& + \sigma_{B} \frac{(1- e^{-\lambda_{B} t_{1}}) e^{-\lambda_{B}
t_{2}} (1-e^{-\lambda_{B} t_{3}})} {\lambda_{B}}\Biggr],
\end{eqnarray}
\noindent where the subscripts $A$ and $B$ label variables referring to, respectively, the parent and the daughter nucleus; the coefficient $f_{AB}$ specifies the fraction of $A$ nuclei decaying to a $B$ nucleus ($f_{AB}=1$, when the contribution from the $\beta$-decay corresponds 100\%); and $(\Delta{N})$ is the total photo-peak area associated with the decays of the daughter and parent isotopes.

If the cross section of formation of the parent isotope is known from experimental data, or if it can be estimated on the basis of other sources, hence the independent cross sections of daughter nuclei can be calculated by the relation \cite{Baba}:
\begin{eqnarray}
\sigma_{B}=&&\frac{\lambda_{B}}{(1-e^{-\lambda_{B}t_{1}})e^{-\lambda_{B}t_{2}}
(1-e^{-\lambda_{B}t_{3}})}\times\nonumber\\
&&\hspace*{-1.5cm}\Biggl[\frac{\Delta{N}}{N_{p}\,N_{n}\,k\,\epsilon\,\eta}-\sigma_{A}\,f_{AB}\,
\frac{\lambda_{A}\,\lambda_{B}}{\lambda_{B}-\lambda_{A}}
\Biggl(\frac{(1-e^{-\lambda_{A} t_{1}}) e^{-\lambda_{A} t_{2}} (1-e^{-\lambda_{A} t_{3}})}
{\lambda^{2}_{A}}-\frac{(1-e^{-\lambda_{B} t_{1}}) e^{-\lambda_{B}
t_{2}}(1-e^{-\lambda_{B} t_{3})}}{\lambda^{2}_{B}}\Biggr)\Biggr],
\end{eqnarray}

The effect of the precursor can be negligible in some limiting cases. Indeed, when either the half-life of the parent nucleus is very long or when its contribution is very small. In the case where the parent and daughter isotopes can not be experimentally separated, the calculated cross sections are classified as cumulative ones (C). It should be mentioned that the use of the induced-activity method imposes several restrictions on the registration of the reaction products. For example, it is not possible to measure stable and very short-living isotopes.

In our experiments, the irradiation time, $t_{1}$, is usually chosen from a few minutes to several hours. The time $t_{2}$ changes from various minutes to several months, and the time of measurements, $t_{3}$, varies from various minutes to several dozens of hours.

The target is consistently measured  several times. The advantage of choosing the time by this scheme is that it allows to measure with high precision and accuracy the cross section of reaction products within a wide range of half-lives, from short-lived (dozens of minutes) to long ones.
The reliability of the obtained data such as, for instance, charge and mass distributions, strongly depends upon the number of independent cross sections measured in the experiment. In such purpose, the formula that takes into account the $\beta$-decay feeding of two precursors, has been elaborated.

Let us consider the scheme of three nuclei decay (or three different states of nuclei) (Diagram 2). For example, $A$ may represent isomeric state and $B$ the ground state of the nucleus under study). After analogous assumptions we will have \cite{Karapetyan}:
\begin{diagram}
A &\rTo^{f_{AC}} &C        & &\\
\dTo^{f_{AB}}  &\ruTo~{f_{BC}} &&\rdTo(1,1)&\\
B  & &&         &\\
&\rdTo(1,1)&&&\\
&&        & &\\\end{diagram}
\begin{eqnarray}
\Delta N(t_{3})\!\!\!\! &=& \!\!\!\!N_{p} N_{n} k \epsilon \eta \Biggl[
\! \sigma_{A} f_{AC} \frac{\lambda_{C}}{\lambda_{C}-\lambda_{A}} e(\lambda_{A}) +(\sigma_{B} + \sigma_{A}) f_{AB}\frac{\lambda_{A}}{\lambda_{A}-\lambda_{B}} \frac{\lambda_{C}}{\lambda_{C}-\lambda_{B}}\times\nonumber\\
&&\hspace*{-1.5cm}\Biggl(f_{BC} + f_{AB}f_{BC} \frac{\lambda_{C}}{\lambda_{C}-\lambda_{B}} \left(1 + \frac{\lambda_{B}}{\lambda_{B}-\lambda_{A}}\right)\Biggr) e(\lambda_{B}) + \nonumber\\
&&\hspace*{-1.5cm}\Biggl(\sigma_{C} + \frac{\lambda_{B}}{\lambda_{B}-\lambda_{C}}\sigma_{B}
\left(f_{BC} + f_{AB}f_{BC} \frac{\lambda_{B}}{\lambda_{B}-\lambda_{C}} \left(1 + \frac{\lambda_{A}}{\lambda_{A}-\lambda_{B}}\right)\right)
 + \sigma_{A} f_{AC}
 \frac{\lambda_{A}}{\lambda_{A}-\lambda_{C}} \Biggr) e(\lambda_{C})\Biggr],
\end{eqnarray}
\noindent where for simplicity we define:
\begin{eqnarray}
e(\lambda_{j}) = \frac{(1- e^{-\lambda_{j} t_{1}}) e^{-\lambda_{j}
t_{2}} (1-e^{-\lambda_{j} t_{3}})} {\lambda_{j}}.
\end{eqnarray}
Finally, one can obtain the formula of production cross section for $C$-nucleus:
\begin{eqnarray}
\sigma_{C}\! &=& \!\frac{1}{e(\lambda_{C})}\Biggl[\frac{\Delta{N}}{N_{p} N_{n} k \epsilon \eta} - 
\sigma_{A} f_{AC}\frac{\lambda_{C}}{\lambda_{C}-\lambda_{A}} e(\lambda_{A}) - (\sigma_{B} + \sigma_{A}) f_{AB}\frac{\lambda_{A}}{\lambda_{A}-\lambda_{B}}\frac{\lambda_{C}}{\lambda_{C}-\lambda_{B}}\times\nonumber\\
&&\hspace*{-1.5cm}\Biggl(f_{BC} + f_{AB}f_{BC} \frac{\lambda_{C}}{\lambda_{C}-\lambda_{B}} \left(1 + \frac{\lambda_{B}}{\lambda_{B}-\lambda_{A}}\right)\Biggr) e(\lambda_{B})\Biggr] -  \frac{\lambda_{B}}{\lambda_{B}-\lambda_{C}} \sigma_{B}\Biggl(f_{BC} + f_{AB}f_{BC} \frac{\lambda_{B}}{\lambda_{B}-\lambda_{C}} \left(1 + \frac{\lambda_{A}}{\lambda_{A}-\lambda_{B}}\right)\Biggr) \nonumber\\&&\qquad\qquad\qquad\qquad\qquad\qquad\qquad\qquad\qquad\qquad\qquad\qquad\qquad\qquad\qquad\qquad- \sigma_{A} f_{AC} \frac{\lambda_{A}}{\lambda_{A}-\lambda_{C}}.
\end{eqnarray}

The application of the induced-activity method allows to study the important features of the reaction in different nuclei and in wide energy ranges, such as the charge, spin and mass distributions of the target residues, and kinematics characteristics as well. It represents a straightforward method to measure the high-spin states of nuclides. Achieved results of the above mentioned method are very important for both theoretical and practical point of view. Moreover, study of the light nuclei structure is very interesting from astrophysical aspects. On the other hand the induced-activity method allows simultaneously to investigate such processes as fission, (multi)fragmentation and evaporation-spallation. Thus, one can reconstruct the whole picture of interaction and investigate different reaction channels on different target and in wide energy ranges.

\section*{3 Multimodal fission approach}

The multimodal fission approach \cite{Brosa} is based on the assumption that experimental mass distributions are a superposition of mass distributions of individual fission modes. These modes are produced by the valley structure of the deformation potential energy surface. It is supposed that there are three distinct fission modes for the heavy nuclei -- symmetric (S) mode and two asymmetric
modes -- Standard 1 (S1) and Standard 2 (S2). S mode fragments are strongly elongated with masses around $A_{f}/2$, where $A_{f}$ is a mass of fissioning nucleus. The S1 mode is characterized by high kinetic energies of fission fragments, spherical mass of heavy fragment $A_{H}\sim132-134$, charge $Z_{H}\sim50$ and number of neutron $N_{H}\sim82$. Kinetic energies of S2 mode fragments are lower than those of S1 mode, heavy slightly deformed fragments has mass $A_{H}\sim138-140$, influenced by the deformed proton shell $Z_{H}\sim52$ and neutron shells $N\sim86-88$.

According to the model \cite{Brosa}, the total mass-yield is obtained by the sum of the 
three Gaussian functions \cite{Younes}:
\begin{align}
 \begin{split}
  \sigma(A)=&
\frac{1}{\sqrt{2\pi}}\bigg[\frac{K_{1AS}}{\sigma_{1AS}}
\exp\left(-\frac{(A-A_S-D_{1AS})^2}{2\sigma^2_{1AS}}\right)+
\frac{K'_{1AS}}{\sigma'_{1AS}}\exp\left(-\frac{(A-A_S+D_{1AS})^{2}}
{2\sigma'^2_{1AS}}\right)+\\
&\frac{K_{2AS}}{\sigma_{2AS}}\exp\left({-\frac{(A-A_S-D_{2AS})^2}
{2\sigma^2_{2AS}}}\right)+
\frac{K'_{2AS}}{\sigma'_{2AS}}\exp\left({-\frac{(A-A_S+D_{2AS})^2}
{2\sigma'^2_{2AS}}}\right)+\frac{K_S}{\sigma_S}\exp\left({-\frac{(A-A_S)^2}
{2\sigma^2_S}}\right)
\bigg],
 \end{split}
\end{align}
where $A$ is the fragment mass number; $A_S$ is the mean mass number, 
which determines the center of the Gaussian functions; and $K_i$, $\sigma_i$, 
and $D_i$ denote the contribution, dispersion and position parameters of 
the $i^{th}$ Gaussian functions. The indexes $AS$ and $S$ respectively designate the 
asymmetric and symmetric components.

\subsection*{3.1 Photo- and proton-induced fission mass distributions}

Competition between symmetric and asymmetric fission is thought to be connected to shell effects in the deformed nucleus. The presence of these shell structures leads to the existence of symmetric and asymmetric fission modes or channels. The contribution of each fission mode that changes with excitation energy as well as shell effects fade with higher excitation energy, leaving  a nucleus which  merely possesses a symmetric fission mode.

Using the electron beams obtained at the synchrophasotron of Yerevan Physics Institute we investigated the photofission reactions in $^{238}$U, $^{232}$Th and $^{181}$Ta targets at the bremsstrahlung end-point energies 50 and 3500 MeV. The details of the measurements can be found in \cite{Nina1, Nina2, Deppman1}. The measurements of radionuclides production yields allowed to determine the distribution in $Z$ and $A$ of the residual nuclei resulting from the interaction. The yields of stable and short-lived nuclei unmeasurable by induced-activity method, were estimated by using the approximation with Gaussian
functions for charge distributions \cite{Kudo}:
\begin{eqnarray}
Y_{A,Z}=\frac{Y_{A}}{(C\pi)^{1/2}}\exp\left(-\frac{(Z-Z_{p})^2}{C}\right),
\end{eqnarray}
\noindent where $Y_{A,Z}$ is the independent yield of the nuclide ($Z, A$). The values $Y_{A}$ (the total chain yield for given mass number $A$),  $Z_{p} $ (the most probable charge for $Z$ distribution
isobars for mass chain $A$) and $C$ (the width parameter of distribution) were considered as a free parameters.

The relative contributions of symmetric and asymmetric channels could be quite different for different fissioning nuclei. 
Fission is well known to be predominantly symmetric for pre-actinides with $A \leq 227$ and for heavy
actinides with $A \geq 257$. However, for actinides $227 < A < 257$, fission is a mixture of two channels (asymmetric and symmetric). Furthermore, the standard channel splits into S1 and S2 channels \cite{Younes}. 
The mass distributions obtained during the fitting procedure for $^{238}$U, $^{232}$Th and $^{181}$Ta at end-point energy $E_{\gamma max} = 50$ MeV are plotted in figs. 2-7. In the case regarding 50 MeV, the mass-yield for $^{238}$U, $^{232}$Th targets were decomposed in assumption of mixture distribution, for $^{181}$Ta target the representation has agreed with the assumption of a symmetric mass distribution.

In the analysis of experiments with bremsstrahlung beam, the latter has a continuous spectrum. Hence, the measurements at high energy include also low-energy photons. Taking into account such aspect, to the decomposition of mass-yield at E$_{\gamma max}$ = 3500 MeV for $^{238}$U, $^{232}$Th targets, we added two additional components, one symmetric and one asymmetric (figs. 5, 6). 
Integrating over the Gaussian and multiplying by a factor 1/2, due to the two fission fragments
in each fission event, it gives an estimate for the total fission yield for each target. The values for the total fission yield ($Y_{tot}$), determined in this way, together with the contribution of symmetric ($Y_{S}$) and asymmetric ($Y_{AS}$) fission to the total mass-yield as well as the values of the asymmetric-to-symmetric ratios ($Y_{AS}/Y_{S}$) are given in table 1.

The decomposition of the mass-yield distribution for $^{238}$U, $^{232}$Th targets allowed to extract the yields of an asymmetric component at intermediate energies, from 50 MeV to 3500 MeV, for $^{238}$U ($Y_{AS} = 49.8\pm7.47$ mb/eq.q.) and $^{232}$Th ($Y_{AS} = 27.3\pm4.1$ mb/eq.q.) targets. From the data obtained it follows that the asymmetric component is stronger in the case of $^{238}$U.
The presence of an asymmetric component in this energy range may be due to the low-energy photons of bremsstrahlung spectrum. 
Comparison of the symmetric fission contribution in energy range from 50 MeV to 3500 MeV showed, that this mode is greater for $^{232}$Th ($Y_{S} = 70\pm10.5$ mb/eq.q.) than for $^{238}$U ($Y_{S} = 68.7\pm10.3$ mb/eq.q.). 
One can observe the clearly manifested growth of symmetric fission, when consider the relative contribution of this component to the total fission yield ($Y_{S}/Y_{tot}$, \%) in energy range 50-3500 MeV: for $^{238}$U this value is 58\%, and for the $^{232}$Th it corresponds to 72\%.

The increment of the symmetric fission contribution with increasing energy of the projectile can be caused due to several factors. On the one hand, an increment of the incident energy increases excitation energy of the fissioning nuclei \cite{Strecker, Gangrsky}, and hence the contribution symmetric binary nuclear decay increases. While the contribution of low-energy fission essentially asymmetric in nature.
From another hand, because of the particle evaporation process that takes place before fission, a set of various fissioning nuclides exists. Each of them has its own fission characteristics, excitation energy and angular momentum distributions. The symmetric component increases with excitation energy, which is coupled to an increment of neutron evaporation resulting in a larger contribution of the
neutron deficient nuclides with a higher fissility parameter. The excitation energy for which
symmetric fission starts to contribute gives rise to an amount of evaporated nucleons, corresponding to a
certain neutron deficient nuclei. These nuclei are responsible for the symmetric contribution growth.

The systematization of symmetric and asymmetric fission cross sections in a wide range of nuclei collected in \cite{Chung1, Chung2} showed that it is possible to use the empirical expression for estimation the probability of different fission modes. In order to characterize this factor quantitatively, the authors introduced a critical value of the fissility parameter in the following form:
\begin{eqnarray}
(Z^{2}/A)_{cr.}=35.5+0.4(Z-90),
\label{crit}
\end{eqnarray}
\noindent where $Z$ is a charge of the fissioning nucleus and $A$ is mass of fissioning nucleus.
 According to \cite{Chung1, Chung2}, for the nuclei with $Z^{2}/A$ values greater than the critical value, the symmetric fission mode was dominant. However, at smaller values of the fissility parameter, the main fission channel leads to an asymmetric fragment distribution. The larger contribution of symmetric fission for $^{232}$Th than for $^{238}$U at high energies one can explain by larger contribution of neutron deficient fissioning systems with high excitation energies. 

The prediction of the multimode-fission model has been used for the first time to analyze the mass distributions of fragments originating from  $^{241}$Am, $^{238}$U and $^{237}$Np targets irradiated by 660-MeV protons \cite{Gaya1, Gaya2} (figures 8-10). The determination of fundamental properties of individual components of the mass distribution and the contributions of these components as well has made it possible to establish a dominant role of the symmetric fission mode (table 2) at this energy range. Since the  neutrons and protons numbers are almost the same for $^{241}$Am and $^{237}$Np, the relative fractions of the different fission components are similar.  
In our measurements the mean masses of the fissioning nuclei at a given proton energy were obtained $A_{f} = 227$, $Z_{f} = 95$ ($Z^{2}/A = 39.76$ for
$^{241}$Am); $A_{f} = 227$, $Z_{f} = 92$ ($Z^{2}/A = 37.29$ for $^{238}$U); $A_{f} = 223.4$, $Z_{f} = 93$ ($Z^{2}/A = 38.72$ for $^{237}$Np), where $A_{f}$ and $Z_{f}$ are the mass and the charge of the fissioning nucleus, respectively. According to the criterion (13), the difference between $Z^{2}/A$ and $(Z^{2}/A)_{cr.}$ is greater for $^{241}$Am and $^{237}$Np nuclei. Therefore symmetric fission in these nuclides is expressed stronger than for $^{238}$U. The presence of the asymmetric fission component may be attributed to a broad mass and excitation energy distributions of fissioning nuclei. 

In the above mentioned photo- and proton-induced fission reactions, we estimated the fissility of actinides under study. 
The fissility ($D$) is defined as the ratio of total fission cross section ($\sigma_{tot}$) and the total inelastic cross section $(\sigma_{in})$. The total fission cross sections were determined by using a fitting based on eq. (11), and the total inelastic cross section values were estimated using cascade-evaporation model \cite{Barashenkov}.
Fig. 11 shows the fissility for $^{241}$Am, $^{238}$U, $^{237}$Np and $^{232}$Th targets from data for photo- and proton-induced fission \cite{Nina1, Nina2, Gaya1, Gaya2}, as a function of the
fissility parameter ($Z^{2}/A$) of the fissioning nucleus. From fig. 11 one can see that the fission probability increases while going to the higher values of the fissility parameter: the fissility of the heavier element is  higher. All fissility values for the targets studied in the case of fission induced by protons and photons at different energies are still below unit. This fact indicates a saturation of the fission probability for heavy nuclei at intermediate energy range.  Moreover, for $^{238}$U target the fissility has the same value in both photo- and proton-induced fission reactions.

The dependence of fissility of $^{241}$Am, $^{238}$U, $^{237}$Np and $^{232}$Th targets for photo- and proton-induced fission \cite{Nina1, Nina2, Gaya1, Gaya2} on excitation energy of fissioning nucleus are shown on fig. 12 together with experimental data of different probes and calculations from \cite{Fukahori}.
The excitation energy values for the reactions under study have been taken from \cite{Barashenkov, Jacobs}. 
 In \cite{Fukahori} a semi-empirical approach for the fission probability of neutron-, proton- and photon-induced fission in the energy range from several tens of MeV to 3 GeV has been proposed. 
As can be realized from fig. 12, regarding data from different experiments  obtained in reactions with proton, neutrons, and photons, the data respective to our work is in satisfactory agreement. This fact confirms the universal mechanism of nuclear fission, which depends on the excitation energy as well as upon mass and charge characteristics of the fissioning nucleus and does not depend on the type of projectile.

\subsection*{3.2 CRISP code calculation for actinide targets.}

Our investigation has been devoted to the analysis of the experimental results obtained in photo- and proton-induced reactions with actinides \cite{Gaya3, Gaya4, Gaya5} using the simulation code CRISP \cite{Deppman2}. CRISP is a Monte Carlo code for simulating nuclear reactions, where the reactions proceed in the two stages: the intranuclear cascade (MCMC code) and the evaporation/fission process (MCEF code). CRISP code calculated spallation yields and neutron multiplicities for reactions induced by high-energy protons, as well as has already been used in studies of the Accelerator Driving System (ADS) nuclear reactors \cite{Dep1, Dep2, Dep3, Pereira, Anefalos1, Anefalos2, Mongelli}.

The main feature of the intranuclear cascade calculation with CRISP is the multicollisional approach \cite{Kodama, Goncalves} where
the full nuclear dynamics is considered in each step of the cascade. In this process the nucleus is modelled by an
infinite square-potential which determines the level structure for protons and neutrons. The effects of the nuclear potential
are present in the transmission of the particles through the nuclear surface or through an effective mass according to the Walecka mean field approximation \cite{Serot}. One important feature of the code, in simulating the
intranuclear cascade, is the Pauli blocking mechanism, which
avoids violation of the Pauli principle. In the CRISP code a
strict verification of this principle is performed at each step
of the cascade, resulting in a more realistic simulation of
the process.
After  nuclear thermalization, the competition 
between fission and evaporation processes, which includes neutrons, protons 
and $\alpha$-particles, is determined by the ratios between their respective 
widths according to the Weisskopf model for evaporation and to the 
Bohr-Wheeler model for fission. These ratios are given by:
\begin{align}
 \dfrac{\Gamma_p}{\Gamma_n} = \dfrac{E_p}{E_n} \exp \left\lbrace 2 \left[ 
(a_p E_p)^{1/2} - (a_n E_n)^{1/2} \right] \right\rbrace ,
 \label{gamap}
\end{align}
and
\begin{equation}
 \dfrac{\Gamma_{\alpha}}{\Gamma_n} = \dfrac{2 E_{\alpha}}{E_n} 
\exp \left\lbrace 2 \left[ (a_{\alpha} E_{\alpha})^{1/2} - (a_n E_n)^{1/2} \right] 
\right\rbrace .
 \label{gamaa}  
\end{equation}
for evaporation and by
\begin{align}
 \dfrac{\Gamma_f}{\Gamma_n} = K_f \exp \left\lbrace 2 \left[ (a_f E_f)^{1/2} - (a_n E_n)^{1/2} \right]  \right\rbrace ,
 \label{gamaf}
\end{align}
where,
\begin{align}
 K_f = K_0 a_n \dfrac{\left[ 2(a_f E_f)^{1/2} - 1 \right]}{(4A^{2/3}a_f E_n)} ,
 \label{Kf}
\end{align}
for fission. The parameters $a_i$ stand for the density levels calculated by 
Dostrovsky's parametrization \cite{Dostrovsky} and $E_i$ is given by
\begin{align}
 \begin{split}
   E_n &= E - B_n, \\
   E_p &= E - B_p - V_p, \\
   E_{\alpha} &= E - B_{\alpha} - V_{\alpha} \\
   E_f &= E - B_f.
 \end{split}
\label{enerEsta}
\end{align}
where $B_n$, $B_p$ and $B_{\alpha}$ are the separation energies for neutrons, 
protons and alphas, respectively, and $B_f$ is the fission barrier. 
$V_i$ stands for the Coulomb potential.
At each $n^{\rm th}$ step of the evaporation, the excitation energy of the 
compound nucleus is modified by
\begin{equation}
E^{n}=E^{n-1}-B-V-\varepsilon,
\end{equation}
where $\varepsilon$ denotes the kinetic energy of the emitted particle.
It should be noted that the CRISP does not describe the (multi)fragmentation process of nuclei at the current stage of calculation, however the initial work of implementation is  already in the progress.

The fission process has been successfully described by the multimodal fission approach (it also called as random neck rupture model (MM-RNRM)) \cite{Brosa}, based on the collective effects of nuclear deformation during fission via a liquid-drop model and takes into account a single-particle effects via microscopic shell model corrections. CRISP code was then adapted to consider the multimodal approach by using the eq.(11). 
Considering the different channels of fission, CRISP reproduced different aspects of photo- and proton-induced fission as the spectrum of fissioning nuclei, charge- and mass-distribution of the fission fragments, calculate the fissility values for actinides under study. The calculation by CRISP allowed us also to distinguish the pre- and post-scission neutrons, and compare this values with experimental values of total neutrons emission.

The results from CRISP together with experimental isobaric distribution
for photo- and proton-induced fission reactions in actinide targets are shown in figs. 13 and 14, respectively \cite{Gaya3, Gaya4}. Qualitatively, the agreement between the calculation and the experimental data is fairly good for both type of reactions. Such agreement is clearly shown for low energies photons (fig. 13 (a, b)), where the calculations reproduce the experimental features quite well. At bremsstrahlung energy 3500 MeV the discrepancy becomes larger. The reason of it could be explained by the continuous character of bremsstrahlung spectrum due to influence of low-energy part. However, in the case of Th we can see more realistic calculations even at high energy of photons. For proton-induced fission (fig. 14), the calculations reproduce the shape of the experimental distributions: position and width of the peaks for symmetric and asymmetric modes are in fair agreement with the experimental data. However, the calculated distributions by CRISP code are systematically below the data for $^{241}$Am and $^{237}$Np targets, indicating that the calculations of total fission cross sections underestimate the experimental ones.

Another very interesting phenomenon of fission, that has been confirmed by CRISP calculation, is the presence of superasymmetric binary fission in $^{238}$U and $^{237}$Np targets at proton energy 660 MeV \cite{Gaya5}.
We have applied the induced-activity method to identify unambiguously in general 115 isotopes as a intermediate-mass fragment (IMF) in the mass range $7 \leq A \leq 69$ u. There is a consideration that the formed nuclides could be produced in a very asymmetric binary decay of heavy nuclei originating from the spallation of heavy targets.

The IMFs, i.e. particles heavier than helium isotopes but lighter than fission residues, appear as reaction products for all target nuclei. As it was discussed in \cite{Ricciardi}, IMFs was observed in the binary decay of a compound nucleus includes fission and evaporation with a natural transition in between, may appear from the first reaction stage excited sufficiently high residual, and it might be called fission in a generalized sense. One of the possible ways of the origin of IMFs is that they represent the products of the fission of a species with $A \sim 120-130$ u \cite{Loveland}. Or, for instance, deep spallation may be accompanied not only by emission of nucleons and light charge particles but emission of IMFs. An alternative explanation of the origin of the products with $A = 30-70$ u was suggested by the intranuclear cascade model \cite{Yariv}, according which these fragments are the result of the fission of species with $A \sim 185$ u that are highly excited. The resulting fission fragments evaporate copious numbers of nucleons resulting in a final fragment mass number $A \sim 30-70$ u.
The main characteristics of the IMFs production process is dependence upon the size of nuclear system in the wide range of the target nuclei from Au to Al at 1 GeV proton energy were investigated in \cite{Kotov}. We conclude that intermediate mass fragments produce mainly as a result of the binary decay of the excited nuclei produced at the first stage of the collisions and the contribution of the (multi)fragmentation process for all targets is small.

To take into account the possibility of a superasymmetric fission, we included the additional asymmetric mode, S3, to the CRISP code, which can be described by the usual Gaussian shape from MM-NRM.
In fig. 15 (a, b) the results of CRISP calculations for $^{238}$U and $^{237}$Np are shown by solid line. The squares denote the experimental values that are corrected using the isobaric chain cross sections determined with the charge distribution from the fit.

The calculations not only give a schematic description of reality, but they enable us to improve the data from previous calculations \cite{Gaya4} and estimate the part of the fission cross section originating from symmetric and asymmetric fission modes. The superasymmetric mode contributes with 0.6\% and 1.2\% of the total fission cross section for the $^{238}$U and $^{237}$Np targets, respectively. 
The most striking feature of the superasymmetric mode is the mass number gap around $A =60$ u with respect to the symmetric fragment for both fissioning nuclei. 
During the calculation, the positions of the heavy ($A_S + D_{3AS} = A_H$) and light ($A_S - D_{3AS} = A_L$) peaks of the S3 mode have been defined on the mass scale.  These quantities are 170.5 u and 56.5 u for $^{238}$U target and 173.7 u and 49.7 u for $^{237}$Np target. Thus, we obtain the masses of the fissioning nuclei $A_{f} = 227$ u and  $A_{f} = 223.4$ u in the case of $^{238}$U and $^{237}$Np targets, respectively, e.g. the same masses which have been obtained in our previous experiments for binary fission \cite{Gaya1, Gaya2}.
Hence, the results of calculations confirm that IMFs at intermediate energies are formed predominantly through a binary process. 
It may be concluded that, at the current stage of knowledge, the experimental signatures in the reactions of this work are consistent with the binary decay of a fully equilibrated compound nucleus. The binary decay of a compound nucleus includes fission and evaporation with a natural transition in between, and it might be called fission in a generalized sense.

\subsection*{3.3 Isomer ratios and angular momentum of the reaction products}

The population of the ground ($\sigma_{g}$) and metastable ($\sigma_{m}$) states of a nucleus depends upon the angular momentum in the entrance channel of the reaction, the excitation energy of the residual nucleus, and the type of particles emitted during its de-excitation. The study of the angular momenta of the reaction fragments can provide insight into an information about the configuration of the nuclear system at high excitation. The information about the primary angular momenta of the fragments can be obtained from the measurements of independent isomer ratios of the reaction products. Thus, the isomer ratio can be used to investigate collective rotational degrees of scission configuration. Usually, such measurements are based on the cross section ratio of high-spin ($I_{h}$) state to low-spin ($I_{l}$) state (isomer ratios, $IR = \sigma(I_{h})/\sigma(I_{l}$)). The results of the measurements represent the overall effect of several processes. It is known that de-excitation of the heavy nucleus produced from the 
primary interaction take place via emission of particles, mainly neutrons or fission. These processes proceed successively in several stages depending on the excitation energy and at each step the reaction remnant can undergo fission or emit neutron \cite{Rubchenya}. Therefore, the primary fission fragment originated from the different fissioning nuclei have a wide range spectrum of angular momenta and excitation energy. De-excitation of primary fragments proceed by evaporation of neutrons and emission a cascade of $\gamma$-quanta until populating of the final states with different spin values. Neutrons and $\gamma$-quanta take certain angular momentum and energy, changing thus the initial distribution of primary fragments. The majority of fragments are produced in states with angular momenta higher than the spin of the fissioning nuclei. It is usually assumed that the deformation of fission fragments caused by twisting and bending is the source of generation of high angular momenta \cite{Vishnevskii, Tanikawa}.

The effect of the excitation energy on the probability of product formation in different spin states may be displayed in two ways. First, the spin of the primary fission fragment increases with the increase of excitation energy due to contribution of higher degrees of freedom into the collective motion; second, the difference in the excitation energy causes the difference in the de-excitation process, so during the evaporation of neutrons and $\gamma$-quanta a broad distribution is obtained for fragments with higher excitation energy. 

The isomer ratio can be derived from formula, which involves the ratio of the areas under the peaks of measured $\gamma$-transitions \cite{Vanska, Kolev}:
\begin{eqnarray}
\frac{\sigma_{m}}{\sigma_{g}} = \left[\frac{\lambda_{g}(1-e^{\lambda_{m}t_{1}})e^{\lambda_{m}t_{2}}(1-e^{\lambda_{m}t_{3}})}{\lambda_{g}(1-e^{\lambda_{m}t_{1}})e^{\lambda_{m}t_{2}}(1-e^{\lambda_{m}t_{3}})} 
\times\left(\frac{k_{m}N_{g}\eta_{m}\epsilon_{m}}{k_{g}N_{m}\eta_{g}\epsilon_{g}} - p\frac{\lambda_{g}}{\lambda_{g}-\lambda_{m}}\right) + p\frac{\lambda_{m}}{\lambda_{g}-\lambda_{m}}\right]^{-1},
\end{eqnarray}
Here, $N$ is the area under the photopeak with energy $E_{\gamma}$. The subscripts $m$ and $g$ label variables referring to the metastable and the ground states, respectively.

The average initial angular momentum of the reaction remnant is deduced from the measured cross sections of isomer by calculations based upon the concepts of the statistical model. The formalism of these calculations was introduced by Huizenga and Vandenbosch \cite{Huizenga} for the calculation of isomeric ratios in spallation reactions where the angular momentum distribution of the nuclei can be calculated. The model considers successive neutrons and $\gamma$-rays emission from the primary fission fragments, leading to the final distribution in the population of high- and low-spin states. The main element in this calculation is the nuclear level density of the spin distribution, which is proportional to the probability of the population the nuclear states.

The probability distribution of initial angular momentum is assumed to be represented by following formula:
\begin{eqnarray}
P(J_{i}) \sim (2J_{i}+1)\exp[-J_{i}(J_{i}+1)/B^{2}],
\end{eqnarray}
\noindent 
where $P(J_{i})$ is the probability of the primary fragment formation with spin value $J$, and $B$ is a parameter which defines the width of the spin distribution. The root-mean-square angular momentum $(\bar{J^{2}})^{1/2}$ of the primary product is equal to $B$ for large values of $J$:
\begin{eqnarray}
(\bar{J^{2}})^{1/2} \cong B.
\end{eqnarray}
\noindent 

The effective excitation energy of residual nucleus after emission of nucleon $E^{*}_{eff}$ is calculated using the following expression:
\begin{eqnarray}
E^{*}_{eff}=E^{*}-E_{Coul}-E_{p}-E_{KE}
\end{eqnarray}
\noindent where $E_{Coul}$ is the Coulomb barrier for the emitted proton, $E_{p}$ is the binding energy of a nucleon inside a nucleus, and $E_{KE}$ is the mean kinetic energy of nucleon. 

According to the evaporation model nucleons are emitted by the excited nucleus with a mean energy $E_{KE}=2T$, where $T$ is the nuclear temperature determined by the expression \cite{Blatt}:
\begin{eqnarray}
aT^{2}-4T = E^{*}_{eff},
\end{eqnarray}
\noindent where parameter $a$ is the level-density parameter ($a = A/10$ MeV$^{-1}$).

At the stage of the process involving the cascade of $\gamma$-transitions eventually leading to the metastable or the ground state, the density of the nuclear level spin distribution determines the probability of population of intermediate nuclear states. In the calculations, we use the spin part of the Bethe-Bloch formula given by: 
\begin{eqnarray}
P(J) \sim (2J+1)\exp[-(J+0.5)^{2}/2\sigma^{2}],
\end{eqnarray}
\noindent where $P(J)$ is the probability distribution of levels with spin $J$ and $\sigma$ is the spin cutoff parameter
which characterizes the angular momentum distribution of the level density and is related to the moment of inertia and the temperature of the excited nucleus, being given by $\sigma^{2}=0.00889\sqrt{aE^{*}_{eff}}A^{2/3}$.

The average energy and number of  $\gamma$-rays emitted from the nucleus of initial excitation energy $E^{*}_{eff}$ can be estimated by means of the formula \cite{Huizenga1}:
\begin{eqnarray}
\bar{E_{\gamma}}=4\left(\frac{E^{*}_{eff}}{a}-\frac{5}{a^{2}}\right)^{1/2}.
\end{eqnarray}
The energy of emitted $\gamma$-ray is determined by using of eq.(18) and the value of $E^{*}_{eff}$ at each stage of calculation.

During calculations we used both dipole $E1$ and quadrupole $E2$ multipolity of $\gamma$-transitions. However, it was shown in \cite{Aumann} the relative share of the quadrupole transitions in the de-excitation process does not exceed 10\%.
The excitation energy $E^{*}_{eff}$ and, accordingly, the energy $E_{\gamma}$ of emitted photons are determined at each stage of the cascade. The last level from which the population of the ground or the isomeric state occurs is characterized by an excitation energy not higher than 2 MeV.

In fig. 16 we represent experimental values of IRs obtained in photofission of $^{238}$U and $^{232}$Th at end-point energy from $E_{\gamma max} = 50$ MeV and 3500 MeV \cite{Gaya-iso}. One can see from the figure the general increase of IRs with fragment mass number $A$. This fact can be related to the excitation energy of fission fragments \cite{Tanikawa}.
The total excitation energy released in fission is well known to be assumed proportional to the fragment mass. The heavier the fragment mass, the higher the excitation energy resulting from the statistical sharing of the total excitation energy gained by the fissioning nucleus.
The only irregularity is in the mass range of $A \sim 131-132$ u, which can be explained by influence of neutron shell effect close to neutron number $N = 82$ with low deformation, low excitation energy and thereby, the low probability of the formation in the high-spin state \cite{Frenne}. 

From fig. 16 it is also apparent that at the increase in yields of fission fragments with the increase in the incident energy, IRs do not practically increase, although all measurements indicate a relative increase in IRs within the accuracy of $\sim$ 10\%. Influence of the increment in the excitation energy and angular momentum of the fissioning nuclei on the value of IRs of the final products has been studies in a number of works in the energy range 10--500 MeV of incident protons \cite{Tanikawa, Saha, Rao, Rudy, Zhuikov}, neutrons with the energy up to 14 MeV \cite{Vishnevskii1, Patronis}, and $\alpha$-particles with the energies 26--42 MeV \cite{Aumann, Warhanek}. The results of the experiments are contradictory. In fact, along with the statement that IR dependence on the excitation energy is weak \cite{Saha, Rao, Rudy, Zhuikov}, there are data indicating the inverse effect \cite{Tanikawa, Vishnevskii1, Aumann, Patronis, Warhanek}. The isomers ratios reported in \cite{Zhuikov} for fission of Ta by 100--500 MeV protons do practically not depend on the incident energy. The authors \cite{Zhuikov} explain this fact by saturation of the population of high-spin states of the fragments and by the competition of various channels at the pre-equilibrium stage of the reaction.

In table 3 we represent experimental and calculated values of IRs together with the angular momentum ($\bar{B}$), determined by using eqs. (21)-(26) and based on the measured IRs, obtained in photofission of $^{238}$U and $^{232}$Th at end-point energy from $E_{\gamma max} = 50$ MeV \cite{Gaya-iso}.
The data of IRs from our work  agrees to the results of measurements performed for fission of uranium targets by low-energy photons and thermal neutrons \cite{Jacobs, Vishnevskii, Aumann}.
Isomer ratios and the average angular momenta of fragments in the fission with fast neutrons, protons, and $\alpha$-particles are relatively higher \cite{Tanikawa, Vishnevskii1, Aumann, Saha, Rao, Rudy}. 
It is explained by the influence of angular momentum of the projectile on the spin distribution of the primary fragments. As it was proved by the cascade-evaporation calculation results \cite{Iljinov}, the forward momentum transmitted to the nucleus at intermediate energy range is higher in the case of proton-, neutron- and $\alpha$-induced reaction, compared with the photonuclear reactions.

\section*{4 Investigation of complex particle-induced reactions}

The study of nucleus-nucleus collisions is a source of experimental data, which are extremely important for scientific and technological application. There are multiple questions addressed in the context of e.g., design of accelerator driven systems of energy amplification and nuclear waste utilization, astrophysical studies, radiological safety, which cannot be answered without knowledge of collision cross section. Experimental data for heavy-ion-induced fission is of particular interest as the knowledge obtained in such experiments can give us the information about the variation of fission cross section and fissility with angular momentum transferred in the entrance channel of reaction for nucleon- and heavy-ion-induced fission. According to the classical rotating-liquid-drop model (RLDM) \cite{Cohen} and various theoretical models, involved different macroscopic and microscopic aspects of excited high-spin nuclear matter \cite{Sierk, Wilczynski, Capurro, Mustafa}, the fission barrier height should decrease monotonically, and eventually vanishes with increasing angular momentum.

Nuclear recoil experiments provide in addition valuable information, such as angular distributions and kinetic energies of product nuclei as well. Furthermore it deepens our understanding about the reaction mechanism. One of the simplest recoil experiments is accomplished by the thick-target thick-catcher method. It has been extensively applied to studies of hadron-induced reactions on various targets over a wide range of incident particle energy \cite{Alexander, Winsberg1}.

The measurements regarding recoil nuclei were transformed into kinematic quantities using the two-step vector model of high energy nuclear reactions \cite{Winsberg1}. According to \cite{Winsberg1}, the first reaction stage involves the formation of a residual cascade nucleus having an excitation energy $E^{*}$ and a velocity $\vec{v}$ (or the momentum $\vec{p}$) along the beam axis. It is assumed that, at the second reaction stage, there occurs the evaporation of nucleons and light nuclei, with the result that the nucleus in question acquires an additional velocity $\vec{V}$. Thus, the velocity $\vec{V_{l}}$, of a recoil nuclide in the laboratory system is taken to be the sum of two vectors $\vec{V_{l}}=\vec{v}+\vec{V}$. The velocity vector $\vec{v}$ results from the fast projectile-target interaction (the ``abrasion" step of the abrasion -- ablation model), while the velocity vector $\vec{V}$, assumed to be isotropic in the moving system, results from the slow de-excitation of the excited primary fragment (the ``ablation step"). The vector $\vec{v}$ is assumed to be constant while the values of the vector $\vec{V}$ are assumed to have a Maxwellian distribution. No correlation is assumed to exist between the two vectors. The vector $\vec{v}$ can be decomposed into its two orthogonal components: parallel and
perpendicular to the beam ($v_{\parallel}$ and $v_{\bot}$). 

According to the model it was assumed that: (1) there is no correlation between the velocity $\vec{v}$ of the excited nucleus and $\vec{V}$; (2) the angular distribution of fragments in the moving frame is isotropic; (3) the mean range in the laboratory system is proportional to the fragment speed; (4) $v_{\bot}$ is zero.

The results of these measurements are the fractions of each nuclide which have recoiled out of a target in the forward
and backward direction, namely
\begin{eqnarray}
F={S_{F}}/(S_{F}+S_{B}+S_{T}),                         \qquad  B={S_{B}}/(S_{F}+S_{B}+S_{T}),
\label{fraction}
\end{eqnarray}
\noi where $S_{F}$ , $S_{B}$, and $S_{T}$ are the photopeak areas associated with the products under study in the catch foils and in the target. The resulting data were used to calculate the forward-backward ($F/B$) anisotropy of product emission and the ranges in the target material ($R$).

The mathematical formalism developed in \cite{Winsberg1} makes it possible to calculate, on the basis of
experimental results for $F$ and $B$, parameters that characterize the first ($v_{\parallel}$, $E^{*}$) and the second ($R$ and $T$) stage of the interaction, where $T$ is the kinetic energy of a fragment and $E^{*}$ is mean excitation energy of the residual cascade nucleus.

In converting product ranges into kinetic energies, one can use the range -- energy tables of Northcliffe and Schlling \cite{Northcliffe}. In order to transform the ranges into the kinetic energy of the reaction products, the relation from \cite{Winsberg1} was used:
\begin{eqnarray}
R =kT^{N/2}                                                                                                             \label{range}                                         
\end{eqnarray}

\noi where parameters $k$ and $N$ are obtained by fitting the range dependence on energy of accelerated ions \cite{Winsberg1}.

For the value of $\eta_{\parallel}(= v_{\parallel}/V$, the ratio of the parallel component of the cascade velocity to the second-step velocity), we have used $\eta_{\parallel}$ as derived from integrating our angular distributions, where 
\begin{eqnarray}
\eta_{\parallel}=v_{\parallel}/V=[(F/B)^{1/2}-1]/[(F/B)^{1/2}+1]  
\label{eta}                                                               
\end{eqnarray}

The relative linear momentum transfer $p_{\parallel}/p_{CN}$, where $p_{CN}$ is the momentum of a hypothetical compound nucleus formed in a complete fusion, is the main feature that can be considered as the sign of a complete or an incomplete fusion. The mean excitation energy of the residual cascade nucleus ($E^{*}$) can be estimated by using the relation \cite{Lagarde}:

\begin{eqnarray}
E^{*}/E_{CN}=0.8p_{\parallel}/p_{CN},                                                                                   
\label{energy}                                 
\end{eqnarray}

\noi where $E_{CN}$ is the excitation energy of a hypothetical compound nucleus formed by fusion of the projectile and the target nucleus.

We measured the cross sections for the binary fission of $^{nat}$Pb induced by $^{7}$Li ions at bombarding energy of 245 MeV \cite{Gaya6, Gaya7}. The recoil technique, using thick targets and thick catchers, combined with a mathematical formalism based on the two-step model, permitted us to determine the kinematic characteristics of the reaction products. 

The experimental isobaric cross sections of fission fragment formed by interaction of $^{7}$Li-ions with natural lead at 245 MeV together with the fit, which assumes symmetric fission mode, are presented in fig. 17.
The total fission cross section in this work was found to be 604.74 $\pm$ 90.71. This value exceeds the fission cross section for proton-induced fission at the same energy of projectile. Comparison the ratio at the same excitation energy for the fission cross sections by Li-ions and protons gives a factor 5--6 \cite{Prokofiev}. The average value of the linear momentum transferred
divided by the full momentum transferred to the target obtained in the present study is $p_{\parallel}/p_{CN} = 0.46\pm0.09$. Our data confirms that a complete fusion is not the dominant process at the energy regarded. 
In this intermediate energy region, the complete fusion process is partially replaced  by other processes.
We can assert that only approximately 40\% of the fission channel at this energy can result from complete fusion. The remained momentum transferred could be connected to incomplete fusion, or a significant pre-equilibrium contribution of the emission of nucleons or light nuclei. We estimated the average angular momentum imparted into fissile system. At energies above the barrier, the formula for calculation the average angular momentum, $< \ell >$, was taken from \cite{Capurro}:
\begin{eqnarray}
< \ell > = \frac{2}{3}\sqrt{\frac{2\mu R^{2}(E_{c.m.}-V_{CB})}{\hslash^{2}}}.                                                                
\end{eqnarray}

Here, $R$ is the maximum distance between two nuclei at which the collision leads to a reaction, $\mu$ is the reduced mass and $V_{CB} = (Z_{1}Z_{2}e^{2})/R$ is the Coulomb energy of the system at distance $R$; $E_{c.m.}$ is at the center-of-mass bombarding energy.
The distance $R$, calculated on the base of experimental total reaction cross sections \cite{Capurro}, expressed as a function of interactive fragment masses $A_{1}$ and $A_{2}$:
\begin{eqnarray}
R=1.36(A_{1}^{1/3}+ A_{2}^{1/3})+0.5 fm.                                                                  
\end{eqnarray}
In our case, the average angular momentum is estimated to be about $< \ell >  = 55-60 \hslash$. The interval is due to the fact that the spin distribution is not a sharp cutoff.

Calculated fission-barrier heights as a function of angular momentum have shown a lowering of the fission barriers with increasing of angular momentum from zero value \cite{Sierk, Wilczynski, Capurro, Mustafa}. It can be seen that fission is expected to play a significant role only above 50 $\hbar$ \cite{Cohen, Blann}, when the fission barrier has dropped to about half of the value it has at zero angular momentum. It was stressed out that the dependence can be neglected in the case of nucleon-nucleus interaction at intermediate energies. This is due to only a small angular momentum (of only a few $\hbar$) being imparted to the nucleus by incident nucleons in the energy region under study. Meanwhile, it becomes important for higher  momentum values for heavy-ion-induced reactions. Comparison with proton-induced fission show that the linear momentum transferred to the fissile system depends on the nature of projectile. This fact demonstrates that fissile systems are formed at intermediate energy with high angular momentum.
This might be the explanation for the high value of the fission cross section for heavy-ion-induced fission.

Spallation reactions with Au target have been investigated in the interaction of 4.4-GeV deuterons \cite{Gaya8, Gaya9}. Investigation of deuteron-nucleus collisions is important since deuteron represents itself a weakly bounded system. And during interaction with heavy nucleus the difference between whole nucleus and distinct nucleons can be derived. 

In the case of the gold target a wealth of data has been accumulated about the individual nuclide production cross section from nucleon-nucleus interaction at energies from thresholds up to 2.6 GeV \cite{Michel, Michel1}. In spite of these studies cover a wider range of energies, they are restricted by the determination of excitation functions of the reaction products. 

The data on the isotopic distributions of residual nuclei produced in proton-induced reactions on $^{179}$Au at 800 MeV have been obtained at the FRS at GSI
\cite{Rejmund, Benlliure1} and satisfactory theoretical interpretations of those data have been published \cite{Benlliure2}. We present additional valuable information based on the induced-activity method which has importance for the models and codes improvement.
In our investigation \cite{Gaya8, Gaya9} we presented a comprehensive work, passing from nucleon to heavier deuteron projectile and to higher energies compared with 2.6 GeV. Application of the thick-target thick-catcher technique for determination of the main kinematic features of the reaction fragments as well as determination its charge and mass distributions, allowed to deep investigate the mechanism of each channel of the reaction and to reconstruct the whole picture of interaction at high energies.

As a result of the measurements it was obtained 110 radioactive nuclide cross sections with mass number $22 \leq A \leq 198$ u. In the present work the analysis of the charge distributions were obtained as the function from \cite{Kaufman1}:
\begin{eqnarray}
\sigma(Z, A)=\sigma(A)\exp\left(-R\left|Z-SA+TA^{2}\right|^{3/2}\right),
\end{eqnarray}
\noindent where $\sigma(Z, A)$ is the independent cross section for a given nuclide production with atomic charge $Z$ and a mass number $A$; $\sigma(A)$ is the total isobaric cross section of the mass chain $A$. The parameter $R$ defines the width parameter of the charge distribution and parameters $S$ and $T$ define the most probable charge ($Z_{p}$) for a given isobar chain $A$.

The mass-yield obtained in this manner is shown as the solid squares in fig. 18. The smooth curve on the fig. 18 indicates the best fit for the mass number $40 \leq A \leq 198$ u based on the experimental data depended on the contribution from different process such as evaporation, fission and (multi)fragmentation. From fig. 18 one can see almost no clear-cut distinction between spallation and fission mass ranges. Integration of the mass-yield curves over mass number gives the cross section for the production of reaction residues $2.20\pm0.44$ b.

We can assume that a few sources having different excitations can take part in the formation of residual nuclei: the heaviest fragments ($A \geq 130$ u) result of spallation-evaporation mechanism; the lightest fragments ($A < 40$ u) can be produced in a asymmetric fission and (multi)fragmentation processes \cite{Bonche1, Bonche2}. In the mass region $120 \leq A \leq 150$ u we can espect existence of all processes, such as fission, (multi)fragmentation and spallation-evaporation \cite{Napolitani}.

In our study the hard sphere model \cite{Heckman} for nucleus-nuclear
interactions was used in order to obtain impact parameter of the collision. This model refers to the overlap of the two sharp
spheres forms of interacting nuclei and the total reaction cross
section is presented in terms of a two-parameters expression:
\begin{eqnarray}
\sigma_{R}=\pi r^{2}_{0}(A^{1/3}_{T}+A^{1/3}_{p}-b_{Tp})^{2}    fm^{2},
\end{eqnarray}
where $A_{T}$ and $A_{p}$ are the mass numbers of the target and
projectile nuclei, respectively; $r_{0}$, is the constant of
proportionality in the expression of geometrical nuclear radius
$r_{i}=r_{0}A^{1/3}_{i}$ and $b_{Tp}$ is the overlap parameter. 
Putting the value of the experimental determined total reaction cross section in expression (34) and calculate the value for $b_{Tp}$, 
the impact parameter $b$ can be estimated using relation:
\begin{eqnarray}
b = r_{0}(A^{1/3}_{T}+A^{1/3}_{p}-b_{Tp})    fm,
\end{eqnarray}
From estimated overlap parameter
$b_{Tp}=0.97$ fm in the present experiment the mean value of the
impact parameter $b$ equals to 8.37 fm was obtained.
Since the value of impact parameter lies in the range $1/2(R_p+R_T)\leq b \leq (R_p+R_T)$ \cite{Morrissey}, we can conclude that the collision is peripheral in most cases. (It can be suggested that in peripheral collisions at large impact parameter the probability of the whole deuteron interaction is very low.) 
Some kinematic features as the $F/B$ values and the recoil parameter $2W(F+B)$ (W is the thickness of the target), which is the mean recoil range of the products, can be measured directly. Others, as the excitation energies ($E^{*}$) of the residual cascade nuclei and the longitudinal cascade velocities of the reaction residues ($v_{\parallel}$) can be estimated on the base of the two-step vector model of high energy nuclear reactions \cite{Winsberg1}.

The longitudinal velocity ($v_{\parallel}$) for the recoiling nuclei are plotted in fig. 19. The presence of a plateau at a wide range of fragment masses can be explained as a possible saturation of the energy-momentum transfer. Our data follow well the general tendency of  values for protons \cite{Kaufman2}. This behavior suggests that a similar process is taking place after the first step for reactions induced by different kinds of incident particles

The value of the longitudinal velocity ($v_{\parallel}$) may be used to determine the average cascade deposition energy (excitation energy, $E^*$) as follows
\cite{schd}:
\begin{eqnarray}
E^{*}=3.253\times10^{-2}k^{'}A_t v_{\parallel} [T_p/(T_p+2)]^{0.5} ,
\end{eqnarray}
\noindent where $E^{*}$ and the bombarding energy $T_p$ are expressed in terms 
of $m_p c^2$. $A_t$ is the target mass and $v_{\parallel}$ is in units of (${\rm MeV/u})^{0.5}$. 
The constant $k'$ has been evaluated by  Scheidemann and Porile \cite{schd} 
based upon Monte Carlo cascade calculations as $k^{'}=0.8$.
The mean excitation energies for the different product mass ranges are: $1416.0\pm283.0$ MeV ($A < 40$ u); $390.0\pm78.0$ MeV ($40 \leq A \leq 120$ u); $264.0\pm53.0$ MeV ($A \geq 131$ u). One can assume that a few sources having different excitations and can take part in the formation of residual nuclei: the lighter nuclides require the largest excitation energy for their formation. The
reaction residues in the mass range of $40 \leq A \leq 120$ u have about the same excitation energy, indicating that fission and probably (multi)fragmentation processes take place in the
same excitation energy regime.
An evidence of a fission process contribution at this mass range has been
confirmed by a number of other experiments \cite{Enqvist, Benlliure1}.
On the other hand, the contribution from the (multi)fragmentation process in the same
mass range is also present \cite{Bondorf}. Actually, the mass yields of
both processes overlap and we cannot clearly isolate the fission
process from (multi)fragmentation. As shown in recent work of
Napolitani $\textit{et al.}$ \cite{Napolitani}, in the vicinity of the (multi)fragmentation threshold, both the fission and (multi)fragmentation processes
can contribute to the formation of the residue in this mass
range.
Fragment with masses $A \geq 131$ u connected to the spallation-evaporation process have excitation energy smaller than fission and (multi)fragmentation products. Obtained data on the excitation energy, $E^{*}$, allow us to 
suggest the existence of a different sources in the formation of the reaction residues in this deuteron-induced reaction.

\section*{5 Concluding remarks and outlook}

We intend to continue the fission and spallation investigations of pre-actinide and actinide
nuclei as well as to do such special nuclear reaction
measurements as detection of high-spin isomers at intermediate- and high-energies of projectile.
Such type of calculation we intend to do soon, for the
reaction induced by deuterons at high energies with gold and lead isotopes.
We plan to deeper investigate  the fission of pre-actinide nuclei.
In the study of fission of pre-actinide nuclei based on the fission-fragment mass and energy distributions, \cite{Itkis} the asymmetric component has been observed in the mass distribution of the fission fragments in the reactions with $\alpha$-particles and protons at low excitation energies. The boundaries of validity of the hypothesis of two distinct fission modes have been estimated at $200 \leq A_{f} \leq 232$ and $82 \leq Z_{f} \leq 92$, where $Z_{f}$ is number of the fissioning nucleus.

The study of reaction mechanism concerning the transition between complete (CF) and incomplete fusion (ICF) processes is important in the view of understanding the interplay between such two dominant modes of the nuclear interaction \cite{Kolata, Hussein}. Exactly such issue we would like to investigate in our further investigation using $^{197}$Au, $^{181}$Ta and $^{209}$Bi pre-actinides targets induced by $^{11}$B-ions at intermediate energy range. The study of the fission cross section and the comparison with calculated data in the frame of different models can give hints about the probability of the compound nucleus production and its suppression for the reason of the break up and incomplete fusion processes. The investigation of the decay of weakly bound projectiles on the different components is very useful since it allows to study of the breakup effect on the different reaction channels. 

Future developments in the calculation by CRISP code concerning of the reaction mechanism may contribute to a better prediction of the multichance fission process, spallation and (multi)fragmentation in wide range of incident energy for different projectiles. So, we intend to modify the code by the inclusion of a new event generator responsible for a nuclear reaction initialized by deuteron and assuming the small binding energy of the deuteron with further development and implementation of heavier nuclei. 

\section*{Acknowledgment}
G. Karapetyan is grateful to FAPESP Grant No. 2011/00314-0 and to International Centre for Theoretical Physics (ICTP) under the Associate Grant Scheme.

\medbreak

\newpage
\begin{figure*}[h!]
\includegraphics[width=16cm]{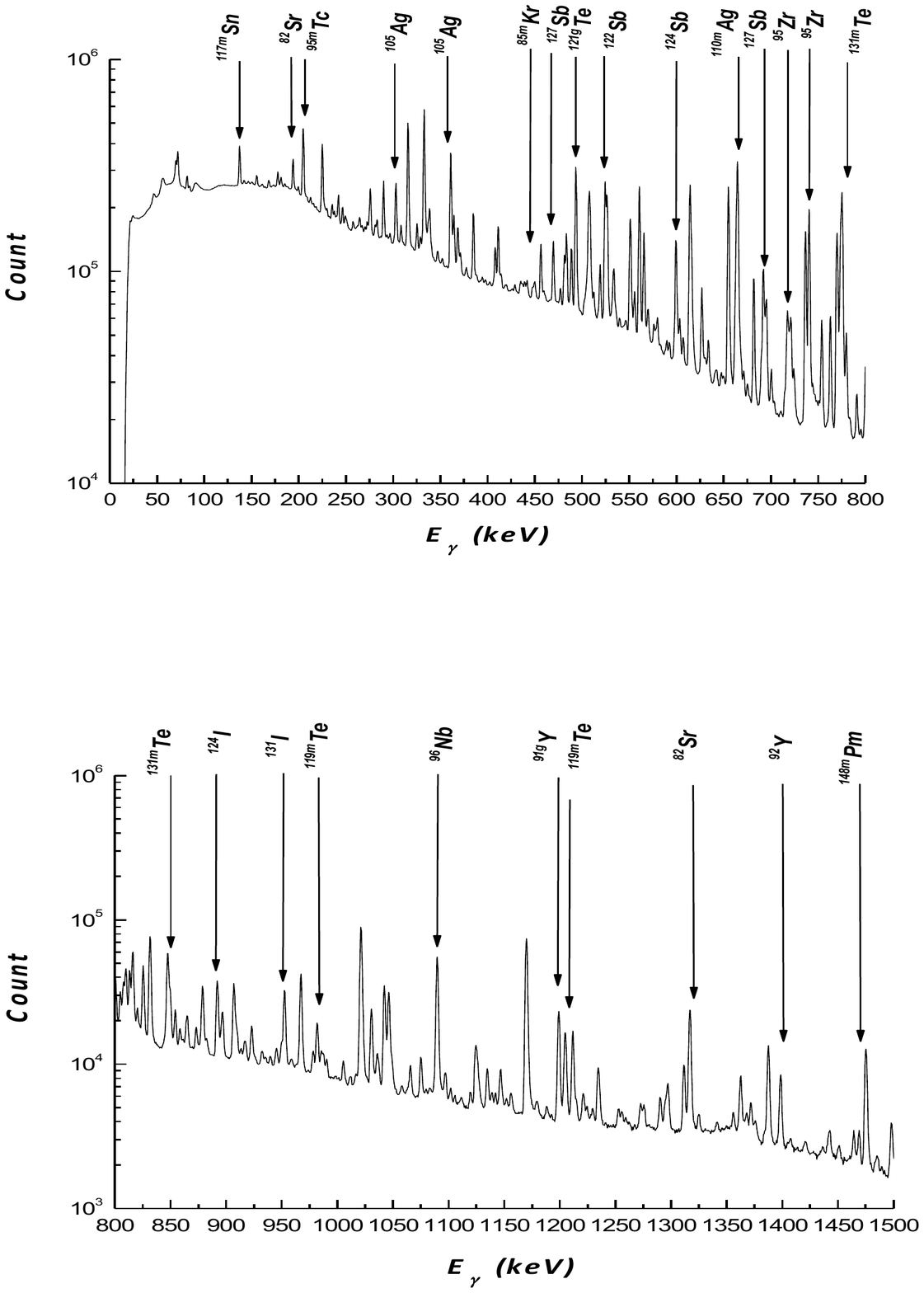}\vspace{-3cm}
\caption{\small $\gamma$-spectrum from the irradiated $^{232}$Th sample at $E_{\gamma max} = 50$ MeV, the measurement time is 30 h.}
\end{figure*}

\begin{figure*}[h!]
\includegraphics[width=10cm]{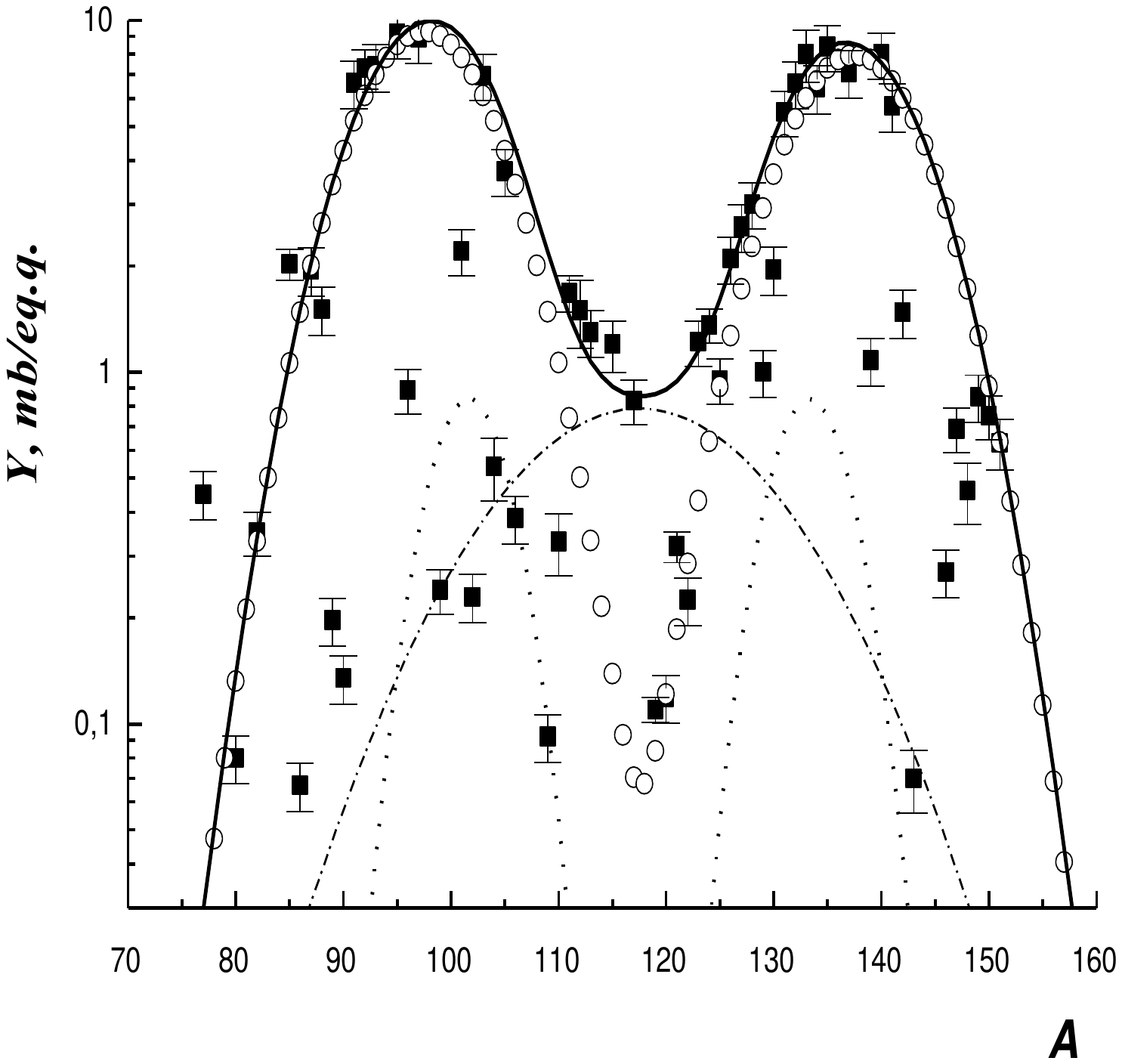}\vspace{-3cm}
\caption{\small Mass-yield distributuion of $^{238}$U photofission  
at end-point energy $E_{\gamma max} = 50$ MeV \cite{Nina1}. S mode indicated as dash-dotted curve, S1 as dotted curve, and S2 as open circles. The total fission yield is represented by the solid curve and experimental data are solid squares, respectively.}
\includegraphics[width=10cm]{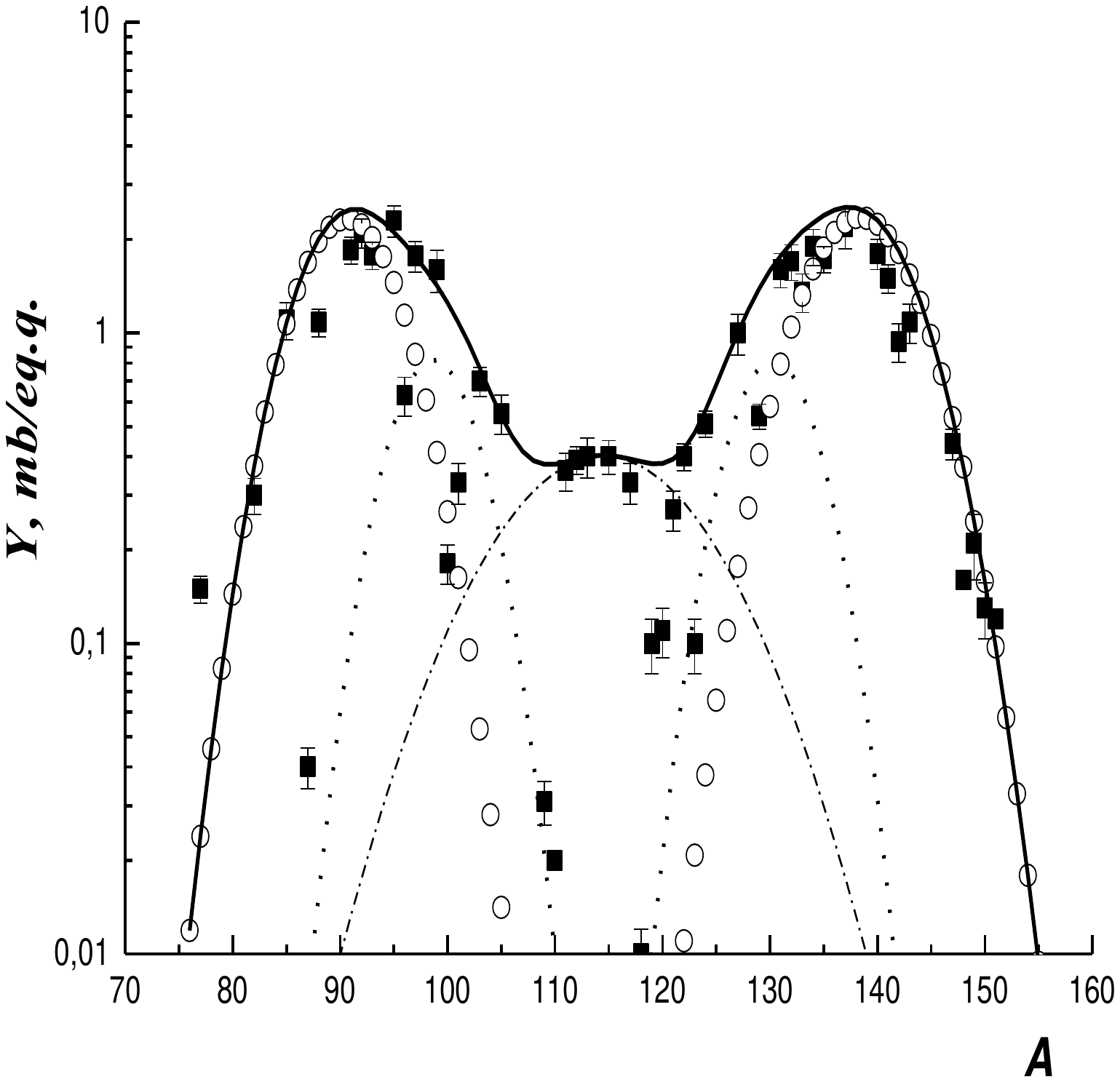}\vspace{-3cm}
\caption{\small Mass-yield distributuion of $^{232}$Th photofission 
at end-point energy $E_{\gamma max} = 50$ MeV \cite{Nina2}. S mode indicated as dash-dotted curve, S1 as dotted curve, and S2 as open circles. The total fission yield is represented by the solid curve and experimental data are solid squares, respectively.}
\end{figure*}

\begin{figure*}[h!]
\includegraphics[width=10cm]{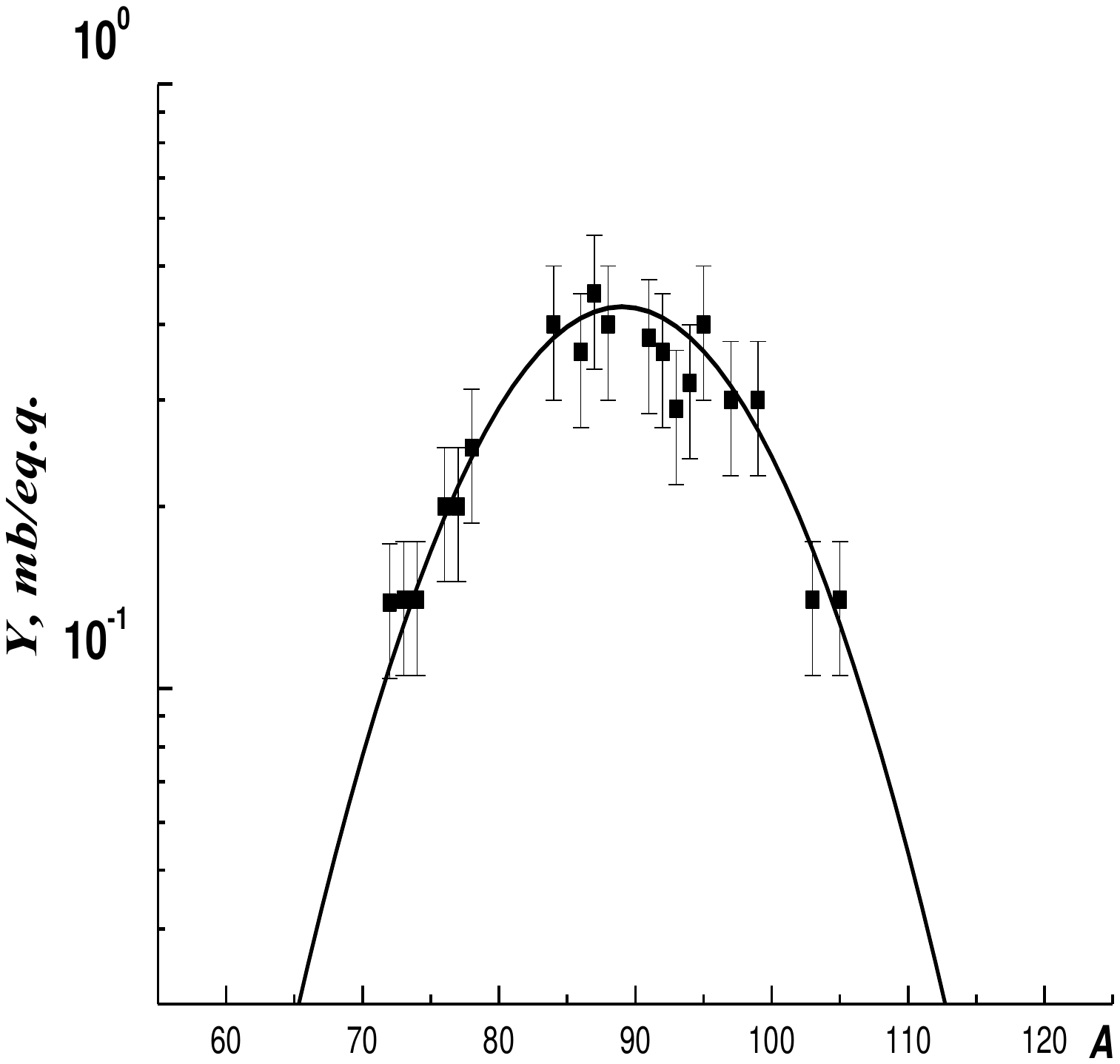}\vspace{-3cm}
\caption{\small Mass-yield distributuion of $^{181}$Ta photofission  
at end-point energy $E_{\gamma max} = 50$ MeV \cite{Deppman1}. The experimental data are shown by 
the solid squares.}
\includegraphics[width=10cm]{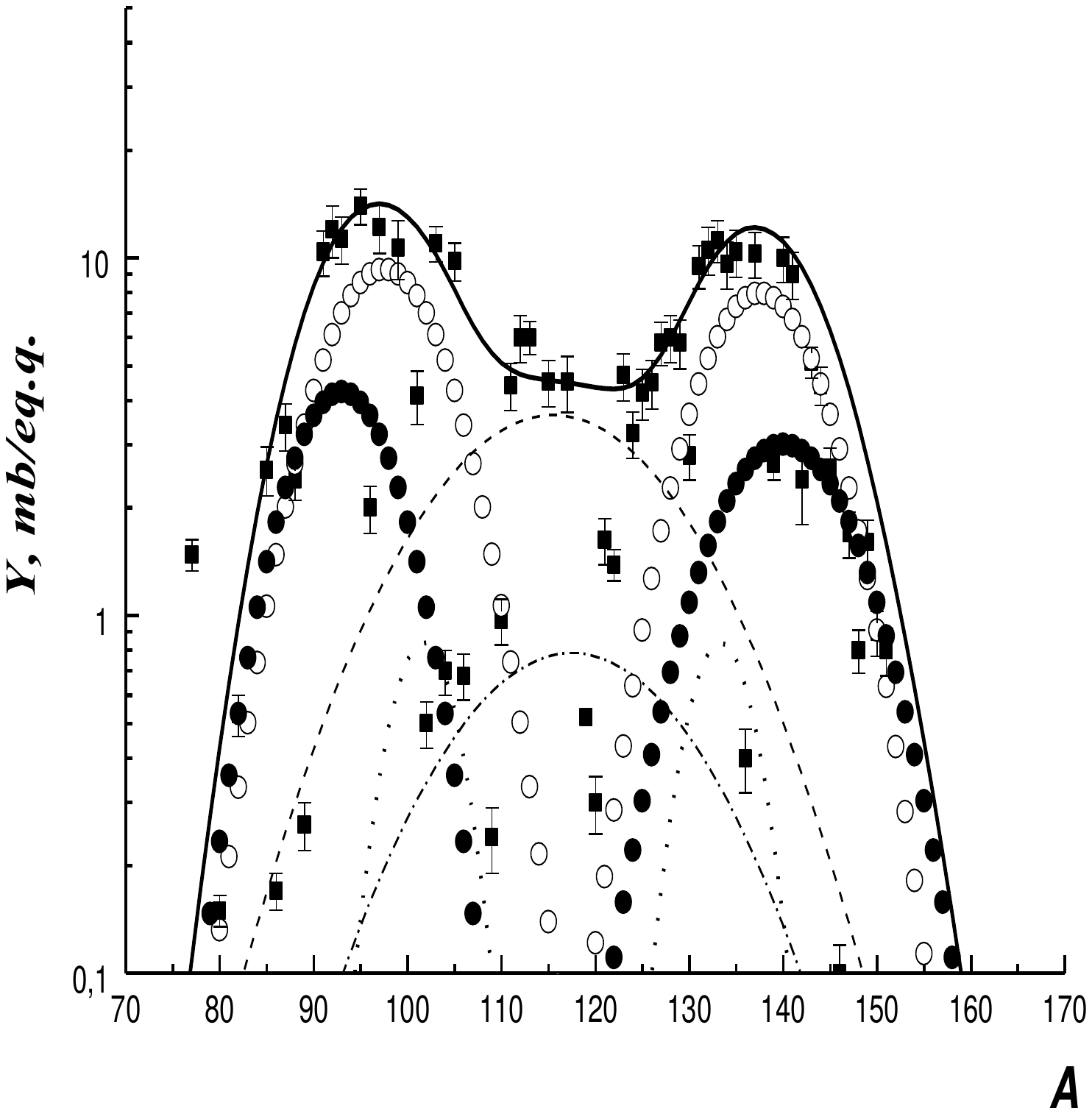}\vspace{-3cm}
\caption{\small Mass-yield distributuion of $^{238}$U photofission 
at end-point energy $E_{\gamma max} = 3500$ MeV \cite{Nina1}. The notation for the yields from low-energy fission is identical to that in Fig. 2. For high-energy fission, the closed circles and dashed curve indicate the additional asymmetric and symmetric modes, respectively.}
\end{figure*}

\begin{figure*}[h!]
\includegraphics[width=10cm]{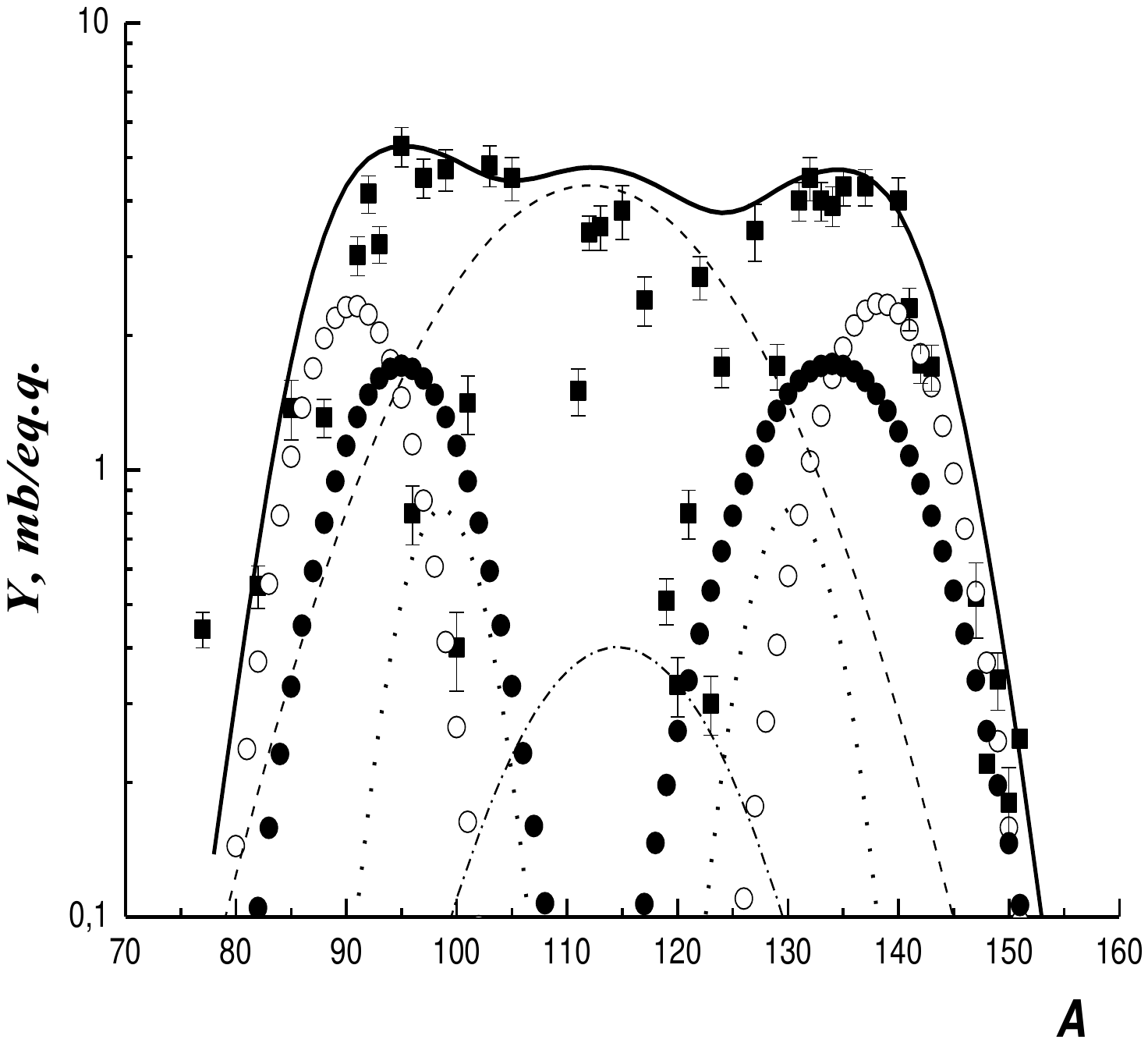}\vspace{-3cm}
\caption{\small Mass-yield distributuion of $^{232}$Th photofission  
at end-point energy $E_{\gamma max} = 3500$ MeV \cite{Nina2}.The notation for the yields from low-energy fission is identical to that in Fig. 3. For high-energy fission, the closed circles and dashed curve indicate the additional asymmetric and symmetric modes, respectively.}
\includegraphics[width=10cm]{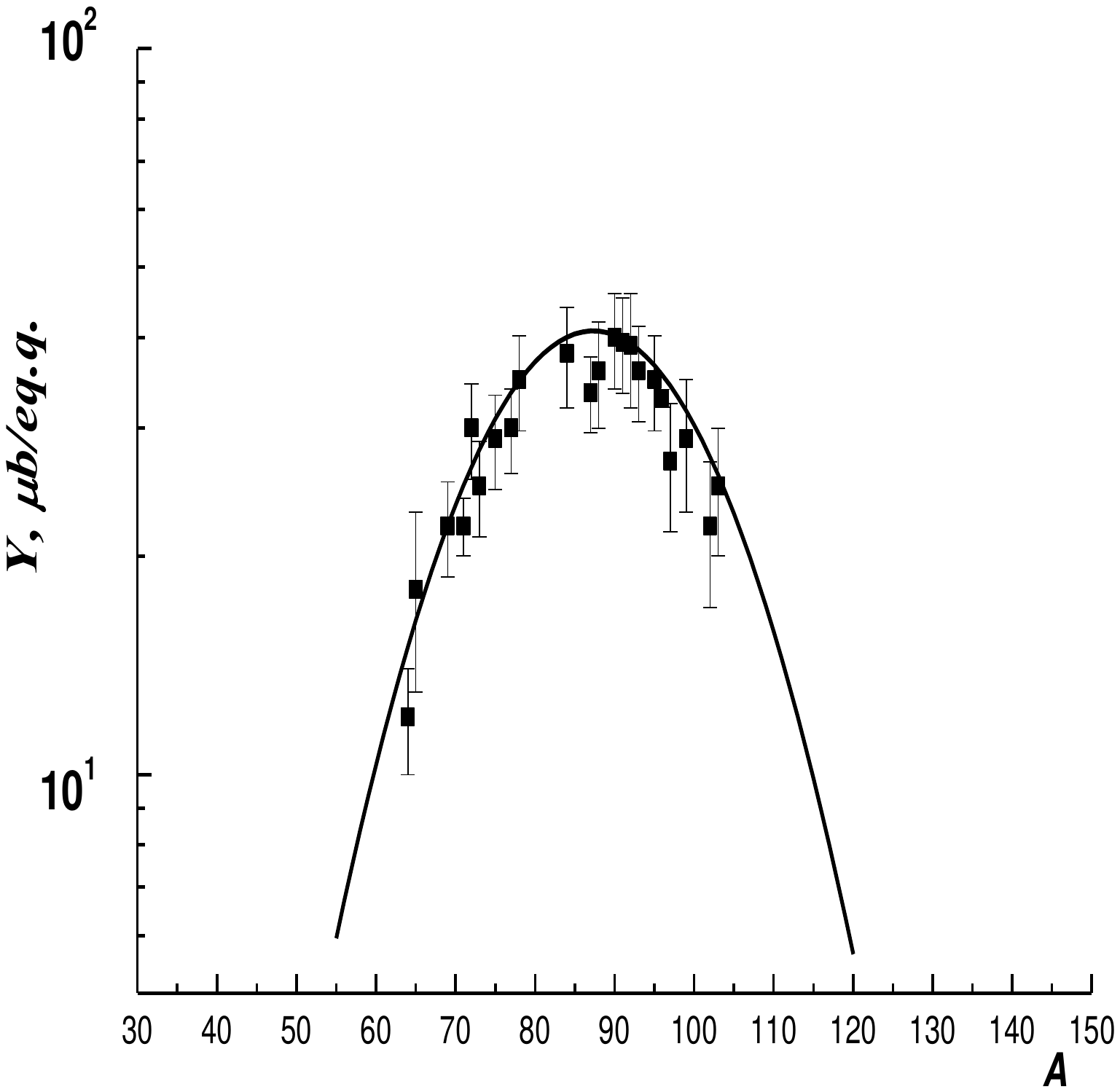}\vspace{-3cm}
\caption{\small Mass-yield distributuion of $^{181}$Ta photofission  
at end-point energy $E_{\gamma max} = 3500$ MeV \cite{Deppman1}. The experimental data are shown by 
the solid squares.}
\end{figure*}

\begin{figure*}[h!]
\includegraphics[width=10cm]{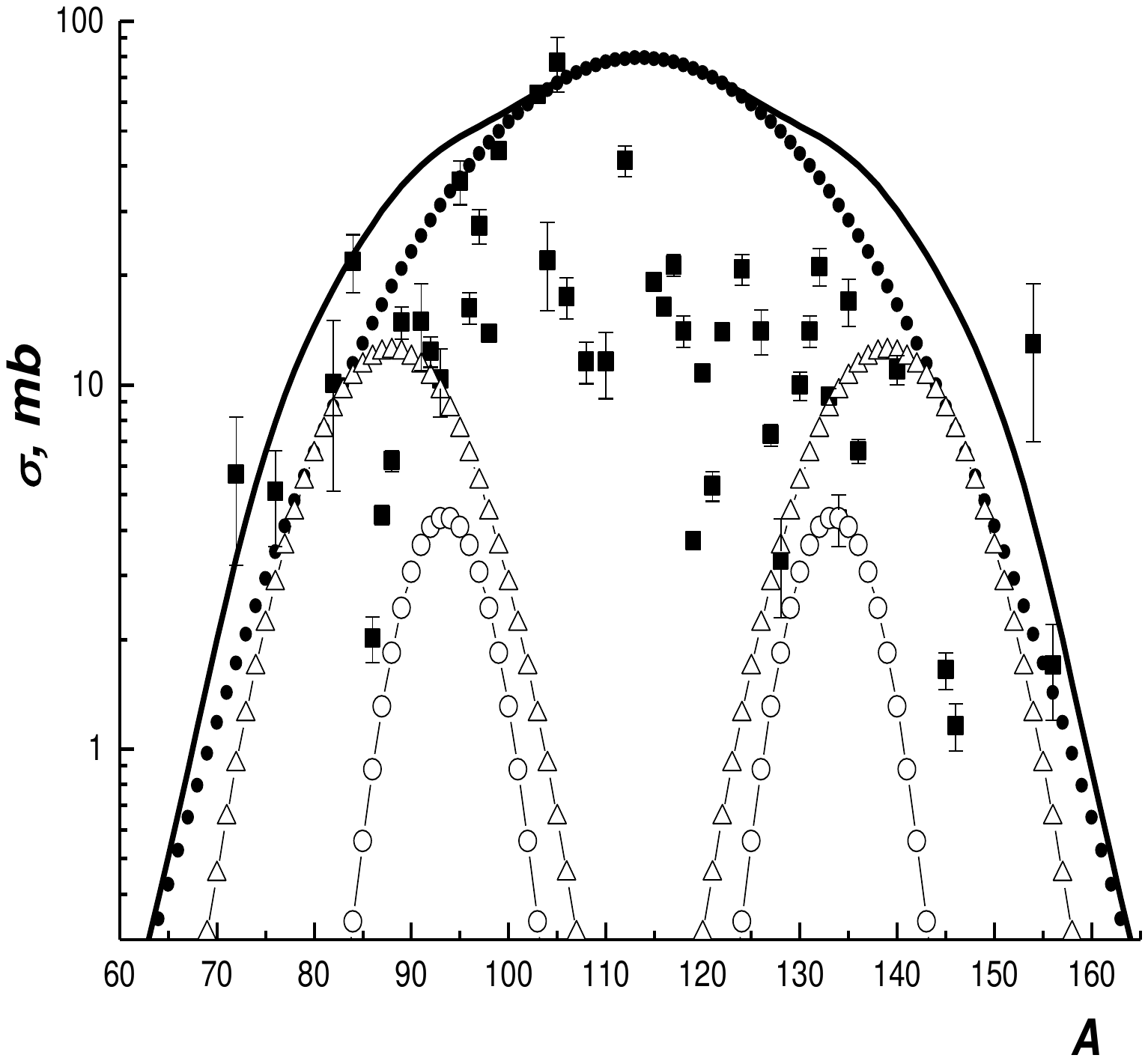}\vspace{-3cm}
\caption{\small Mass-yield distributuion of proton-induced fission of $^{241}$Am 
at the energy 660 MeV \cite{Gaya1}. S mode indicated as closed circles, S1 as open circles, and S2 as open triangles. The total fission yield is represented by the solid curve and experimental data by solid circles, respectively.}
\includegraphics[width=10cm]{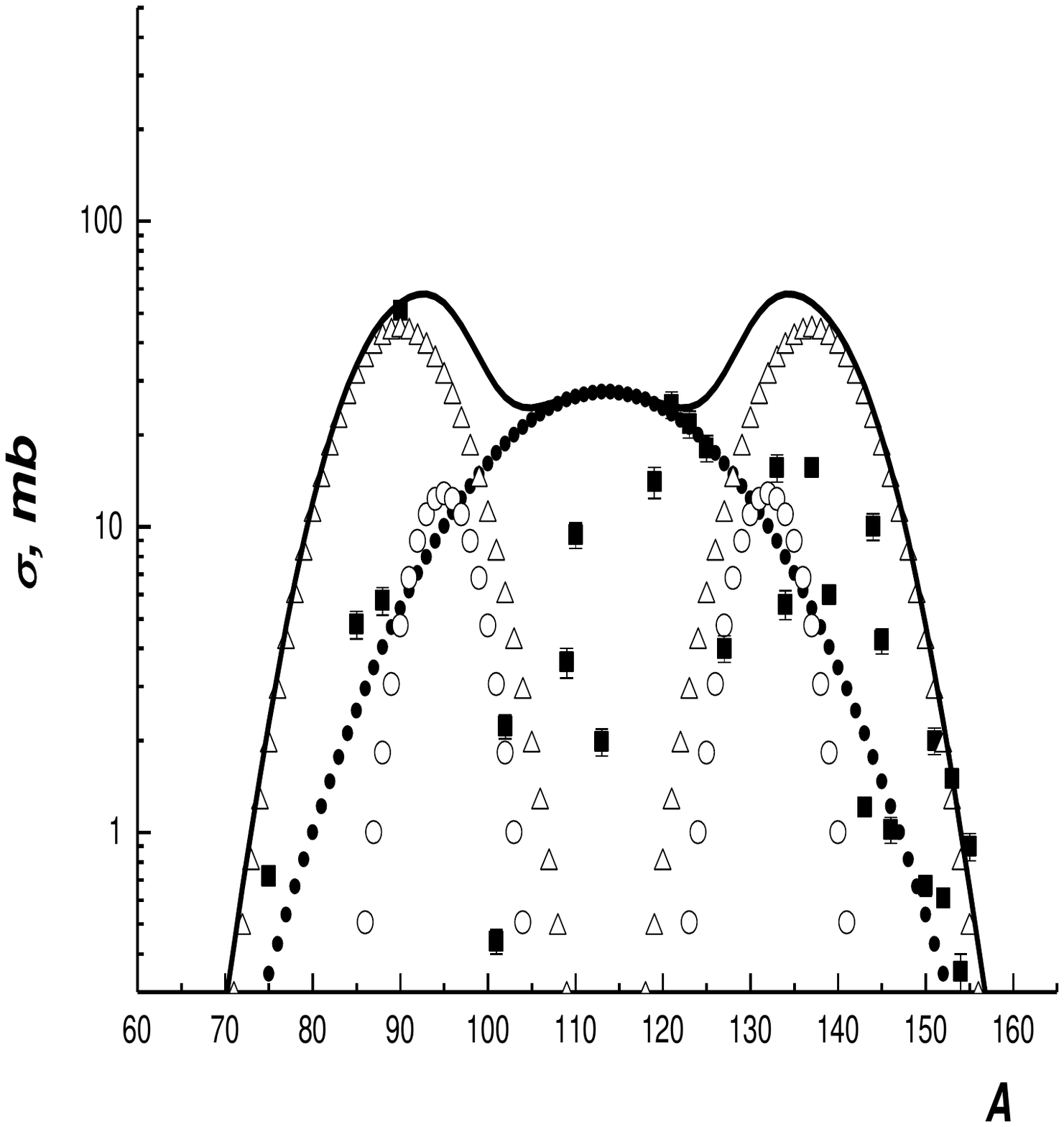}\vspace{-3cm}
\caption{\small Mass-yield distributuion of proton-induced fission of $^{238}$U 
at the energy 660 MeV \cite{Gaya2}. S mode indicated as closed circles, S1 as open circles, and S2 as open triangles. The total fission yield is represented by the solid curve and experimental data by solid circles, respectively.}
\end{figure*}

\newpage
\begin{figure*}[h!]
\includegraphics[width=12cm]{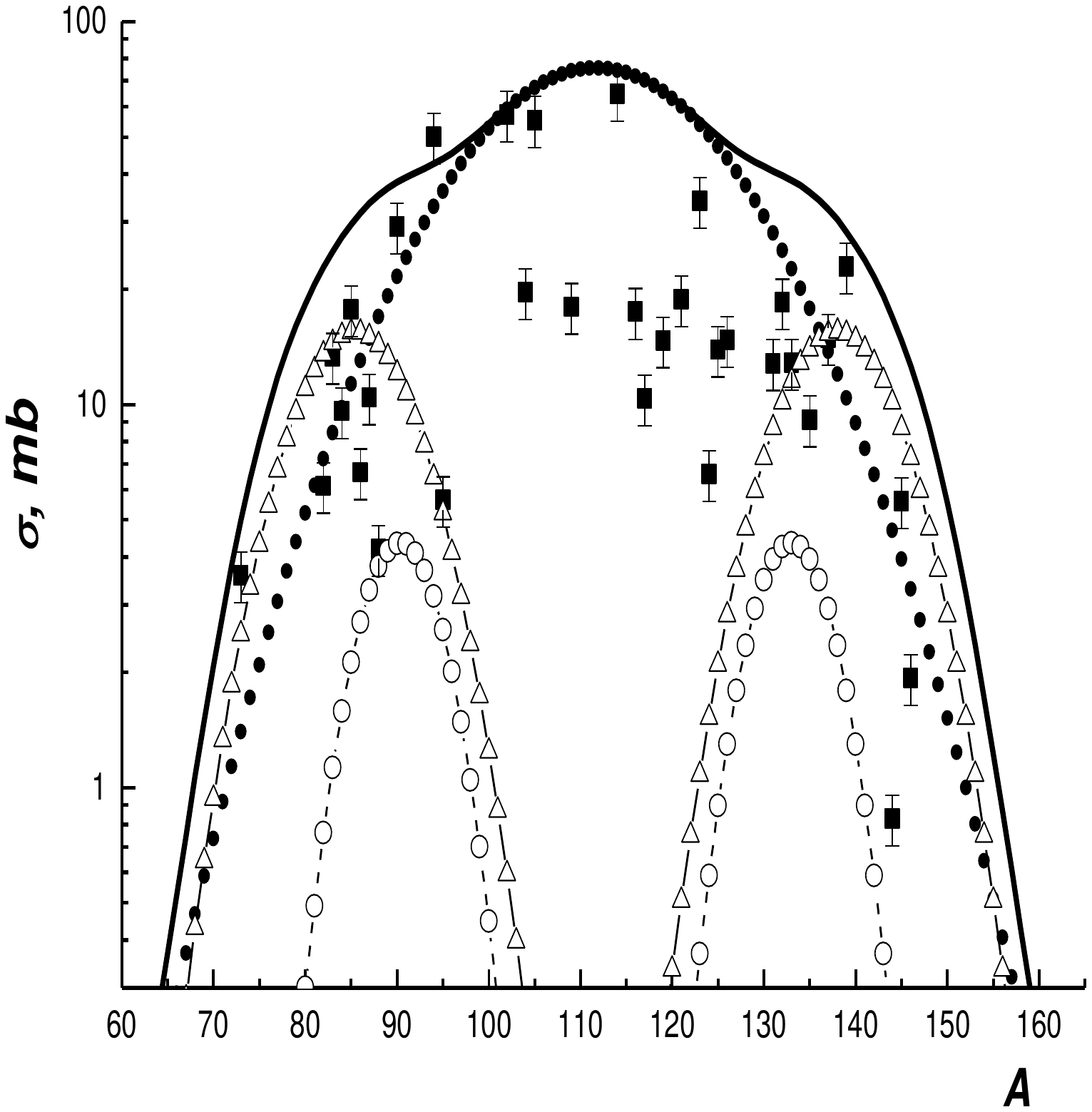}\vspace{-3cm}
\caption{\small Mass-yield distributuion of proton-induced fission of $^{237}$Np 
at the energy 660 MeV \cite{Gaya1}. S mode indicated as closed circles, S1 as open circles, and S2 as open triangles. The total fission yield is represented by the solid curve and experimental data by solid circles, respectively.}
\end{figure*}

\newpage
\begin{figure*}[h!]
\includegraphics[width=10cm]{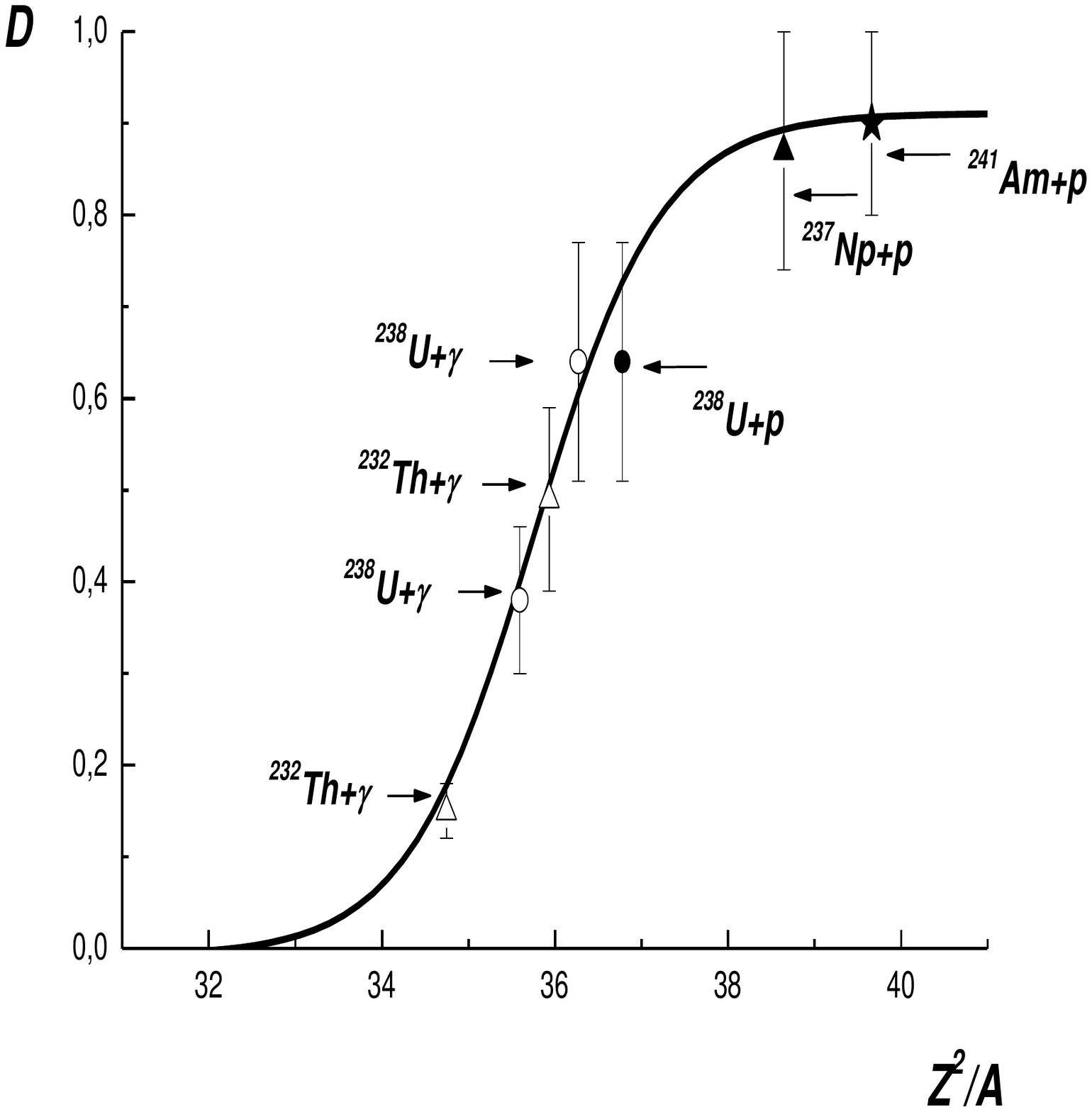}\vspace{-3cm}
\caption{\small Fissility, $D$, as a function of $Z^{2}/A$ for p+$^{237}$Np (solid triangle), p+$^{238}$U (solid circles), p+$^{241}$Am (solid asterisks) at 660 MeV \cite{Gaya1, Gaya2}, and $\gamma$+$^{238}$U (open circles) and $\gamma$+$^{232}$Th (open triangles) at $E_{\gamma max} = 50$ MeV and 3500 MeV\cite{Nina1, Nina2}. The solid line is to guide the eye through the experimental points.}
\includegraphics[width=10cm]{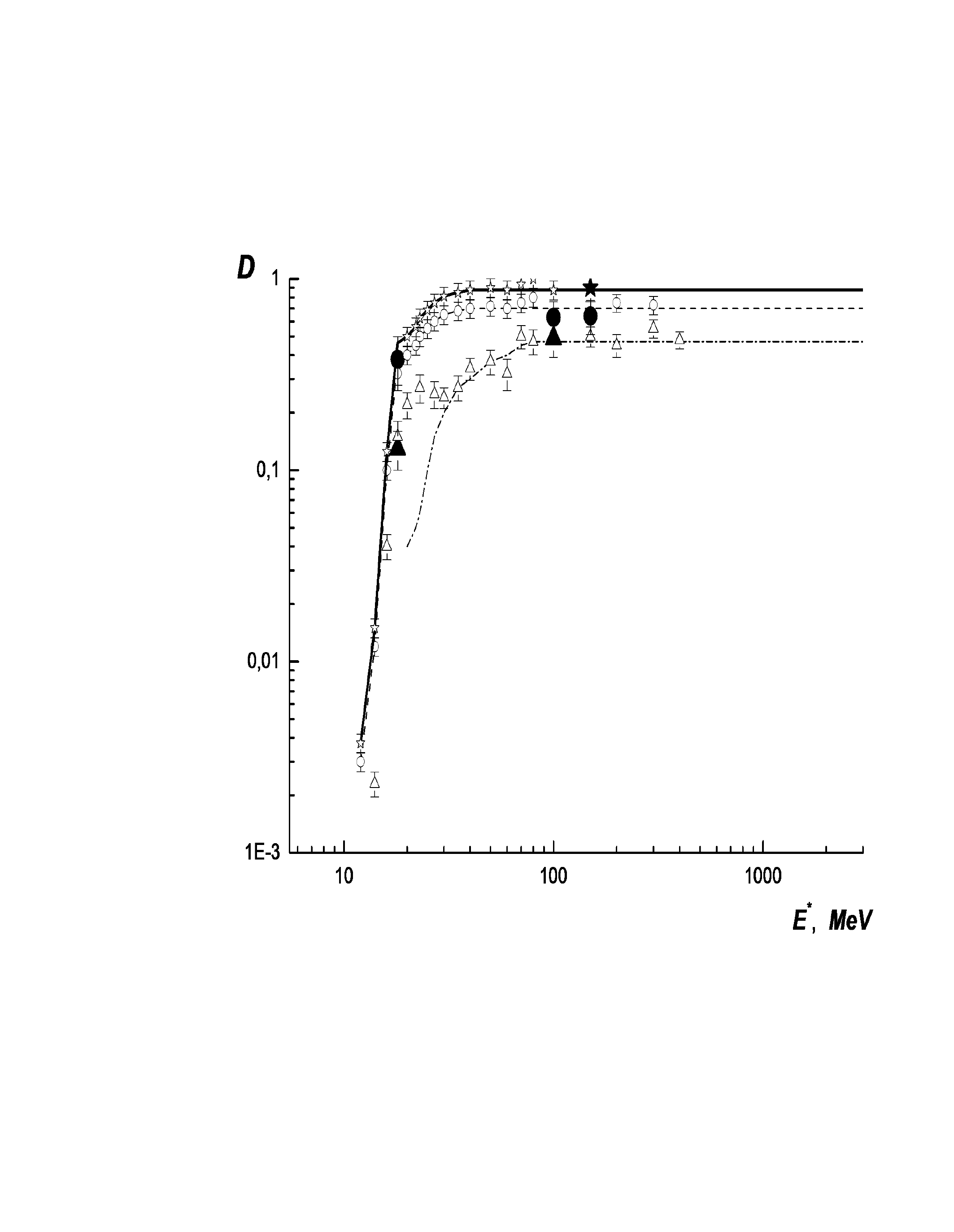}\vspace{-3cm}
\caption{\small Dependence of fissility, $D$, for $^{241}$Am, $^{238}$U, $^{237}$Np and $^{232}$Th targets for photo- and proton-induced fission \cite{Nina1, Nina2, Gaya1, Gaya2} on the excitation energy of fissioning nucleus $E^{*}$. Calculations from \cite{Fukahori}: solid curve is $^{237}$ and $^{237}$Np, dashed curve is $^{238}$U, dash-dot curve is $^{232}$Th. Experimental points: data for $^{241}$Am and $^{237}$Np indicated as solid asterisks (\cite{Gaya1}), $^{238}$U indicated as solid circles (\cite{Nina1, Gaya2}), $^{232}$Th indicated as solid triangles (\cite{Nina2}). Open asterisks, circles and squares are the systematization of experimental data for $^{241}$Am/$^{237}$Np, $^{238}$U and $^{232}$Th, respectively, from \cite{Fukahori}.}
\end{figure*}

\newpage
\begin{figure*}[h!]
\includegraphics[width=16cm]{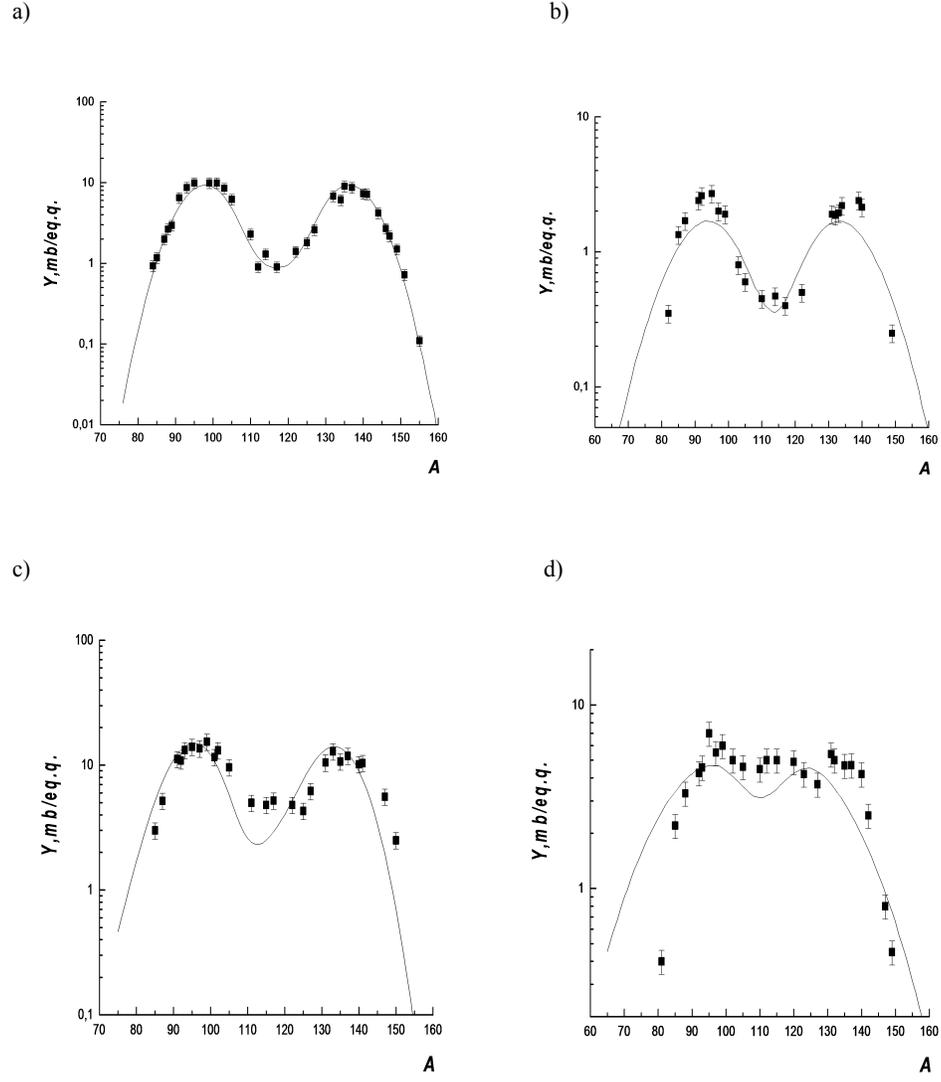}\vspace{-3cm}
\caption{\small Mass-yield distributuion of photofission for $^{238}$U (a, c) and $^{232}$Th (b, d) 
at end-point energies $E_{\gamma max} = 50$ MeV (a, b) and 3500 MeV (c, d). The experimental isobaric cross sections are shown by the solid squares. The solid line is the result of a CRISP calculation \cite{Gaya3}.}
\end{figure*}

\newpage
\begin{figure*}[h!]
\includegraphics[width=16cm]{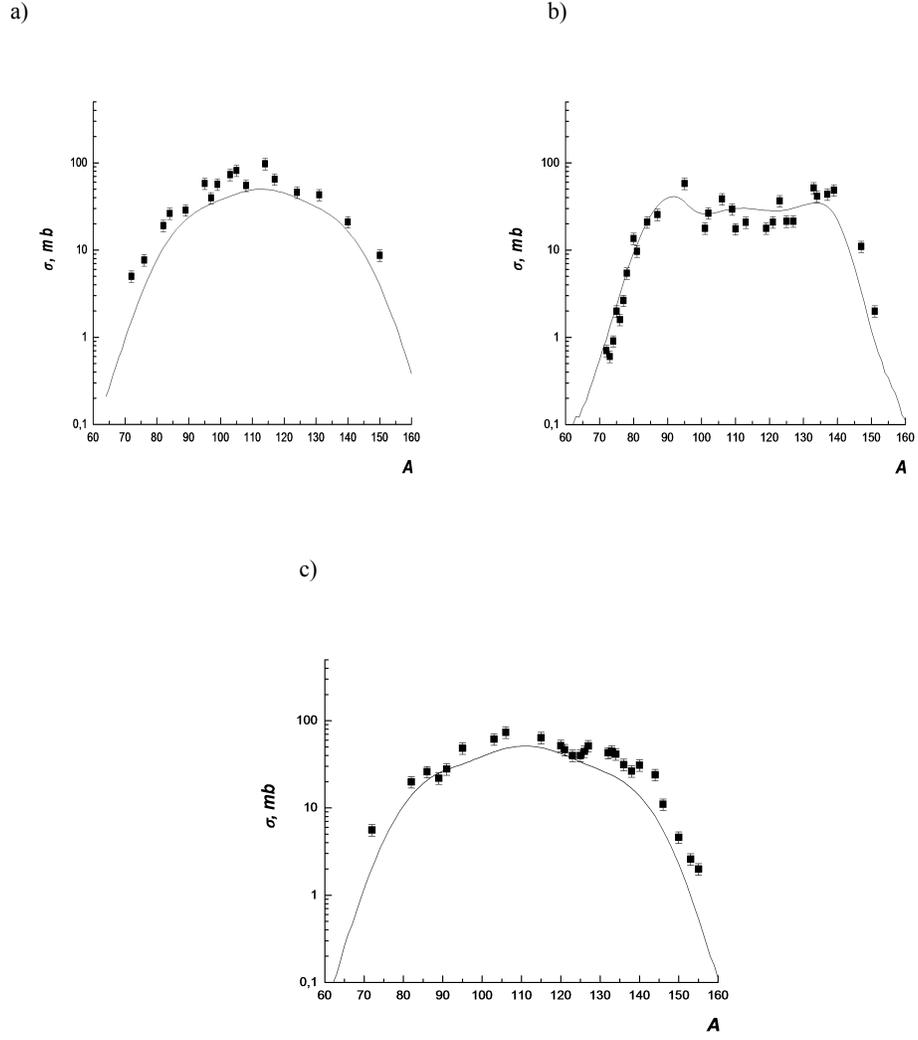}\vspace{-3cm}
\caption{\small Mass-yield distributuion of proton-induced fission for $^{241}$Am (a), $^{238}$U (b) and $^{237}$Np (c) at 660 MeV. The experimental isobaric cross sections are shown by the solid squares. The solid line is the result of a CRISP calculation \cite{Gaya4}.}
\end{figure*}

\newpage
\begin{figure*}[h!]
\includegraphics[width=16cm]{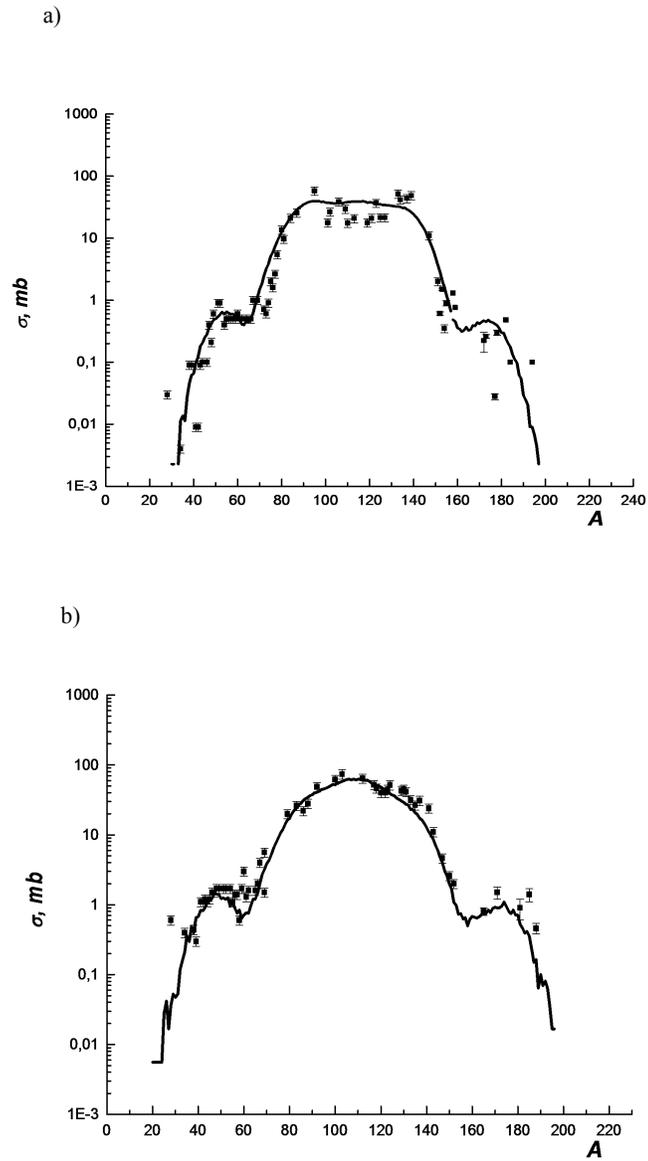}\vspace{-2cm}
\caption{\small Mass-yield distributuion of proton-induced fission for $^{238}$U (a) and $^{237}$Np (b) at 660 MeV. Experimental data for isobaric cross sections are taken from \cite{Gaya1, Gaya2, Gaya5} (solid squares). The solid line is the result of a CRISP calculation \cite{Gaya5}.}
\end{figure*}

\begin{figure*}[h!]
\includegraphics[width=10cm]{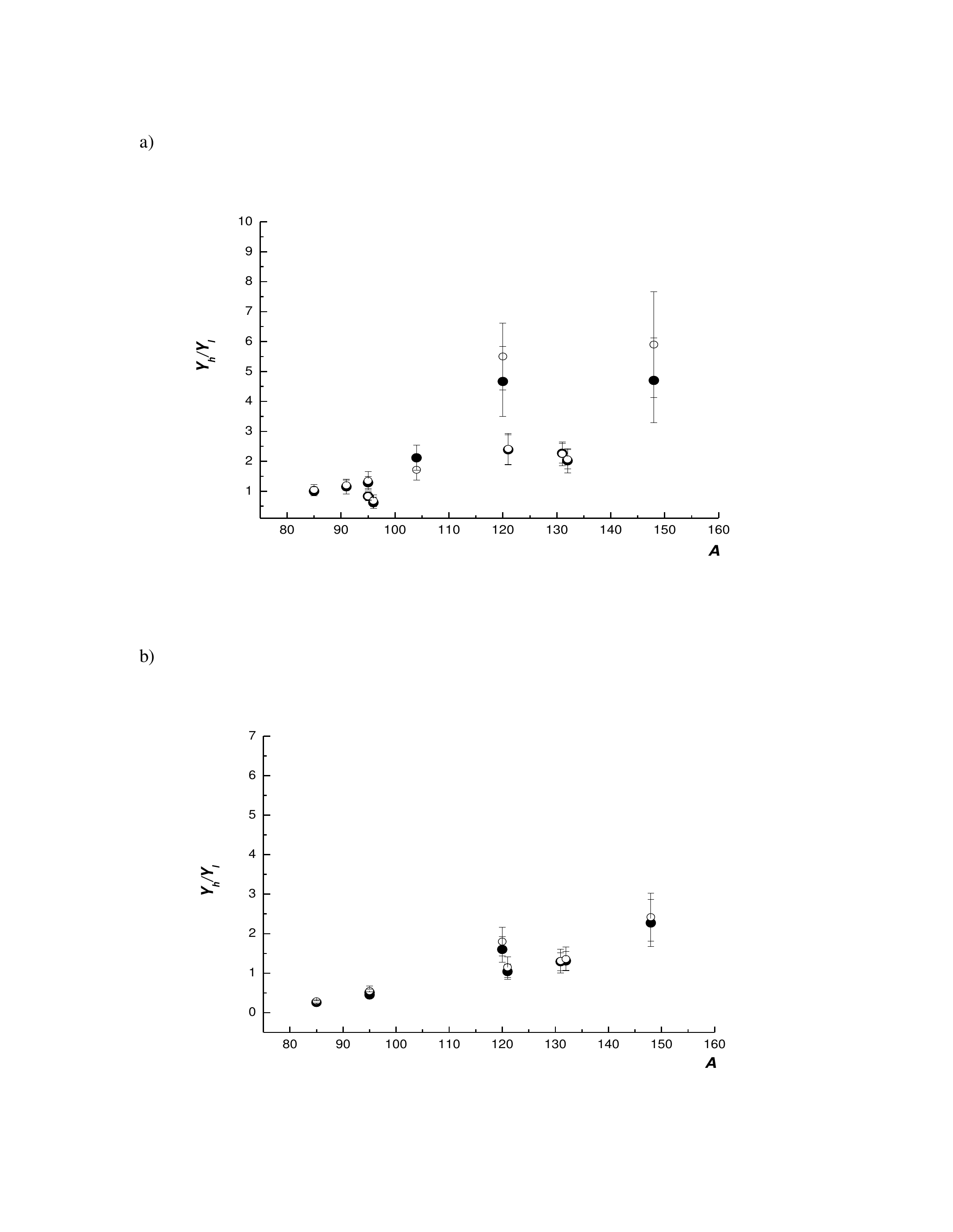}\vspace{-2cm}
\caption{\small Dependence of isomer ratios ($Y_{h}/Y_{l}$) on fragments mass $A$ from $^{238}$U (a) and $^{232}$Th (b) targets at end-point energies $E_{\gamma max} = 50$ MeV (solid circles) and 3500 MeV (open circles) \cite{Gaya-iso}.}
\includegraphics[width=10cm]{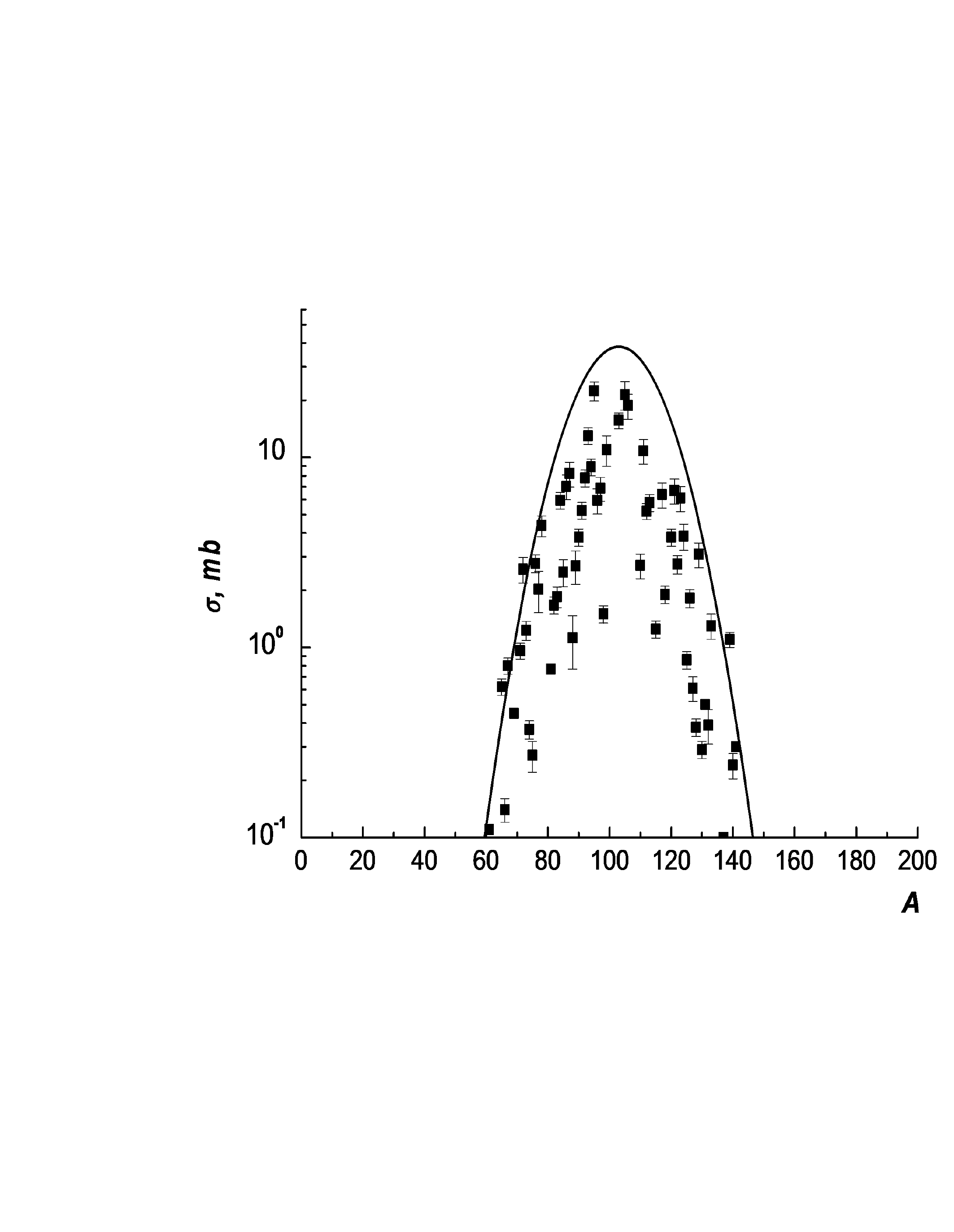}\vspace{-3cm}
\caption{\small Fission-product mass yields from 245 MeV $^{7}$Li-induced fission on $^{nat}$Pb: the total fission cross section (thick black continuous curve), experimental data are shown by 
the solid square symbol \cite{Gaya6, Gaya7}.}
\end{figure*}

\begin{figure*}[h!]
\includegraphics[width=10cm]{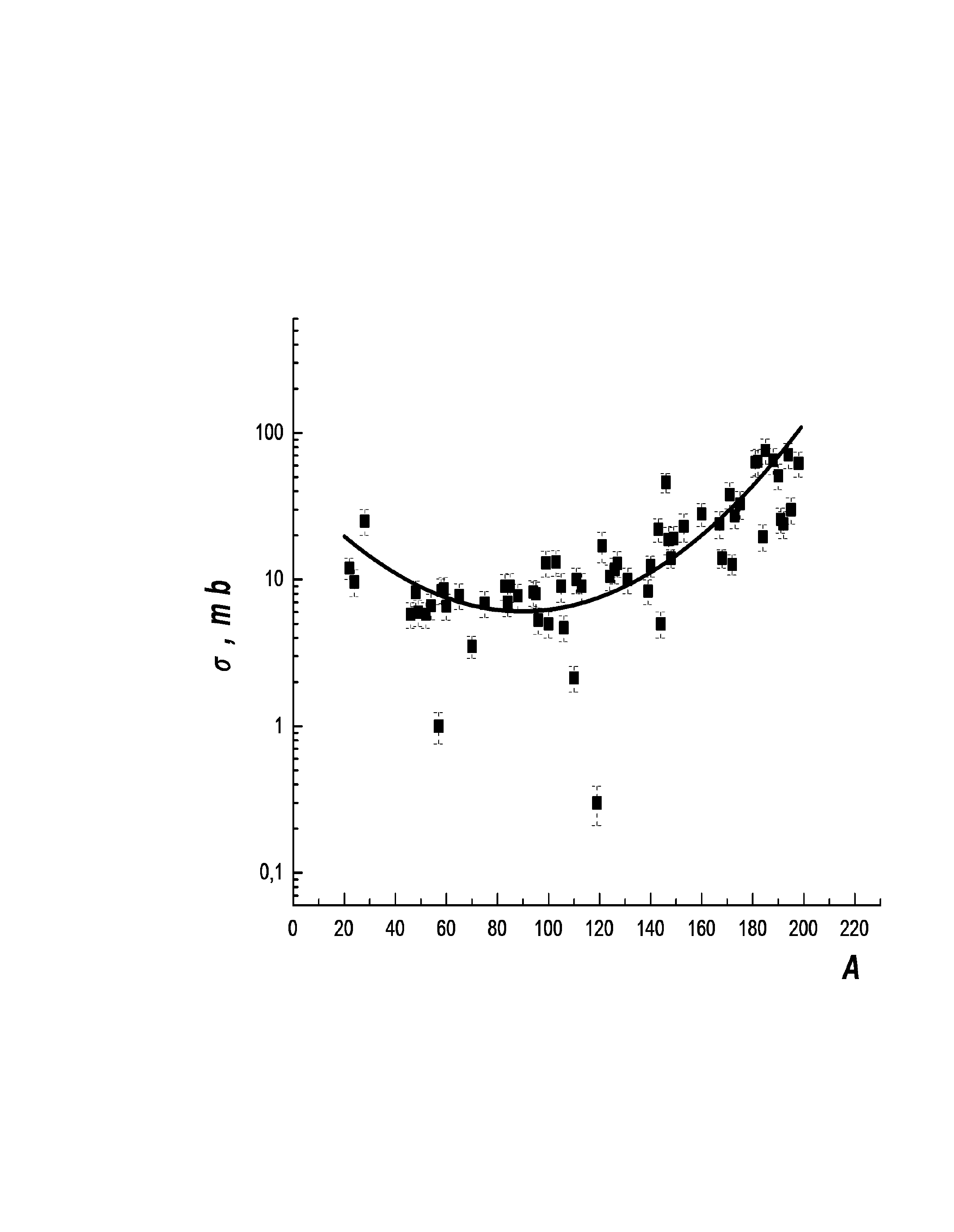}\vspace{-3cm}
\caption{\small Mass-yield distribution of the isobars produced by the
4.4 GeV deuteron-induced reaction on $^{197}$Au \cite{Gaya8}. The solid line shows the
general trend of the isobaric cross sections of the reaction fragments, experimental data are shown by 
the solid square symbol.} 
\includegraphics[width=10cm]{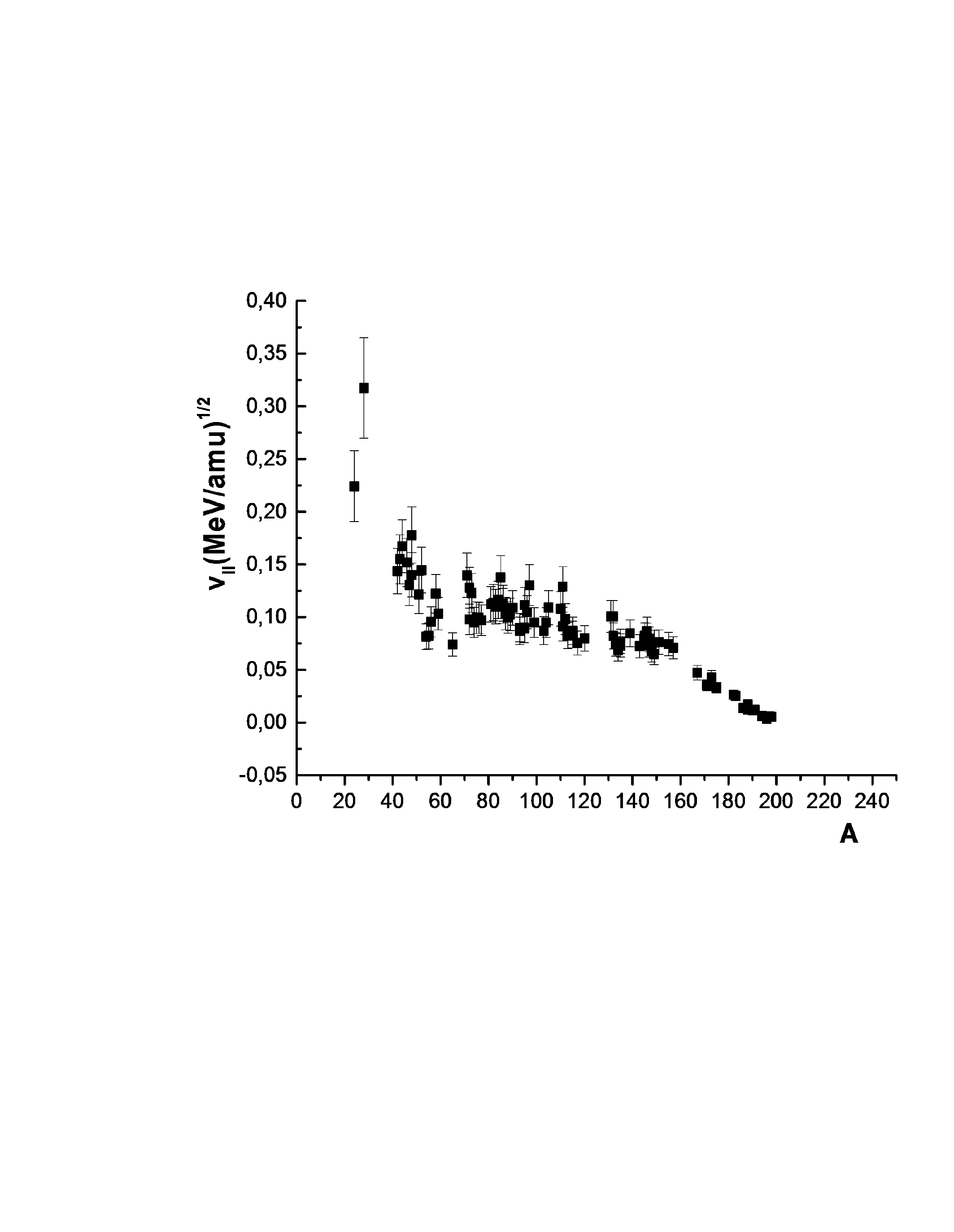}\vspace{-3cm}
\caption{\small The longitudinal cascade velocity, $v_{\parallel}$, as a
function of the fragment mass number, $A$, for the 4.4 GeV deuteron-induced reaction on $^{197}$Au \cite{Gaya9}.} 
\end{figure*}

\newpage
\begin{table}[!ht]\centering
\caption{Experimental values obtained for the
mass-yield distributions of $^{238}$U, $^{232}$Th, and $^{181}$Ta in the interaction of bremsstrahlung at 50 MeV and 3500 MeV \cite{Nina1, Nina2, Deppman1}. The total fission yield ($Y_{tot}$); the contribution of symmetric ($Y_{S}$) and asymmetric ($Y_{AS}$) fission to the total mass-yield; the values of the asymmetric-to-symmetric ratios ($Y_{AS}/Y_{S}$).}
\begin{tabular}{|c|c|c|c|c|c|c|}  \hline
Yield, mb/eq.q.&\multicolumn{2}{|c|}{$^{238}$U}&\multicolumn{2}{|c|}{$^{232}$Th}&\multicolumn{2}{|c|}{$^{181}$Ta}\\
\cline{2-7} &$E_{max}=50$ MeV&$E_{max}=3500$ MeV&$E_{max}=50$ MeV&$E_{max}=3500$ 
MeV&$E_{max}=50$ MeV&$E_{max}=3500$ MeV\\
\hline Y$_{tot}$& 131.6$\pm$19.5 & 250.1$\pm$37.5 & 40.2$\pm$6.0 & 137.5$\pm$20.6 &0.0054$\pm$0.0011 &0.77$\pm$0.11\\
\hline Y$_{S}$& 10.6$\pm$1.5 & 79.3$\pm$11.9 & 4.50$\pm$0.68 & 74.50$\pm$11.17 &0.0054$\pm$0.0011 &0.77$\pm$0.11\\
\hline Y$_{AS}$& 121.0$\pm$18 & 170.8$\pm$25.6 & 35.70$\pm$5.36 & 63.0$\pm$9.45 &-- &--\\
\hline Y$_{AS}/Y_{S}$&11.41$\pm$1.71 & 2.16$\pm$0.4 & 7.93$\pm$1.59 & 0.84$\pm$0.17 &-- &--\\
\hline
\end{tabular}
\end{table}

\begin{table}[!ht]\centering
\caption{Experimental values obtained for the
mass-yield distributions of $^{241}$Am, $^{238}$U and $^{237}$Np in the interaction of 660 MeV protons \cite{Gaya1, Gaya2}. The total fission cross section ($\sigma_{tot}$); the contribution of symmetric ($\sigma_{S}$) and asymmetric ($\sigma_{AS}$) fission to the total mass-yield; the values of the asymmetric-to-symmetric ratios ($\sigma_{S}/\sigma_{AS}$).}
\begin{tabular}{|c|c|c|c|c|}  \hline
Target & \multicolumn{3}{|c|}{Cross section, mb} & $\sigma_{S}/\sigma_{AS}$\\
\cline{2-4} &$\sigma_{tot}$&$\sigma_{S}$&$\sigma_{AS}$& \\
\hline $^{241}$Am& 1763.7$\pm$265.0 & 1487.7$\pm$223.0 & 276.0$\pm$41.0 & 5.4$\pm$1.0\\
\hline $^{238}$U& 1226.5$\pm$183.9 & 698.3$\pm$104.7 & 528.2$\pm$79.2 & 1.3$\pm$0.2\\
\hline $^{237}$Np& 1600.2$\pm$240.0 & 1298.0$\pm$195.0 & 302.2$\pm$45.0 & 4.3$\pm$1.0\\
\hline
\end{tabular}
\end{table}

\begin{table}[!ht]\centering
\caption{Experimental and calculation values of IR and angular momentum ($\bar{B}$) obtained in photofission of $^{238}$U and $^{232}$Th at end-point energy E$_{\gamma max} = 50$ MeV \cite{Gaya-iso}.}
\begin{tabular}{|c|c|c|c|c|c|c|c|}  \hline
Element & State & \multicolumn{3}{|c|}{$^{238}$U}&\multicolumn{3}{|c|}{$^{232}$Th}\\
\cline{3-8}& &$(Y_{h}/Y_{l})_{exp}$ &$(Y_{h}/Y_{l})_{cal}$& B($\hbar$)&$(Y_{h}/Y_{l})_{exp}$ &$(Y_{h}/Y_{l})_{cal}$& $\bar{B}$($\hbar$)\\
\hline $^{85}$Sr& g(${(9/2)^+}$) & 0.60$\pm$0.09 & 0.57$\pm$0.2 & 2.4$\pm$0.5 & 0.26$\pm$0.04 & 0.24$\pm$0.06 & 2.5$\pm$0.4\\
& m(${(1/2)^+}$) &  &  &  &  &  & \\
\hline $^{91}$Y& g(${(1/2)^-}$) & 0.69$\pm$0.15 & 0.72$\pm$0.30 & 2.6$\pm$0.3 & -- & -- & --\\
& m(${(1/2)^+}$) &  &  &  &  &  & \\
\hline $^{95}$Nb& g(${(9/2)^+}$) & 0.50$\pm$0.09 & 0.49$\pm$0.15 & 2.3$\pm$0.4 & 0.46$\pm$0.08 & 0.51$\pm$0.08 & 3.0$\pm$0.5\\
& m(${(1/2)^-}$) &  &  &  &  &  & \\
\hline $^{95}$Tc& g(${(9/2)^+}$) & 0.77$\pm$0.12 & 0.72$\pm$0.30 & 2.6$\pm$0.3 & 0.51$\pm$0.09 & 0.60$\pm$0.09 & 2.8$\pm$0.4\\
& m(${(1/2)^-}$) &  &  &  &  &  & \\
\hline $^{96}$Tc& g(${7^+}$) & 0.37$\pm$0.11 & 0.39$\pm$0.10 & 4.9$\pm$0.2 & -- & -- & --\\
& m(${4^+}$) &  &  &  &  &  & \\
\hline $^{104}$Ag& g(${5^+}$) & 1.27$\pm$0.2 & 1.34$\pm$0.30 & 5.3$\pm$0.2 & -- & -- & --\\
& m(${2^+}$) &  &  &  &  &  & \\
\hline $^{120}$I& g(${2^+}$) & 2.80$\pm$0.70 & 2.60$\pm$0.39 & 5.8$\pm$0.9 & 1.60$\pm$0.32 & 1.61$\pm$0.40 & 6.4$\pm$1.3\\
& m(${(4--8)^+}$) &  &  &  &  &  & \\
\hline $^{121}$Te& g(${(1/2)^+}$) & 1.43$\pm$0.30 & 1.35$\pm$0.18 & 4.2$\pm$0.5 & 1.04$\pm$0.19 & 1.16$\pm$0.23 & 4.3$\pm$0.9\\
& m(${(11/2)^-}$) &  &  &  &  &  & \\
\hline $^{131}$Te& g(${(3/2)^+}$) & 1.36$\pm$0.20 & 1.39$\pm$0.30 & 4.0$\pm$0.7 & 1.29$\pm$0.23 & 1.31$\pm$0.22 & 4.7$\pm$0.8\\
& m(${(11/2)^-}$) & 1.08$\pm$0.25 \cite{Jacobs} &  & 4.1$\pm$1.5 \cite{Jacobs}&  &  & \\
&  & 1.38$\pm$0.21 \cite{Vishnevskii} &  & 5.1$\pm$0.4 \cite{Vishnevskii}&  &  & \\
\hline $^{132}$I& g(${4^+}$) & 1.21$\pm$0.24 & 1.2$\pm$0.14 & 7.2$\pm$0.2 & 1.31$\pm$0.24 & 1.60$\pm$0.32 & 7.5$\pm$1.5\\
& m(${8^-}$) & 1.08$\pm$0.13 \cite{Jacobs} &  & 6.9$\pm$1.4 \cite{Jacobs}&  &  & \\
\hline $^{148}$Pm& g(${1^-}$) & 2.82$\pm$0.85 & 3.02$\pm$0.60 & 7.5$\pm$1.5 & 2.27$\pm$0.60 & 2.21$\pm$0.44 & 7.8$\pm$1.6\\
& m(${6^-}$) & 2.6$\pm$0.9 \cite{Aumann} &  & 10.0$\pm$2.5 \cite{Aumann}&  &  & \\
\hline
\end{tabular}
\end{table}

\end{document}